\documentclass{amsart} 

\usepackage{graphicx}
\usepackage{amsmath}
\usepackage{hyperref}
\usepackage{gensymb} 

\usepackage{natbib}
\let\vec=\mathbf 

\newcommand{\unit}[1]{\hat{\vec{#1}}} 

\newcommand{\tensor}[1]{\mathcal{#1}} 

\begin{document} 
	
	\title[Representation of light pressure resultant force\dots]{Representation of light pressure resultant force and moment as a tensor series}

	%
	\author{Nikolay Nerovny} 
	
	\address{Bauman Moscow State Technical University, 5 stroenie 1, 2-ya Baumanskaya st., Moscow, 105005, Russian Federation}
	\email{nick.nerovny@bmstu.ru}
	
	\author{Vladimir Zimin} 
	
	\address{Bauman Moscow State Technical University, 5 stroenie 1, 2-ya Baumanskaya st., Moscow, 105005, Russian Federation}
	\email{zimin@bmstu.ru}

	\author{Sergey Fedorchuk} 
	
	\address{Astro Space Center of P.N. Lebedev Physical Institute, 84/32 Profsoyuznaya st., Moscow, GSP-7, 117997, Russian Federation}
	\email{fedorchuk@asc.rssi.ru}

	\author{Evgeny Golubev} 
	
	\address{Astro Space Center of P.N. Lebedev Physical Institute, 84/32 Profsoyuznaya st., Moscow, GSP-7, 117997, Russian Federation}
	\email{golubev.ev@asc.rssi.ru}

	

	\begin{abstract}
		In this article we address a problem of determination of light pressure upon space structures of the complex geometric shape. For the surface element, we wrote a condition that this element can interact with light only from the front side, not from the back side. This condition To in the form of series of Chebyshev polynomials of the first kind. Chebyshev expansion lets us move to the series of tensors of increasing rank for the problem of determination of force and moment. We obtained expressions for the fiber method for determination of light pressure on space structures of complex geometry taking into account self--shadowing and reflections within the structure. We also give the expressions for tensor parametrization by the specularity coefficient in case of specular-diffuse reflection. For these expressions we calculated the principal moment and force upon two-sided flat solar sail, spherical and cylindrical bodies, and approximated light pressure upon a perspective space observatory Millimetron. The proposed expressions can be used in the ballistic analysis of solar sails and other space objects, which are significantly affected by the radiation pressure. Also, these results can be used to analyze the dynamics of movement of large-scale space structures around the center of gravity under the light pressure. 
	\end{abstract} 

	\maketitle

\allowdisplaybreaks

	\section{Introduction} \label{intro} 
	
	This work\footnote{This work was prepared during development of BMSTU-Sail Space Experiment~\cite{bmstu_iac_2011,bmstu_knts}} is related to the calculation of the light pressure force on space structures. The phenomenon of light pressure with radiations by J. Maxwell based on his theory of electromagnetism \cite{maxwell_1873}. The light radiation pressure was found by P. Lebedev \cite{lebedev_1901} in the series of experiments. The first one who suggested using the force of light pressure to fly in space was F. Tsander in the mid 20-s of the 20 century \cite{tsander}.
	Regarding solar sails, there is a sufficiently complete overview of the current state of solar sails development in the book \cite{mcinnes_solar_2004}. 
	Up to the present moment several experiments regarding solar sail technology were conducted in space, including Znamya-1 and Znamya-2~\cite{raikunov_2009}, Nanosail-D2~\cite{alhorn_nanosail-d_2011,johnson_nanosail-d:_2011}, Lightsail-1~\cite{ridenoure2015lightsail}, IKAROS~\cite{tsuda_flight_2011,ikaros1}.
	
	Solar radiation pressure (SRP) is affecting celestial bodies providing various effects~\cite{burns_radiation_1979,burns_radiation_2014}.
	One of these effects is the phenomenon called Yarkovsky effect~\cite{neiman_ivan_1965,opik_collision_1951,beekman_i.o._2006}.
	It is originated from the anisotropy of heat radiation emission from heated surface of celestial body~\cite{radzievskii_about_1952} and provides the additional moment which causes acceleration or deceleration of spinning~\cite{paddack_rotational_1969,rubincam_radiative_2000} and also affecting the orbital parameters~\cite{rubincam_asteroid_1995}.
	For small bodies (e.g. asteroids) it can increase the spinning velocity up to the disintegration by centrifugal forces~\cite{radzievskii_mechanism_1954}.
	There are several good analytical models of SRP on celestial bodies, including Vokrouhlick\'{y}'s model~\cite{vokrouhlicky_yarkovsky_1998,vokrouhlicky_complete_1999,vokrouhlicky_yorp-induced_2002}, other models by~\cite{katasev_physical_1981,hartmann_reviewing_1999,william_f._bottke_yarkovsky_2006} etc.

	It is necessary to study the effects of SRP for the precise orbit propagation.
	There are several studies related to research of solar radiation effects on GPS satellites~\cite{bar-sever_new_1997,bar-sever_new_2004,rodriguez-solano_adjustable_2012,fliegel_global_1992,fliegel_solar_1996,springer_new_1999}
	It is important to analyze the angular movement around the center of inertia for space structures in distant space far from Earth's atmosphere in low gravity gradient conditions because moment from light radiation can affect the attitude of spacecrafts~\cite{kinzel_managing_2005}.

	It is possible to create a more exact model of SRP by numerical integration of SRP equations around the surface represented as a mesh of simple elements~\cite{ziebart_generalized_2004}.
		
	During the development of the theory of light pressure acting on space objects some authors got the result according to which in the equations for the resultant force and moment of light pressure on arbitrary body it is possible to separate the description of structure's surface from its orientation \cite{rios-reyes_applications_2004,rios_reyes_generalized_2005,rios-reyes_2006,rios-reyes_solar-sail_2007,scheeres_dynamical_2007,mcmahon_new_2010,mcmahon_general_2014,mcmahon_improving_2015}. These authors developed a so-termed tensor approach for the description of light pressure. Tensor approach is also being developed in the works \cite{jing_curved_2012,jing_solar_2014}. This work was done using the tensor approach for the description of the light pressure force, and it is a continuation of the author's previous research \cite{Nerovny_Zimin_2014,Zimin_Nerovny_2015_izvuz,Zimin_Nerovny_2016_bmstu}. 
	
	\section{Light pressure on an infinitesimal surface element} \label{sec:1} 
	
	As it was shown in work \cite{rios-reyes_applications_2004} it is possible to write analytical expressions for the resultant force and moment of light pressure forces on the spacecraft surface under certain assumptions. Now we will derive the expression for the resultant vector and resultant moment of an arbitrary convex structure. 
	
	Let us write the expression for the light pressure force on the infinitesimal surface element in general form: \begin{equation} d\vec{F}=P(R)\left[ -a_0\unit{n}-a_1(\unit{n}\cdot\unit{s})\unit{s} + a_2(\unit{n}\cdot\unit{s})\unit{n} -2a_3(\unit{n}\cdot\unit{s})^2\unit{n} \right]dA,\label{eq:dF_full} \end{equation} where generalized optical parameters are given as the following: \begin{align*} &a_0 = \frac{\varepsilon B \sigma T^4}{P(R)};\\ &a_1 = 1 - \rho s;\\ &a_2 = B\rho(1-s);\\ &a_3 = \rho s. \end{align*} 
	
	In this case, the following symbols are used: \begin{itemize} \item $P(R)=q_0/c$ --- light pressure on a flat diffuse area normal to the incident radiation at a distance $R$ from the Sun, which is assumed to be a point source, and the light flux at the distance $R$ is $q_0$, where $c$ is a speed of light in vacuum; \item $\unit{n}$ --- local unit vector of normal to the surface of space structure; \item $\unit{s}$ --- unit vector defining the direction from the light source to the surface in the spacecraft coordinate system; \item $\varepsilon$ --- emissivity of the surface element; \item $\sigma$ --- Boltzmann constant; \item $T$ --- temperature of surface element; \item $B$ --- coefficient equal to $2/3$ for diffuse scattering surface (makes sense only for the materials without anisotropy of optical parameters depending on direction along the surface), it is often called as Lambertian coefficient; \item $\rho$ --- total reflectivity of the surface material (spectral reflectivity integrated by wavelength from 0 to $\infty$); \item $s$ --- specularity coefficient showing which part of the radiation is reflected glossy, and what -- diffusely. \end{itemize} 
	
	In the other studies (e.g. \cite{rios-reyes_applications_2004}) there was an implicit assumption that the lightning configuration does not change with time i.e. there is no change in the set of lighted and shadowed surfaces of spacecraft. In the subsequent derivations for the convex structure we assume that the structure can take an arbitrary orientation relative to the incident solar radiation. Taking into account of this condition is similar to replacement in~\eqref{eq:dF_full} of expression $\unit{n}\cdot\unit{s}$ by the following relationship $V(\unit{n},\unit{s})$: \[ V(\unit{n},\unit{s}) = \frac{\unit{n}\cdot\unit{s} - |\unit{n}\cdot\unit{s}|}{2}. \] 
	
	The above statement can be interpreted as a visibility function in the works \cite{scheeres_dynamical_2007}. \cite[equation (15)]{mcmahon_new_2010}. 
	
	In point of fact, if $\unit{n}\cdot\unit{s}>0$, it turns out that the area is illuminated from the internal side which cannot be. Let us write an expression~\eqref{eq:dF_full} with this new relationship: \begin{align*} d\vec{F}=\frac{P(R)}{2}\bigg( -2a_0\unit{n} - a_1(\unit{n}\cdot\unit{s} - |\unit{n}\cdot\unit{s}|)\unit{s} + a_2 (\unit{n}\cdot\unit{s} - |\unit{n}\cdot\unit{s}|) \unit{n} \\ - a_3 (\unit{n}\cdot\unit{s} - |\unit{n}\cdot\unit{s}|)^2\unit{n} \bigg)dA, \end{align*} and after transformations, \begin{align} d\vec{F}=\frac{P(R)}{2}\bigg( -2a_0\unit{n} - a_1(\unit{n}\cdot\unit{s})\unit{s} + a_2(\unit{n}\cdot\unit{s})\unit{n} - 2a_3(\unit{n}\cdot\unit{s})^2\unit{n} + a_1|\unit{n}\cdot\unit{s}|\unit{s} \nonumber\\ - a_2|\unit{n}\cdot\unit{s}|\unit{n} + 2a_3(\unit{n}\cdot\unit{s})|\unit{n}\cdot\unit{s}|\unit{n} \bigg)dA.\label{eq:dF_abs} \end{align} 
	
	Since the range of the $\unit{n}\cdot\unit{s}$ is $[-1;1]$, then $|\unit{n}\cdot\unit{s}|$ may be expanded in the well-known series of Chebyshev polynomials of the first kind: \begin{align} |\unit{n}\cdot\unit{s}| = \frac{2}{\pi} - \frac{4}{\pi}\sum\limits_{n=1}^{\infty}\frac{(-1)^n T_{2n}(\unit{n}\cdot\unit{s})}{-1+4n^2}=\nonumber\\ =-\frac{4}{\pi}\sum\limits_{n=1}^{\infty}\sum\limits_{k=0}^{n-1}\frac{(-1)^n(-1)^k n (2n-k-1)!}{(-1+4n^2)k!(2n-2k)!}4^{n-k}(\unit{n}\cdot\unit{s})^{2(n-k)}. \label{eq:chebyshev_abs} \end{align} 
	
	The main purpose of Chebyshev expansion is the representation of non-analytical visibility function as a power series of $\unit{n}\cdot\unit{s}$. After this, we can utilize the method originated from works by D.J. Scheeres, L. Rios-Reyes, and others~\cite{rios-reyes_applications_2004} by transforming of multiple scalar and vector products into the assembly of dyadic for $\unit{n}$ and several scalar products of $\unit{s}$. The original method of D.J. Scheeres and others provides an analytical separation of attitude from the description of the surface, so one can calculate some tensor parameters once in the beginning and then calculate the resultant force and moment using simple scalar products by orientation vector which allows using this method in the restricted calculation environments (e.g. in spacecraft's onboard computer) or as a part of dynamics model of spacecraft.
	
	After transformations we can write the series \eqref{eq:chebyshev_abs} as follows: \begin{equation} |\unit{n}\cdot\unit{s}| = \sum\limits_{m=1}^{\infty} B_m (\unit{n}\cdot\unit{s})^{2m},\label{eq:chebyshev_short} \end{equation} where \begin{equation} B_m = -\frac{(-1)^m 4^{m+1}}{\pi(2m)!}\sum\limits_{n=m}^{\infty}\frac{n(n+m-1)!}{(-1+4n^2)(n-m)!}.\label{eq:chebyshev_b} \end{equation} 
	
	The series in the expression \eqref{eq:chebyshev_b} diverge, however for any finite number of terms of the series in the final summation \eqref{eq:chebyshev_short} we have a convergent sequence. Hereafter for the infinite sums we will consider that they either constrained by the maximum number of terms $N_{\max}$, or renormalization has been made if appropriate. Analysis of possible renormalization of these series is beyond the scope of this article. If we constrain the number of terms in the series \eqref{eq:chebyshev_short} by $N_{\max}$, the upper bound for $B_m$ series \eqref{eq:chebyshev_b} will be as follows: \[ N_{\max B} = \left\lfloor \frac{N_{\max}-1}{2}\right\rfloor, \] where $\lfloor x \rfloor$ is a floor function of real $x$. We can write the approximate equation: \begin{equation} B_m \approx -\frac{(-1)^m 4^{m+1}}{\pi(2m)!}\sum\limits_{n=m}^{\left\lfloor \frac{N_{\max}-1}{2}\right\rfloor}\frac{n(n+m-1)!}{(-1+4n^2)(n-m)!}.\label{eq:chebyshev_b_approx} \end{equation}
	
	By substituting expression \eqref{eq:chebyshev_short} in \eqref{eq:dF_abs} we can obtain an expression for the infinitesimal force of light pressure as a series form: \begin{align} d\vec{F} = \frac{P(R)}{2}\bigg( -2a_0\unit{n} - a_1(\unit{n}\cdot\unit{s})\unit{s} + a_2 (\unit{n}\cdot\unit{s})\unit{n} - 2a_3 (\unit{n}\cdot\unit{s})^2\unit{n} +\nonumber\\ a_1\sum\limits_{m=1}^{{N_{\max}}}B_m(\unit{n}\cdot\unit{s})^{2m}\unit{s} - a_2\sum\limits_{m=1}^{{N_{\max}}}B_m(\unit{n}\cdot\unit{s})^{2m}\unit{n} + 2a_3\sum\limits_{m=1}^{{N_{\max}}}B_m(\unit{n}\cdot\unit{s})^{2m+1}\unit{n}\bigg) dA.\label{eq:dF_series} \end{align} 
	
	\section{The resultant vector and moment of light pressure forces} \label{sec:2} 
	
	Using the approach from work~\cite{rios-reyes_applications_2004} we can write the following relations: \begin{align*} &\underbrace{(\unit{n}\cdot\unit{s})\ldots(\unit{n}\cdot\unit{s})}_{p}\unit{n}=(\underbrace{\unit{n}\otimes\unit{n}\otimes\ldots\otimes\unit{n}}_{p+1})\cdot\underbrace{\unit{s}\cdot\ldots\cdot\unit{s}}_{p} = \tensor{J}_A^{p+1}\cdot\underbrace{\unit{s}\cdot\ldots\cdot\unit{s}}_{p};\\ &\underbrace{(\unit{n}\cdot\unit{s})\ldots(\unit{n}\cdot\unit{s})}_{p}\unit{s}=(\underbrace{\unit{n}\otimes\unit{n}\otimes\ldots\otimes\unit{n}}_{p}\otimes\tensor{E}^2)\cdot\underbrace{\unit{s}\cdot\ldots\cdot\unit{s}}_{p+1} = \tensor{J}_B^{p+2}\cdot\underbrace{\unit{s}\cdot\ldots\cdot\unit{s}}_{p+1}, \end{align*} where $\tensor{J}_A^{p+1}$ и $\tensor{J}_B^{p+2}$ -- some tensors in Euclidean space with a rank of $p+1$ и $p+2$, respectively. 
	
	Let us write \eqref{eq:dF_series} in the new tensor notation: \begin{align} d\vec{F} = \frac{P(R)}{2}\bigg(-2a_0\unit{n} - a_1\tensor{J}_B^3\cdot\unit{s}\cdot\unit{s} + a_2\tensor{J}_A^2\cdot\unit{s}-2a_3\tensor{J}_A^3\cdot\unit{s}\cdot\unit{s}+\nonumber\\ a_1\sum\limits_{m=1}^{{N_{\max}}}B_m\tensor{J}_B^{2m+2}\cdot\underbrace{\unit{s}\cdot\ldots\cdot\unit{s}}_{2m+1} - a_2\sum\limits_{m=1}^{{N_{\max}}}B_m\tensor{J}_A^{2m+1}\cdot\underbrace{\unit{s}\cdot\ldots\cdot\unit{s}}_{2m}+\nonumber\\ 2a_3\sum\limits_{m=1}^{{N_{\max}}}B_m\tensor{J}_A^{2m+2}\cdot\underbrace{\unit{s}\cdot\ldots\cdot\unit{s}}_{2m+1}\bigg)dA.\label{eq:dF_tensor_series} \end{align} 
	
	By grouping members of infinite sums in \eqref{eq:dF_tensor_series} with each other and by introducing additional notation we can obtain an expression for the radiation pressure force on unit area which receives only the radiation that falls on the front side: \begin{equation} d\vec{F} = P(R)\bigg( \tensor{J}^1 + \sum\limits_{n=2}^{{N_{\max}}}\tensor{J}^n\cdot\underbrace{\unit{s}\cdot\ldots\cdot\unit{s}}_{n-1} \bigg)dA,\label{eq:dF_final} \end{equation} where \begin{align*} &\tensor{J}^1 = -a_0\unit{n};\\ &\tensor{J}^2 = \frac{1}{2}a_2\tensor{J}_A^2;\\ &\tensor{J}^3 = \frac{1}{2}\bigg( -a_1\tensor{J}_B^3 -2a_3\tensor{J}_A^3 - B_1a_2\tensor{J}_A^3 \bigg);\\ &\tensor{J}^n = \frac{1}{2}\bigg( -B_{\frac{n-1}{2}}\frac{1-(-1)^n}{2}a_2\tensor{J}_A^n + B_{\frac{n-2}{2}}\frac{1+(-1)^n}{2}\left( a_1\tensor{J}_B^n + 2a_3\tensor{J}_A^n \right) \bigg),\ n>3;\\ &\tensor{J}_A^n = \underbrace{\unit{n}\otimes\ldots\otimes\unit{n}}_{n};\\ &\tensor{J}_B^n = \underbrace{\unit{n}\otimes\ldots\otimes\unit{n}}_{n-2}\otimes\tensor{E}^2, \end{align*} where $\tensor{E}^2$ --- unit tensor of second rank. 
	
	Let introduce the tensor $\tensor{R}^2$ as follows: \begin{equation} \vec{r}\times\mathbf{a}=\tensor{R}^2\cdot\mathbf{a}.\label{R} \end{equation} 
	
	Obviously, the tensor $\tensor{R}^2$ from (\ref{R}) is skew-symmetric and can be represented in the following matrix form: \[ \tensor{R}^2=\begin{pmatrix} 0 &-r_3 &r_2 \\ r_3 &0 &-r_1 \\ -r_2 &r_1 &0 \end{pmatrix}. \] 
	
	For the infinitisemal moment we will use the following equations: \begin{align*} &\underbrace{(\unit{n}\cdot\unit{s})\ldots(\unit{n}\cdot\unit{s})}_{p}(\tensor{R}^2\cdot\unit{n})=(\underbrace{\unit{n}\otimes\unit{n}\otimes\ldots\otimes\unit{n}}_{p}\otimes\tensor{R}^2\cdot\unit{n})\cdot\underbrace{\unit{s}\cdot\ldots\cdot\unit{s}}_{p} = \tensor{L}_A^{p+1}\cdot\underbrace{\unit{s}\cdot\ldots\cdot\unit{s}}_{p};\\ &\underbrace{(\unit{n}\cdot\unit{s})\ldots(\unit{n}\cdot\unit{s})}_{p}(\tensor{R}^2\cdot\unit{s})=(\underbrace{\unit{n}\otimes\unit{n}\otimes\ldots\otimes\unit{n}}_{p}\otimes\tensor{R}^2)\cdot\underbrace{\unit{s}\cdot\ldots\cdot\unit{s}}_{p+1} = \tensor{L}_B^{p+2}\cdot\underbrace{\unit{s}\cdot\ldots\cdot\unit{s}}_{p+1}. \end{align*} 
	
	By writing \eqref{eq:dF_series} in the new tensor notation we can get: \begin{align} d\vec{M} =\vec{r}\times d\vec{F} = \tensor{R}^2\cdot d\vec{F}= \frac{P(R)}{2}\bigg(-2a_0\tensor{R}^2\cdot\unit{n} - a_1\tensor{L}_B^3\cdot\unit{s}\cdot\unit{s} + a_2\tensor{L}_A^2\cdot\unit{s}-\nonumber\\ 2a_3\tensor{L}_A^3\cdot\unit{s}\cdot\unit{s}+a_1\sum\limits_{m=1}^{{N_{\max}}}B_m\tensor{L}_B^{2m+2}\cdot\underbrace{\unit{s}\cdot\ldots\cdot\unit{s}}_{2m+1} - a_2\sum\limits_{m=1}^{{N_{\max}}}B_m\tensor{L}_A^{2m+1}\cdot\underbrace{\unit{s}\cdot\ldots\cdot\unit{s}}_{2m}+\nonumber\\ 2a_3\sum\limits_{m=1}^{{N_{\max}}}B_m\tensor{L}_A^{2m+2}\cdot\underbrace{\unit{s}\cdot\ldots\cdot\unit{s}}_{2m+1}\bigg)dA,\label{eq:dM_tensor_series} \end{align} where \begin{align} &\tensor{L}_A^n = \underbrace{\unit{n}\otimes\ldots\otimes\unit{n}}_{n-1}\otimes\tensor{R}^2\cdot\unit{n}; \\ &\tensor{L}_B^n = \underbrace{\unit{n}\otimes\ldots\otimes\unit{n}}_{n-2}\otimes\tensor{R}^2. \end{align} 
	
	After transformations similar to those we did in the derivation of the expression of the force, we can obtain an expression for the moment from infinitesimal force of light pressure: \begin{equation} d\vec{M}= P(R)\bigg( \tensor{L}^1 + \sum\limits_{n=2}^{{N_{\max}}}\tensor{L}^n\cdot\underbrace{\unit{s}\cdot\ldots\cdot\unit{s}}_{n-1} \bigg)dA,\label{eq:dM_final} \end{equation} where \begin{align*} &\tensor{L}^1 = -a_0(\tensor{R}^2\cdot\unit{n});\\ &\tensor{L}^2 = \frac{1}{2}a_2\tensor{L}_A^2;\\ &\tensor{L}^3 = \frac{1}{2}\bigg( -a_1\tensor{L}_B^3 -2a_3\tensor{L}_A^3 - B_1a_2\tensor{L}_A^3 \bigg);\\ &\tensor{L}^n = \frac{1}{2}\bigg( -B_{\frac{n-1}{2}}\frac{1-(-1)^n}{2}a_2\tensor{L}_A^n + B_{\frac{n-2}{2}}\frac{1+(-1)^n}{2}\left( a_1\tensor{L}_B^n + 2a_3\tensor{L}_A^n \right) \bigg),\ n>3; \end{align*} 
	
	By integrating expression \eqref{eq:dF_final} and \eqref{eq:dM_final} over the entire surface let us introduce the following values: \begin{align} &\tensor{I}^n = \int\limits_A\tensor{J}^n dA;\\ &\tensor{K}^n = \int\limits_A\tensor{L}^n dA, \end{align} where $n>0$. 
	
	Finally we can get the expressions for the resultant force and moment of light pressure on a body with convex shape: 
	
	\begin{align} &\vec{F} = P(R)\bigg( \tensor{I}^1 + \sum\limits_{n=2}^{{N_{\max}}}\tensor{I}^n\cdot\underbrace{\unit{s}\cdot\ldots\cdot\unit{s}}_{n-1} \bigg);\label{eq:F_full}\\ &\vec{M} = P(R)\bigg( \tensor{K}^1 + \sum\limits_{n=2}^{{N_{\max}}}\tensor{K}^n\cdot\underbrace{\unit{s}\cdot\ldots\cdot\unit{s}}_{n-1} \bigg).\label{eq:M_full} \end{align} 
	
	\section{Light pressure on the structure of the arbitrary shape} \label{sec:3} 
	
	Considering the problem of the calculation of the radiation pressure force the structure with complex geometry taking into account self-shadowing and reflections in structure, we should note the fact that this problem is related to the problem of radiative heat transfer in a complex structure, which has been shown multiple times by different authors, it cannot be solved in unified general form for the structure of arbitrary configuration.~\cite{howell_thermal_2015}. Nevertheless, there are some approximate methods to calculate the radiative heat transfer in complex structures from the view factor algebra to the Monte--Carlo methods (ibid.). 
	
	This problem is also close to the problems faced in computer graphics in the analysis of global illumination \cite{liu_all-frequency_2004,tsai_all-frequency_2006,kristensen_precomputed_2005,maki-patola_precomputed_2003,sloan_precomputed_2002}. In these works authors used a method of calculation of global lightning with precomputed bidirectional radiation distribution function (BRDF) for each element of the surface. After the BRDF calculation, one can dynamically change the ambient light luminance map and calculate the brightness of each surface element. Moreover, if the construction of BRDF is a complex computational problem, the process of restoring of luminance of surface of the object with existing BRDF is less resource consuming. 
	
	Given the above, we present the main assumptions used for the development of the calculation method for determination of the light pressure on the structure of complex geometry. 
	
	\newtheorem{mydef}{Assumption} 
	
	\begin{mydef} We assume that the expressions of infinitesimal force and moment on the body with the complex shape can also be written in a form similar to \eqref{eq:dF_final} and~\eqref{eq:dM_final} respectively. \end{mydef} This statement will be briefly called as the assumption of similarity of notation. 
	
	\begin{mydef} Since some surface elements can be illuminated not by direct sunlight but by light reflected and re-emitted by structure, we will assume that the vector of direction from the sun $\vec{s}$ is not a constant unit vector and it is dependent on the position of infinitesimal area $dA$. \end{mydef} This statement will be briefly called as an assumption of variable light flux. Vector~$\vec{s}$ will be called as a vector of a local light load.
	
	\begin{mydef} BRDF will be approximated by a second-rank tensor $\vec{s}=\tensor{D}^2\cdot\unit{s}$, where the unit vector $\unit{s}$ is constant for the entire surface. \end{mydef} This proposal will be called as a distribution assumption. 
	
	\begin{mydef} During calculation of BRDF we assume that the vector $\vec{s}$ is zero if there are no such vector $\unit{s}$ for which the area $dA$ is been illuminated. \end{mydef} 
	
	\begin{mydef} When taking into account the light pressure from thermal radiation from structure we assume that the magnitude of force is also associated with the BRDF, $\vec{f}\propto\tensor{D}^2\cdot\unit{n}$. \end{mydef} 
	
	In the distribution assumption it is possible to introduce the high--order approximation, however, we decided to simplify the problem by limiting the approximation by the tensor of a second rank. Thus, the need to use higher-order approximation remains still open. It should be noted that the expressions \eqref{eq:dF_final} and~\eqref{eq:dM_final} are already contain features, according to which the area element $dA$ does not receive radiation incident from the internal side. 
	
	By analogy with section \ref{sec:2} we can transform the vector expressions in \eqref{eq:dF_final} and~\eqref{eq:dM_final} assuming approximation \begin{equation} \vec{s}=\tensor{D}^2\cdot\unit{s}. \end{equation} 
	
	We can obtain the following expressions: \begin{align*} (\unit{n}\cdot(\tensor{D}^2\cdot\unit{s}))^{p}\unit{n}=(\underbrace{\underbrace{\unit{n}\cdot\tensor{D}^2}\otimes\ldots\otimes\underbrace{\unit{n}\cdot\tensor{D}^2}}_{p}\otimes\unit{n})\cdot\underbrace{\unit{s}\cdot\ldots\cdot\unit{s}}_{p}=\\ =\tilde{\tensor{J}}_A^{p+1}\cdot\underbrace{\unit{s}\cdot\ldots\cdot\unit{s}}_{p};\\ (\unit{n}\cdot(\tensor{D}^2\cdot\unit{s}))^{p}(\tensor{D}^2\cdot\unit{s})=(\underbrace{\underbrace{\unit{n}\cdot\tensor{D}^2}\otimes\ldots\otimes\underbrace{\unit{n}\cdot\tensor{D}^2}}_{p}\otimes\tensor{D}^2)\cdot\underbrace{\unit{s}\cdot\ldots\cdot\unit{s}}_{p+1}=\\ =\tilde{\tensor{J}}_B^{p+2}\cdot\underbrace{\unit{s}\cdot\ldots\cdot\unit{s}}_{p+1};\\ (\unit{n}\cdot(\tensor{D}^2\cdot\unit{s}))^{p}(\tensor{R}^2\cdot\unit{n})=(\underbrace{\underbrace{\unit{n}\cdot\tensor{D}^2}\otimes\ldots\otimes\underbrace{\unit{n}\cdot\tensor{D}^2}}_{p}\otimes\tensor{R}^2\cdot\unit{n})\cdot\underbrace{\unit{s}\cdot\ldots\cdot\unit{s}}_{p}=\\ =\tilde{\tensor{L}}_A^{p+1}\cdot\underbrace{\unit{s}\cdot\ldots\cdot\unit{s}}_{p};\\ (\unit{n}\cdot(\tensor{D}^2\cdot\unit{s}))^{p}(\tensor{R}^2\cdot(\tensor{D}^2\cdot\unit{s}))=(\underbrace{\underbrace{\unit{n}\cdot\tensor{D}^2}\otimes\ldots\otimes\underbrace{\unit{n}\cdot\tensor{D}^2}}_{p}\otimes\tensor{D}^2\cdot\tensor{R}^2)\cdot\underbrace{\unit{s}\cdot\ldots\cdot\unit{s}}_{p+1}=\\ =\tilde{\tensor{L}}_B^{p+2}\cdot\underbrace{\unit{s}\cdot\ldots\cdot\unit{s}}_{p+1}. \end{align*} 
	
	Since the tensors $\tilde{\tensor{J}}_A$, $\tilde{\tensor{J}}_B$, $\tilde{\tensor{L}}_A$, $\tilde{\tensor{L}}_B$ are introduced as well as tensors $\tensor{J}_A$, $\tensor{J}_B$, $\tensor{L}_A$, $\tensor{L}_B$ respectively, the final expressions for the infinitesimal force and moment for the surface area element $dA$ will look like the follows: \begin{align} d\vec{F} = P(R)\bigg( \tilde{\tensor{J}}^1 + \sum\limits_{n=2}^{{N_{\max}}}\tilde{\tensor{J}}^n\cdot\underbrace{\unit{s}\cdot\ldots\cdot\unit{s}}_{n-1} \bigg)dA;\label{eq:dF_approx}\\ d\vec{M} = P(R)\bigg( \tilde{\tensor{L}}^1 + \sum\limits_{n=2}^{{N_{\max}}}\tilde{\tensor{L}}^n\cdot\underbrace{\unit{s}\cdot\ldots\cdot\unit{s}}_{n-1} \bigg)dA,\label{eq:dM_approx} \end{align} where \begin{align} &\tilde{\tensor{J}}^1 = -a_0\tensor{D}^2\cdot\unit{n};\label{eq:method_start}\\ &\tilde{\tensor{J}}^2 = \frac{1}{2}a_2\tilde{\tensor{J}}_A^2;\\ &\tilde{\tensor{J}}^3 = \frac{1}{2}\bigg( -a_1\tilde{\tensor{J}}_B^3 -2a_3\tilde{\tensor{J}}_A^3 - B_1a_2\tilde{\tensor{J}}_A^3 \bigg);\\ &\tilde{\tensor{J}}^n = \frac{1}{2}\bigg( -B_{\frac{n-1}{2}}\frac{1-(-1)^n}{2}a_2\tilde{\tensor{J}}_A^n + B_{\frac{n-2}{2}}\frac{1+(-1)^n}{2}\left( a_1\tilde{\tensor{J}}_B^n + 2a_3\tilde{\tensor{J}}_A^n \right) \bigg),\ n>3;\\ &\tilde{\tensor{J}}_A^n = \underbrace{\underbrace{\unit{n}\cdot\tensor{D}^2}\otimes\ldots\otimes\underbrace{\unit{n}\cdot\tensor{D}^2}}_{n-1}\otimes\unit{n};\\ &\tilde{\tensor{J}}_B^n = \underbrace{\underbrace{\unit{n}\cdot\tensor{D}^2}\otimes\ldots\otimes\underbrace{\unit{n}\cdot\tensor{D}^2}}_{n-2}\otimes\tensor{D}^2;\\ &\tilde{\tensor{L}}^1 = -a_0(\tensor{D}^2\cdot\tensor{R}^2\cdot\unit{n});\\ &\tilde{\tensor{L}}^2 = \frac{1}{2}a_2\tilde{\tensor{L}}_A^2;\\ &\tilde{\tensor{L}}^3 = \frac{1}{2}\bigg( -a_1\tilde{\tensor{L}}_B^3 -2a_3\tilde{\tensor{L}}_A^3 - B_1a_2\tilde{\tensor{L}}_A^3 \bigg);\\ &\tilde{\tensor{L}}^n = \frac{1}{2}\bigg( -B_{\frac{n-1}{2}}\frac{1-(-1)^n}{2}a_2\tilde{\tensor{L}}_A^n + B_{\frac{n-2}{2}}\frac{1+(-1)^n}{2}\left( a_1\tilde{\tensor{L}}_B^n + 2a_3\tilde{\tensor{L}}_A^n \right) \bigg),\ n>3;\\ &\tilde{\tensor{L}}_A^n = \underbrace{\underbrace{\unit{n}\cdot\tensor{D}^2}\otimes\ldots\otimes\underbrace{\unit{n}\cdot\tensor{D}^2}}_{n-1}\otimes\tensor{R}^2\cdot\unit{n};\\ &\tilde{\tensor{L}}_B^n = \underbrace{\underbrace{\unit{n}\cdot\tensor{D}^2}\otimes\ldots\otimes\underbrace{\unit{n}\cdot\tensor{D}^2}}_{n-2}\otimes\tensor{D}^2\cdot\tensor{R}^2;\\ &B_m = -\frac{(-1)^m 4^{m+1}}{\pi(2m)!}\sum\limits_{n=m}^{\left\lfloor \frac{N_{\max}-1}{2}\right\rfloor}\frac{n(n+m-1)!}{(-1+4n^2)(n-m)!}. \end{align} 
	
	By integrating \eqref{eq:dF_approx} and \eqref{eq:dM_approx} across the surface $A$ we can obtain the final expression for the resultant force and moment of light pressure on the spacecraft with complex geometric shape: \begin{align} &\vec{F} = P(R)\bigg( \tilde{\tensor{I}}^1 + \sum\limits_{n=2}^{{N_{\max}}}\tilde{\tensor{I}}^n\cdot\underbrace{\unit{s}\cdot\ldots\cdot\unit{s}}_{n-1} \bigg);\label{eq:F_final}\\ &\vec{M} = P(R)\bigg( \tilde{\tensor{K}}^1 + \sum\limits_{n=2}^{{N_{\max}}}\tilde{\tensor{K}}^n\cdot\underbrace{\unit{s}\cdot\ldots\cdot\unit{s}}_{n-1} \bigg),\label{eq:M_final} \end{align} where \begin{align} &\tilde{\tensor{I}}^n = \int\limits_A\tilde{\tensor{J}}^n dA;\label{eq:method_end_0}\\ &\tilde{\tensor{K}}^n = \int\limits_A\tilde{\tensor{L}}^n dA,\label{eq:method_end} \end{align} where $n>0$. 
	
	Expressions \eqref{eq:method_start} --- \eqref{eq:method_end} should be supplemented by the procedure of calculation of tensor approximation $\tensor{D}^2$: \begin{equation} \tensor{D}^2 = \tensor{D}^2(x_1,x_2,x_3).\label{eq:D_definition} \end{equation} 
	
	In case of convex shape $\tensor{D}^2 = \tensor{E}^2$.
	In this work, we will not discuss the calculation techniques for determination of BRDF, some of them can be derived from works by~\cite{kristensen_precomputed_2005} and~\cite{tsai_all-frequency_2006} by the transition from spherical harmonics representation of BRDF to the tensor representation. The above definitions are aimed to create the equations of resultant force and moment in which the parameters of the structure are analytically separated from its attitude towards the sun, eq.~\eqref{eq:F_final} and~\eqref{eq:M_final}. For the given equations we will develop the approximation method of SRP in the section~\ref{sec:4} and show the numerical example results in the section~\ref{seq:numerical}.
	
	Expressions \eqref{eq:method_start} --- \eqref{eq:D_definition} give us a full statement of the approximate method for determination of resultant vector and moment of light radiation pressure upon a structure with complex geometry taking into account self-shadowing, multipath reflection, a variability of the optical parameters and temperature over the surface of the structure. The reduced pressure from reflected light is incorporated into the BRDF as far as tensor $\tensor{D}^2$ can generate the non-unit vector $\vec{s}$ of light load. The self-shadowing is considering in two ways: in the Chebyshev expansion of visibility function and BRDF $\tensor{D}^2$. The surface properties can be specular, diffuse, specular diffuse, as well as other rotationally symmetrical scattering law other than the Lambert law can be used. 
	
	Note that due to the physical meaning of the BRDF it cannot create infinite local light load vector from the finite sun vector load as this would mean the infinite concentration of energy in the structure which is impossible to achieve. The components of the tensor $D ^ 2_{ij}$ always have a finite value for the whole area $A$. On the other hand the approximation tensor $\tensor{D}^2$ is included in the expression for the $\tilde{\tensor{J}}_A$, $\tilde{\tensor{J}}_B$, $\tilde{\tensor{L}}_A$, $\tilde{\tensor{L}}_B$ in a variable number of times, which depends on the rank of approximation. Therefore, the question of convergence of the series in terms \eqref{eq:F_final} and \eqref{eq:M_final} remains still open.  the  
	
	\section{The method of approximating of the light pressure by tensor series} \label{sec:4} 
	
	The resulting expressions for the resultant force and moment in the series form allows us to offer a method of approximation of light radiation pressure. 
	
	Let us assume that in some way (e.g. by Monte--Carlo ray tracing) we obtained the values of normalized force $\vec{f}^{(i)}=\vec{F}^{(i)}/P(R)$ and normalized moment $\vec{m}^{(i)}=\vec{M}^{(i)}/P(R)$ for the given set of different orientations $\hat{\mathbf{s}}^{(i)}$ in the amount of $N$. Superscript $(i)$ indicates the number of orientation case for which $\vec{F}$ and $\vec{M}$ were calculated by the different method (e.g. by Monte--Carlo), from $1$ to $N$. We need to get approximated values of tensor components in the expansion for $\vec{F}$ and $\vec{M}$. To do this, we define the vectors for the unknown components of the tensors $\tilde{\tensor{I}}^n$ и $\tilde{\tensor{K}}^n$ (hereinafter we indicate vector and matrix dimensions in brackets): \begin{align} &\underset{\small{\left[\left(\frac{3}{2}(3^M-1)\right)\times1\right]}}{\mathbf{j}} =\left( \tilde{I}^1_1\ \tilde{I}^2_{11}\ \tilde{I}^2_{21}\ \tilde{I}^2_{31}\ \tilde{I}^3_{111}\ \ldots\ \tilde{I}^M_{\footnotesize{\underbrace{3\ldots3}_{M-1}}1}\ \tilde{I}^1_2\ \ldots\ \tilde{I}^M_{\footnotesize{\underbrace{3\ldots3}_{M-1}}2}\ \tilde{I}^1_3\ \ldots\right.\nonumber\\ &\ldots\ \left.\tilde{I}^M_{\footnotesize{\underbrace{3\ldots3}_{M}}} \right)^T;\\ &\underset{\small{\left[\left(\frac{3}{2}(3^M-1)\right)\times1\right]}}{\mathbf{k}}=\left( \tilde{K}^1_1\ \tilde{K}^2_{11}\ \tilde{K}^2_{21}\ \tilde{K}^2_{31}\ \tilde{K}^3_{111}\ \ldots\ \tilde{K}^M_{\footnotesize{\underbrace{3\ldots3}_{M-1}}1}\ \tilde{K}^1_2\ \ldots\ \tilde{K}^M_{\footnotesize{\underbrace{3\ldots3}_{M-1}}2}\ \tilde{K}^1_3\ \ldots\right.\nonumber\\ &\ldots\ \left.\tilde{K}^M_{\footnotesize{\underbrace{3\ldots3}_{M}}} \right)^T, \end{align} where $M$ -- number of terms in the expansion of force and moment (for simplicity we assume that this number is the same for the calculation of force and moment). In the above two equations subscript is a vector component, superscript is the order of series expansion. 
	
	Let us define the vectors of free terms: \begin{align} \underset{\small{[3N\times1]}}{\mathbf{f}}=\left(f_1^{(1)}\ f_1^{(2)}\ \ldots\ f_1^{(N)}\ f_2^{(1)}\ f_2^{(2)}\ \dots\  f_2^{(N)}\ f_3^{(1)}\ f_3^{(2)}\ \dots\ f_3^{(N)}\right)^T;\label{eq:f_approximated}\\ \underset{\small{[3N\times1]}}{\mathbf{m}}=\left(m_1^{(1)}\ m_1^{(2)}\ \ldots\ m_1^{(N)}\ m_2^{(1)}\ m_2^{(2)}\ \dots\ m_2^{(N)}\ m_3^{(1)}\ m_3^{(2)}\ \dots\ m_3^{(N)}\right)^T.\label{eq:m_approximated} \end{align} 
	
	The matrix of coefficients will be as the follows: \begin{align} \underset{\small{\left[3n\times\left(\frac{3}{2}(3^M-1)\right)\right]}}{S}=\hspace{10cm}\nonumber\\ \small{\begin{pmatrix}1 &\cdots &1 &0 &\cdots &0 &0 &\cdots &0\\ s_1^{(1)} &&s_1^{(N)} &&&&&&\\ s_2^{(1)} &&s_2^{(N)} &&&&&&\\ s_3^{(1)} &&s_3^{(N)} &\vdots &&\vdots &\vdots &&\vdots \\ s_1^{(1)}s_1^{(1)} &&s_1^{(N)}s_1^{(N)} &&&&&&\\ \vdots &&\vdots &&&&&&\\ \underbrace{s_3^{(1)}\ldots s_3^{(1)}}_{M-1} &\cdots &\underbrace{s_3^{(N)}\ldots s_3^{(N)}}_{M-1} &0 &\cdots &0 &0 &\cdots &0\\ 0 &\cdots &0 &1 &\cdots &1 &0 &\cdots &0\\ &&&s_1^{(1)} &&s_1^{(N)} &&&\\ &&&s_2^{(1)} &&s_2^{(N)} &&&\\ \vdots &&\vdots &s_3^{(1)} &&s_3^{(N)} &\vdots &&\vdots \\ &&&s_1^{(1)}s_1^{(1)} &&s_1^{(N)}s_1^{(N)} &&&\\ &&&\vdots &&\vdots &&&\\ 0 &\cdots &0 &\underbrace{s_3^{(1)}\ldots s_3^{(1)}}_{M-1} &\cdots &\underbrace{s_3^{(N)}\ldots s_3^{(N)}}_{M-1} &0 &\cdots &0\\ 0 &\cdots &0 &0 &\cdots &0 &1 &\cdots &1\\ &&&&&&s_1^{(1)} &&s_1^{(N)}\\ &&&&&&s_2^{(1)} &&s_2^{(N)}\\ \vdots &&\vdots &\vdots &&\vdots &s_3^{(1)} &&s_3^{(N)}\\ &&&&&&s_1^{(1)}s_1^{(1)} &&s_1^{(N)}s_1^{(N)}\\ &&&&&&\vdots &&\vdots \\ 0 &\cdots &0 &0 &\cdots &0 &\underbrace{s_3^{(1)}\ldots s_3^{(1)}}_{M-1} &\cdots &\underbrace{s_3^{(N)}\ldots s_3^{(N)}}_{M-1} \end{pmatrix}}^T. \end{align} 
	
	For vectors $\mathbf{f}$ and $\mathbf{m}$ let us introduce the following overdefined system of linear equations: \begin{align} &S\mathbf{j}=\mathbf{f};\\ &S\mathbf{k}=\mathbf{m}. \end{align} 
	
	For $\mathbf{j}$ and $\mathbf{k}$ we can find approximations $\tilde{\mathbf{j}}$ and $\tilde{\mathbf{k}}$ by the least squares method: \begin{align} &||S\mathbf{j}-\mathbf{f}||^2\rightarrow \min,\ \tilde{\mathbf{j}}=\left(S^TS\right)^+S^T\mathbf{f};\label{eq:approx_j}\\ &||S\mathbf{k}-\mathbf{m}||^2\rightarrow \min,\ \tilde{\mathbf{k}}=\left(S^TS\right)^+S^T\mathbf{m},\label{eq:approx_k} \end{align} where $^+$ is a pseudo--inverse operator for square matrix. 
	
	After calculation of approximated values of $\mathbf{f}$ and $\mathbf{m}$ we can reconstruct the approximated tensors $\tilde{\tensor{I}}^n$ and $\tilde{\tensor{K}}^n$ using \eqref{eq:f_approximated} and \eqref{eq:m_approximated} respectively. After this it is possible to determine the approximated value of the resultant force and moment for the arbitrary orientation $\unit{s}$ of light source around structure using the expressions~\eqref{eq:F_final} and~\eqref{eq:M_final} respectively. 
	
	\section{Parametrization of the resultant vector and moment in case of specular-diffuse reflection} \label{sec:5} 
	
	For the ballistic analysis, it is often necessary to study the impact of the change of parameters. That is necessary to make a parametrization for our problem. Parametrization problem can be viewed in two ways: as a problem of geometric parameterization and as a problem of parametrization of the optical parameters. In this work, we will make parametrization by specularity coefficient $s$. 
	
	Suppose we obtain an approximation for the tensors describing the force and moment of light pressure in pure diffuse and pure specular cases. For the tensors \eqref{eq:method_end_0} of the first rank, we write the expression for diffuse and specular cases, taking the temperature distribute on the surface constant: \begin{align} \tilde{\tensor{I}}^1 \approx \tilde{\tensor{I}}^1|_{s=0} \approx \tilde{\tensor{I}}^1|_{s=1}.\label{eq:param_start} \end{align} 
	
	For the tensors \eqref{eq:method_end_0} of second rank we will get: \begin{align*} &\tilde{\tensor{I}}^2|_{s=0} = \frac{1}{2}B\rho\int\limits_A\tilde{\tensor{J}}_A^2 dA;\\ &\tilde{\tensor{I}}^2|_{s=1} = 0, \end{align*} therefore, assuming that the tensors $\tilde{\tensor{J}}_A^2$ are constant at different $s$, we obtain \begin{align} \tilde{\tensor{I}}^2 \approx (1-s)\tilde{\tensor{I}}^2|_{s=0}. \end{align} 
	
	For the tensors \eqref{eq:method_end_0} of third rank let us consider diffuse and specular cases: \begin{align*} &\tilde{\tensor{I}}^3|_{s=0} = \frac{1}{2}\int\limits_A\left( -\tilde{\tensor{J}}_B^3 - B_1B\rho\tilde{\tensor{J}}_A^3 \right)dA;\\ &\tilde{\tensor{I}}^3|_{s=1} = \frac{1}{2}\int\limits_A\left( -(1-\rho)\tilde{\tensor{J}}_B^3 - 2\rho\tilde{\tensor{J}}_A^3 \right)dA, \end{align*} then approximating expression for the third rank will be as follows: \begin{align} \tilde{\tensor{I}}^3 \approx (1-s)\tilde{\tensor{I}}^3|_{s=0} + s\tilde{\tensor{I}}^3|_{s=1}. \end{align} 
	
	In the general case of arbitrary rank we will get: \begin{align*} &\tilde{\tensor{I}}^n|_{s=0} = \frac{1}{2}\int\limits_A\left( -B_{\frac{n-1}{2}}\frac{1-(-1)^n}{2}B\rho\tilde{\tensor{J}}_A^n + B_{\frac{n-2}{2}}\frac{1+(-1)^n}{2} \tilde{\tensor{J}}_B^n \right)dA;\\ &\tilde{\tensor{I}}^3|_{s=1} = \frac{1}{2}B_{\frac{n-2}{2}}\frac{1+(-1)^n}{2}\int\limits_A\left( (1-\rho)\tilde{\tensor{J}}_B^n + 2\rho\tilde{\tensor{J}}_A^n \right)dA, \end{align*} where we get \begin{align} \tilde{\tensor{I}}^n \approx (1-s)\tilde{\tensor{I}}^n|_{s=0} + s\tilde{\tensor{I}}^n|_{s=1}, n>3. \end{align} 
	
	Similarly, we obtain expressions for the tensors \eqref{eq:method_end}: \begin{align} &\tilde{\tensor{K}}^1 \approx \tilde{\tensor{K}}^1|_{s=0} \approx \tilde{\tensor{K}}^1|_{s=1};\\ &\tilde{\tensor{K}}^2 \approx (1-s)\tilde{\tensor{K}}^2|_{s=0};\\ &\tilde{\tensor{K}}^n \approx (1-s)\tilde{\tensor{K}}^n|_{s=0} + s\tilde{\tensor{K}}^n|_{s=1}, n>2,\label{eq:param_end} \end{align} the resultant force and moment will be determined by the expressions \eqref{eq:F_final} and \eqref{eq:M_final}. 
	
	The expressions \eqref{eq:param_start} --- \eqref{eq:param_end} represent the approximation of tensors $\tilde{\tensor{I}}$ and $\tilde{\tensor{K}}$ for specular-diffuse surface when we know the components of tensors $\tilde{\tensor{I}}$ and $\tilde{\tensor{K}}$ in two cases: where the reflection is completely diffuse ($s=0$) and completely specular ($s=1$). 
	
	\section{Analytical examples} 
	
	In the analytical examples below the light source orientation vector $\unit{s}$ is defined by two angles $\alpha$ and $\beta$ as follows: \begin{itemize} \item $\alpha\in[0,2\pi]$ --- angle between unit vector $\unit{e}_1$ of axis $Ox_1$ and projection of vector $\unit{s}$ on the plane $Ox_1x_3$; \item $\beta\in[-\frac{\pi}{2};\frac{\pi}{2}]$ --- angle between plane $Ox_1x_3$ and vector $\unit{s}$. \end{itemize} The components of vector $\unit{s}$ can be written as follows: \begin{equation*} \unit{s} = (\cos\alpha \cos\beta, \sin\beta, \sin\alpha \cos\beta)^T. \end{equation*} 
	
	\subsection{Two-sided specular-diffuse flat solar sail} 
	
	Let us consider a flat solar sail with specular--diffuse surface properties which are different on the opposite sides. The coordinate system origin is placed in the geometric center of the solar sail. The front side (side 1) has reflection coefficient $\rho_1$ and specularity parameter $s_1$. The other side (side 2) has reflectivity $\rho_2$ and specularity $s_2$. The area of the solar sail is $A$. 
	
	The components of tensors $\tensor{I}$ and $\tensor{K}$ can be calculated by the following equations: \begin{align*} &\tensor{I} = A \left( \tensor{J}(\unit{n}_1,\rho_1,s_1)+\tensor{J}(\unit{n}_2,\rho_2,s_2) \right);\\ &\tensor{K} = A \left( \tensor{L}(\unit{n}_1,\vec{r}_1,\rho_1,s_1) + \tensor{L}(\unit{n}_2,\vec{r}_2,\rho_2,s_2) \right);\\ &\unit{n}_1 = \left( 0, 0, 1 \right)^T;\\ &\unit{n}_2 = \left( 0, 0, -1 \right)^T;\\ &\vec{r}_1 = \vec{r}_2 = \left( 0, 0, 0 \right)^T. \end{align*} 
	
	By setting the maximum number of terms in the series as $N_{\max}=6$ we can obtain the following analytical expressions for the components of the principal force and moment of light pressure: \begin{align*} &F_1 = \frac{P(R)A}{30\pi}\left( -6(-2 + \rho_1 s_1 + \rho_2 s_2) + \cos\beta \sin\alpha \left( 15\pi (\rho_1 s_1 - \rho_2 s_2) + \right.\right.\\ &\left.\left. +8 (-2 + \rho_1 s_1 + \rho_2 s_2) \cos\beta \sin\alpha \left( -9 + 4 \cos^2\beta \sin^2\alpha \right) \right) \right)\cos\alpha\cos\beta ;\\ &F_2 = \frac{P(R)A}{30\pi} \left( -6(-2 + \rho_1 s_1 + \rho_2 s_2) + \cos\beta \sin\alpha \left( 15\pi (\rho_1 s_1 - \rho_2 s_2) + \right.\right.\\ &\left.\left. +8 (-2 + \rho_1 s_1 + \rho_2 s_2) \cos\beta \sin\alpha \left( -9 + 4 \cos^2\beta \sin^2\alpha \right) \right) \right)\sin\beta ;&\\ &F_3 = \frac{P(R)A}{90\pi} \left( 24 (\rho_2(1-s_2) - \rho_1(1-s_1)) + \cos\beta\sin\alpha \right.\\ &\left. \left( 6(5\pi (\rho_1(1-s_1) +\rho_2(1-s_2))    + 3(2+\rho_1 s_1 \rho_2 s_2)) + \cos\beta\sin\alpha \right.\right.\\ &\left.\left. \left( -9(16 \rho_1 (1-s_1) + 5\pi\rho_1 s_1 -16\rho_2(1-s_2) - 5\pi\rho_2 s_2 ) + 8\cos\beta\sin\alpha \right.\right.\right.\\ &\left.\left.\left. \left( 27(2+\rho_1 s_1 + \rho_2 s_2) + 4\cos\beta\sin\alpha (2(\rho_1(1-s_1) - \rho_2(1-s_2))- \right.\right.\right.\right.\\ &\left.\left.\left.\left. -3 (2+\rho_1 s_2 + \rho_2 s_2)\cos\beta\sin\alpha ) \right) \right) \right) \right);\\ &\vec{M} = 0. \end{align*} 
	
	To check the correctness of these statements we made a comparison with the known solutions for specular and diffuse solar sails \cite{forward_1989,mcinnes_solar_2004,polyakhova_2011}. 
	
\begin{figure} \begin{minipage}[h]{0.49\linewidth} \center{\includegraphics[width=1\linewidth]{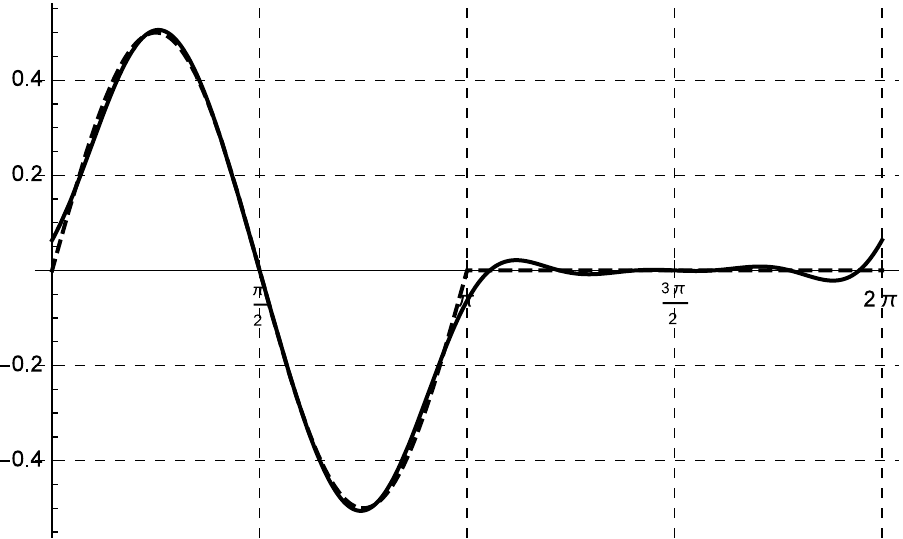}} a \end{minipage} \hfill
		\begin{minipage}[h]{0.49\linewidth} \center{\includegraphics[width=1\linewidth]{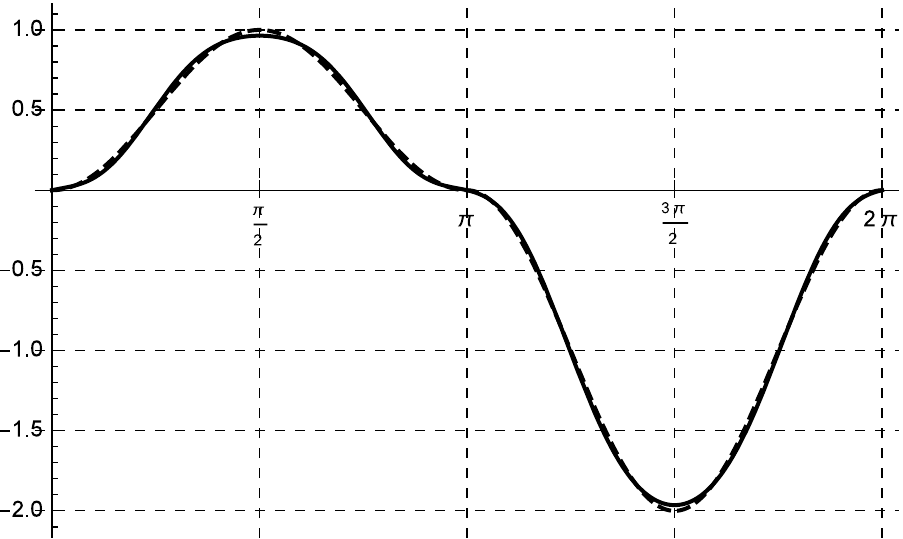}} b \end{minipage}
		
		\caption{Projection on $Ox_1$ (a) and on $Ox_3$ (b) of resultant force of light pressure upon two-sided specular solar sail with unit area, $N_{\max}=6$. Solid line -- approximate solution, dashed line --- exact solution. Values are divided by $P(R)$.} \label{fig:mirror_fx_fz}
	\end{figure}

\begin{figure} \begin{minipage}[h]{0.49\linewidth} \center{\includegraphics[width=1\linewidth]{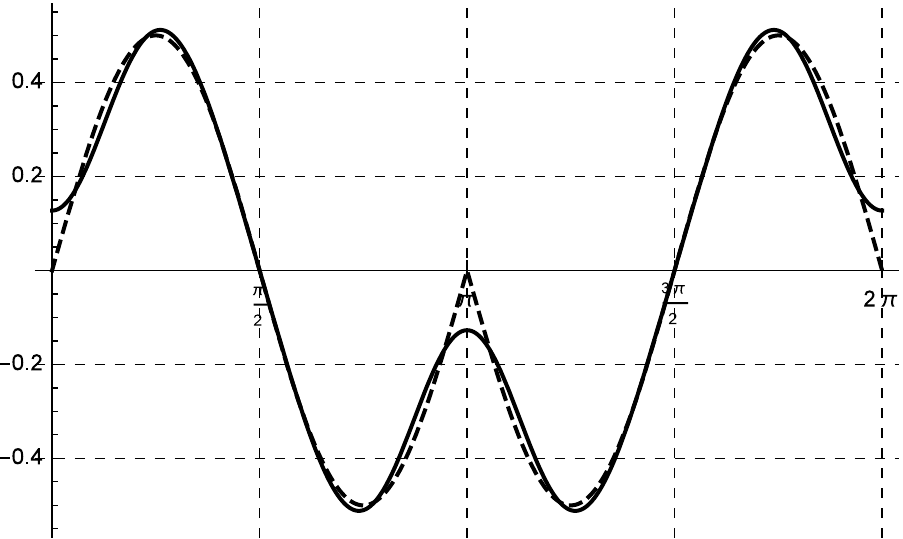}} a \end{minipage} \hfill
		\begin{minipage}[h]{0.49\linewidth} \center{\includegraphics[width=1\linewidth]{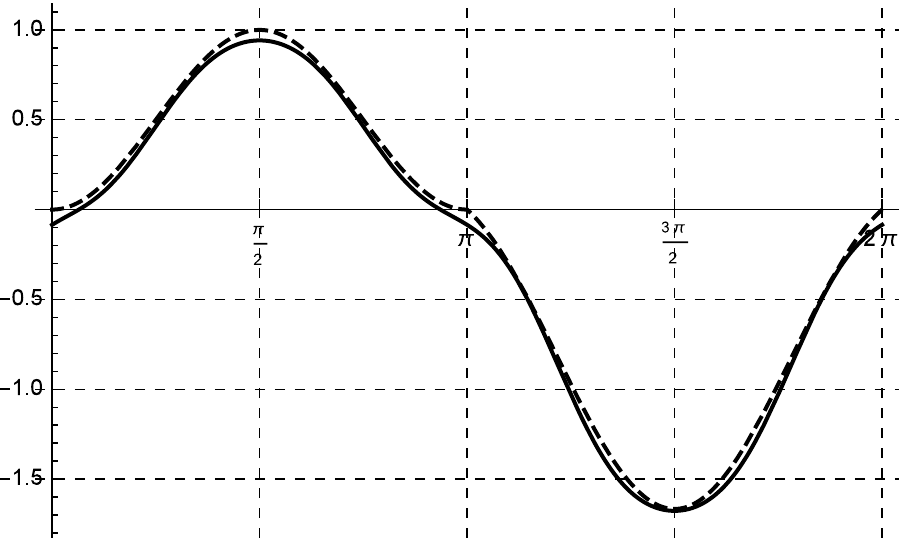}} b \end{minipage}
		
		\caption{Projection on $Ox_1$ (a) and on $Ox_3$ (b) of resultant force of light pressure upon two-sided diffuse solar sail with unit area, $N_{\max}=6$. Solid line -- approximate solution, dashed line -- exact solution. Values are divided by $P(R)$.} \label{fig:diffuse_fx_fz}
	\end{figure}

\begin{figure} \begin{minipage}[h]{0.49\linewidth} \center{\includegraphics[width=1\linewidth]{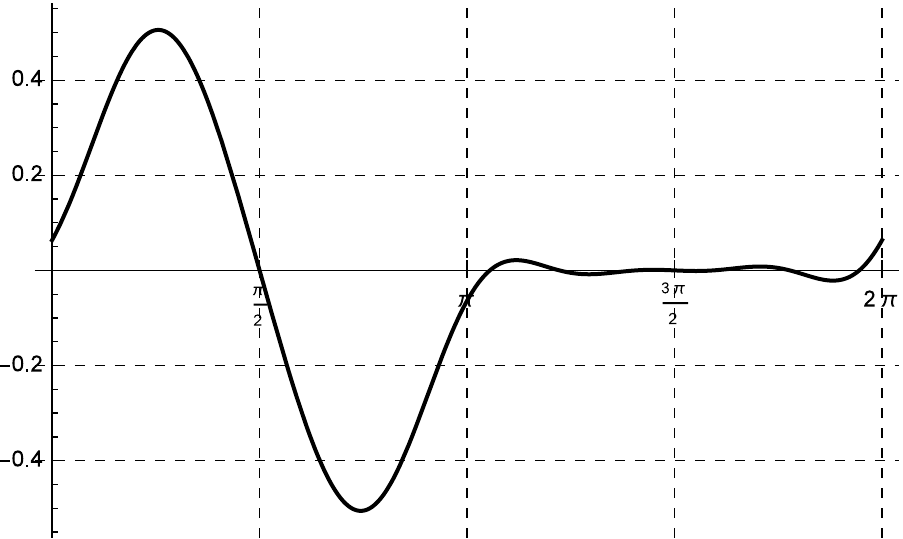}} a \end{minipage} \hfill
		\begin{minipage}[h]{0.49\linewidth} \center{\includegraphics[width=1\linewidth]{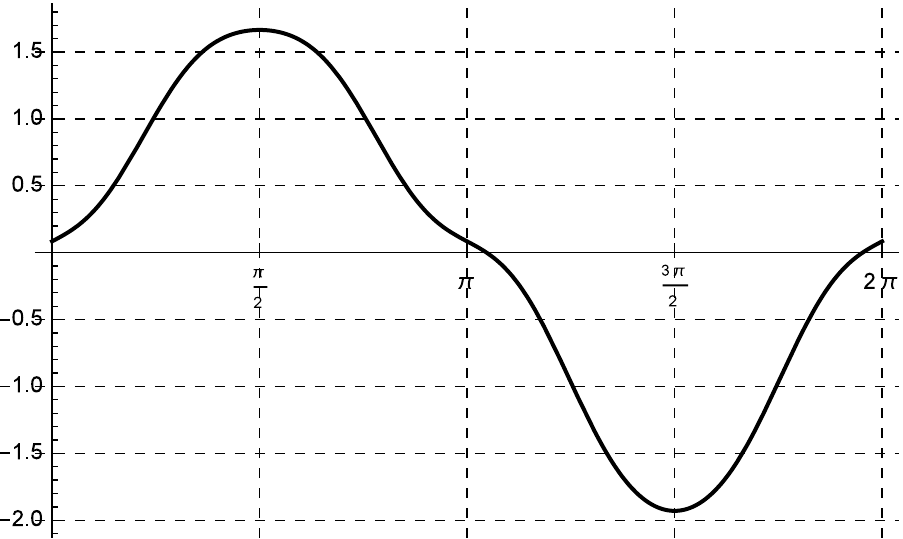}} b \end{minipage}
		
		\caption{Projection on $Ox_1$ (a) and on $Ox_3$ (b) of resultant force of light pressure upon two-sided specular-diffuse solar sail with unit area, $N_{\max}=6$. Values are divided by $P(R)$.} \label{fig:plane_fx_fz}
	\end{figure}

	At the first, we selected a two-sided solar sail, one side of which is completely specular, the other is absorbing, i.e. $\rho_1=1,\ \rho_2=0,\ s_1=s_2=1$. Solar sail area was set to 1. We calculated the non-zero components of the resultant vector of light pressure on a solar sail, namely projections on axises $Ox_1=Ox$ and $Ox_3=Oz$, depending on the angle of inclination $\alpha$ considering $\beta=0$. The calculation results are shown in Fig. \ref{fig:mirror_fx_fz}. As can be seen from the figure, there is a fairly good convergence of the approximate solution to the exact solution. On this and subsequent figures (fig.~\ref{fig:mirror_fx_fz}, \ref{fig:diffuse_fx_fz} and \ref{fig:plane_fx_fz}) the range of angles $\alpha\in[0;\pi]$ corresponds to the lightning of side 2, and range $\alpha\in[\pi;2\pi]$ corresponds to the lightning of side 1, and for $\alpha=\pi/2$ and $\alpha=3\pi/2$ the light strikes perpendicular to the surface.

	Then we selected the two-sided solar sail, one side of which is diffuse, $B=2/3$, the other -- absorbing, i.e. $\rho_1=1$, $\rho_2=0$, $s_1=s_2=0$. Solar sail area was set to 1. We calculated the non-zero components of the resultant vector of light pressure on a solar sail, namely projections on axises $Ox_1=Ox$ and $Ox_3=Oz$, depending on the angle of inclination $\alpha$ considering $\beta=0$. The calculation results are shown in Fig. \ref{fig:diffuse_fx_fz}. As can be seen from the figure, there is a fairly accurate convergence of the approximate solutions to the exact solution.

	For the solar sail with unit area with parameters $\rho_1=1,\ \rho_2=1,\ s_1=1,\ s_2=0$ we calculated the projections of principal force and moment on the axises $Ox_1=Ox$ and $Ox_3=Oz$ considering $\beta=0$ ( fig.~\ref{fig:plane_fx_fz}). At the indicated chart values of ​​force are divided by $P(R)$. 
	
	\subsection{Specular--diffuse sphere} 
	
	Let us consider a sphere of radius $R_0$ with a homogeneous specular-diffusive surface, the reflection coefficient of which is equal to $\rho$ and the degree of specular reflection is $s$.  
	
	The expressions for the components of tensors $\tensor{I}$ and $\tensor{K}$ have the following form: \begin{align*} &\tensor{I} = R_0^2\int\limits_0^{2\pi}\int\limits_{-\frac{\pi}{2}}^{\frac{\pi}{2}}\tensor{J}(\unit{n},\rho,s)d\theta d\phi;\\ &\tensor{K} = R_0^2\int\limits_0^{2\pi}\int\limits_{-\frac{\pi}{2}}^{\frac{\pi}{2}}\tensor{L}(\unit{n},\vec{r},\rho,s)d\theta d\phi;\\ &\unit{n} = \left( \cos\phi \cos\theta, \sin\phi \cos\theta, \sin\theta \right)^T;\\ &\vec{r} = \left( R_0 \cos\phi \cos\theta, R_0 \sin\phi \cos\theta, R_0 \sin\theta \right)^T. \end{align*} 
	
	We obtained the following analytical expressions considering $N_{\max}=6$: \begin{align*} &F_1 = P(R)\frac{4}{1575}\left( 175\pi \rho (1-s) + 3 (413 + \rho s) \right)R_0^2\cos\alpha\cos\beta;\\ &F_2 = P(R)\frac{4}{1575}\left( 175\pi \rho (1-s) + 3 (413 + \rho s) \right)R_0^2\sin\beta;\\ &F_3 = P(R)\frac{4}{1575}\left( 175\pi \rho (1-s) + 3 (413 + \rho s) \right)R_0^2\cos\beta\sin\alpha;\\ &\vec{M} = 0. \end{align*} 
	
	For $\rho=1$ and $s=1$ we get: \begin{align} &F_1 \approx P(R)\pi R_0^2\cos\alpha\cos\beta;\\ &F_2 \approx P(R)\pi R_0^2\sin\beta;\\ &F_3 \approx P(R)\pi R_0^2\cos\beta\sin\alpha. \end{align} 
	
	\subsection{Specular--diffuse cylinder}

	Let us consider the cylinder with following parameters: \begin{itemize} \item $\rho_0$ --- reflectance of the envelope; \item $\rho_1$ --- reflectance of the butt surface $+x_3$; \item $\rho_2$ --- reflectance of the butt surface $-x_3$; \item $s_0$ --- specularity coefficient of the envelope; \item $s_1$ --- specularity coefficient of the butt surface $+x_3$; \item $s_2$ --- specularity coefficient of the butt surface $-x_3$; \item $R_1$ --- radius of the cylinder; \item $H$ --- height of the cylinder. \end{itemize} 
	
	The expressions for the components of tensors $\tensor{I}$ and $\tensor{K}$ have the following form: \begin{align*} &\tensor{I} = \tensor{J}(\unit{n}_1,\rho_1,s_1)\pi R_1^2 + \tensor{J}(\unit{n}_2,\rho_2,s_2)\pi R_1^2 + H R_1\int\limits_0^{2\pi}\tensor{J}(\unit{n}_0,\rho_0,s_0)d\phi;\\ &\tensor{K} = \tensor{L}(\unit{n}_1,\vec{r}_1,\rho_1,s_1)\pi R_1^2 + \tensor{L}(\unit{n}_2,\vec{r}_2,\rho_2,s_2)\pi R_1^2 + H R_1\int\limits_0^{2\pi}\tensor{L}(\unit{n}_0,\vec{r}_0,\rho_0,s_0)d\phi;\\ &\unit{n}_1 = (0,0,1)^T;\\ &\unit{n}_2 = (0,0,-1)^T;\\ &\unit{n}_0 = \left( \cos\phi,\sin\phi,0 \right)^T;\\ &\vec{r}_1 = (0,0,H/2)^T;\\ &\vec{r}_2 = (0,0,-H/2)^T;\\ &\vec{r}_0 = \left( R_1 \cos\phi, R_1 \sin\phi, 0 \right)^T. \end{align*} 
	
	For the number of terms of the series $N_{\max}=6$ we obtained analytical dependences for the principal force and moment of light radiation pressure: 
	
	\begin{align*} &F_1 = \frac{P(R)R_1}{30}\cos\alpha \cos\beta (-8 H (3 + 2 \rho_0 s_0) \cos^4\alpha \cos^4\beta +\\ &+4 H \cos^2\alpha \cos^2\beta (12 + 5 \rho_0 s_0 + (6 + 4 \rho_0 s_0) \cos2\beta) +\\ &+R_1 \cos\beta \sin\alpha (15 \pi (\rho_1 s_1 - \rho_2 s_2) + \\ &+8 (-2 + \rho_1 s_1 + \rho_2 s_2) \cos\beta \sin\alpha (-9 + 4 \cos^2\beta \sin^2\alpha)) + \\ &+2 (H (6 - 5 \pi \rho_0 (-1 + s_0)) - 3 R_1 (-2 + \rho_1 s_1 + \rho_2 s_2) + \\ &+2 H (15 + 7 \rho_0 s_0 + (3 + 2 \rho_0 s_0) \cos2\beta) \sin^2\beta));\\ &F_2 = \frac{P(R) R_1}{30} \sin\beta (-8 H (3 + 2 \rho_0 s_0) \cos^4\alpha \cos^4\beta + \\ &+4 H \cos^2\alpha \cos^2\beta (12 + 5 \rho_0 s_0 + (6 + 4 \rho_0 s_0) \cos2 \beta) + \\ &+R_1 \cos\beta \sin\alpha (15 \pi (\rho_1 s_1 - \rho_2 s_2) + \\ &+8 (-2 + \rho_1 s_1 + \rho_2 s_2) \cos\beta \sin\alpha (-9 + 4 \cos^2\beta \sin^2\alpha)) + \\ &+2 (H (6 - 5 \pi \rho_0 (-1 + s_0)) - 3 R_1 (-2 + \rho_1 s_1 + \rho_2 s_2) + \\ &+2 H (15 + 7 \rho_0 s_0 + (3 + 2 \rho_0 s_0) \cos2 \beta) \sin^2\beta));\\ &F_3 = \frac{P(R)R_1}{90} (24 R_1 (\rho_2(1-s_2) - \rho_1 (1 - s_1) ) + \\ &+\frac{3}{8} \cos\beta (-363 H (-1 + \rho_0 s_0) + 16 R_1 (5 \pi (\rho_1 + \rho_2 - \rho_1 s_1 - \rho_2 s_2) + \\ &+3 (2 + \rho_1 s_1 + \rho_2 s_2)) + 3 H (-1 + \rho_0 s_0) (44 \cos2\beta + 5 \cos4\beta)) \sin\alpha - \\ &-9 R_1 (\rho_1 (16 + (-16 + 5 \pi) s_1) + 16 \rho_2 (-1 + s_2) - 5 \pi \rho_2 s_2) \cos^2\beta \sin^2\alpha + \\ &+64 R_1 (\rho_1 - \rho_1 s_1 + \rho_2 (-1 + s_2)) \cos^4\beta \sin^4\alpha + \\ &+\frac{9}{4} \cos^3\beta (96 R_1 (2 + \rho_1 s_1 + \rho_2 s_2) \sin^3\alpha - H (-1 + \rho_0 s_0) (13 + 5 \cos2\beta) \sin3\alpha) + \\ &+\frac{3}{2} \cos^5\beta (-64 R_1 (2 + \rho_1 s_1 + \rho_2 s_2) \sin^5\alpha + 3 H (-1 + \rho_0 s_0) \sin5\alpha));\\ &M_1 = \frac{P(R)H R_1^2}{60} (6 \rho_1 s_1 - 6 \rho_2 s_2 + \cos\beta \sin\alpha (15 \pi (-2 \rho_0 s_0 + \rho_1 s_1 + \rho_2 s_2) + \\ &+8 (\rho_1 s_1 - \rho_2 s_2) \cos\beta \sin\alpha (-9 + 4 \cos^2\beta \sin^2\alpha))) \sin\beta;\\ &M_2 = \frac{P(R) H R_1^2}{60} (-6 \rho_1 s_1 + 6 \rho_2 s_2 +\cos\beta \sin\alpha (15 \pi (2 \rho_0 s_0 - \rho_1 s_1 - \rho_2 s_2) + \\ &+8 (\rho_1 s_1 - \rho_2 s_2) \cos\beta \sin\alpha (9 - 4 \cos^2\beta \sin^2\alpha)))\cos\alpha \cos\beta;\\ &M_3 = 0. \end{align*} 
	
\begin{figure} \begin{minipage}[h]{0.49\linewidth} \center{\includegraphics[width=1\linewidth]{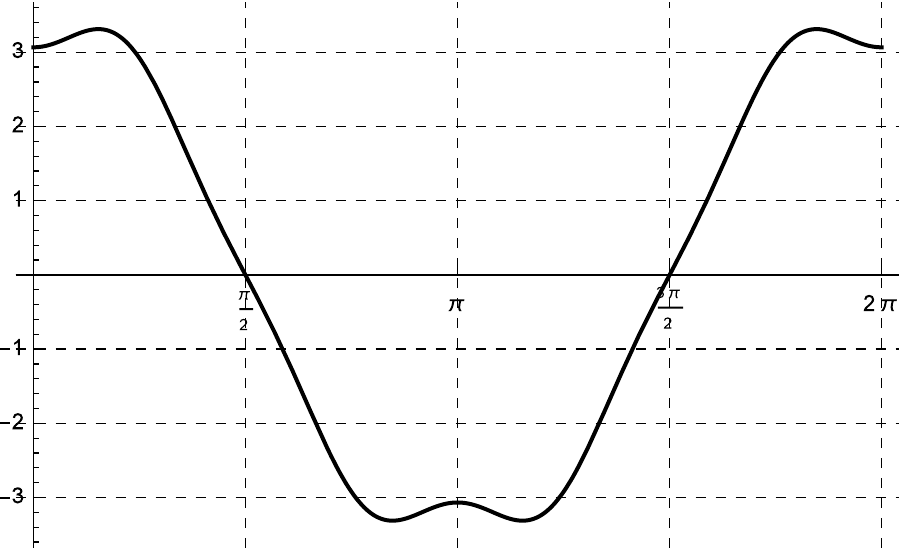}} a \end{minipage} \hfill
		\begin{minipage}[h]{0.49\linewidth} \center{\includegraphics[width=1\linewidth]{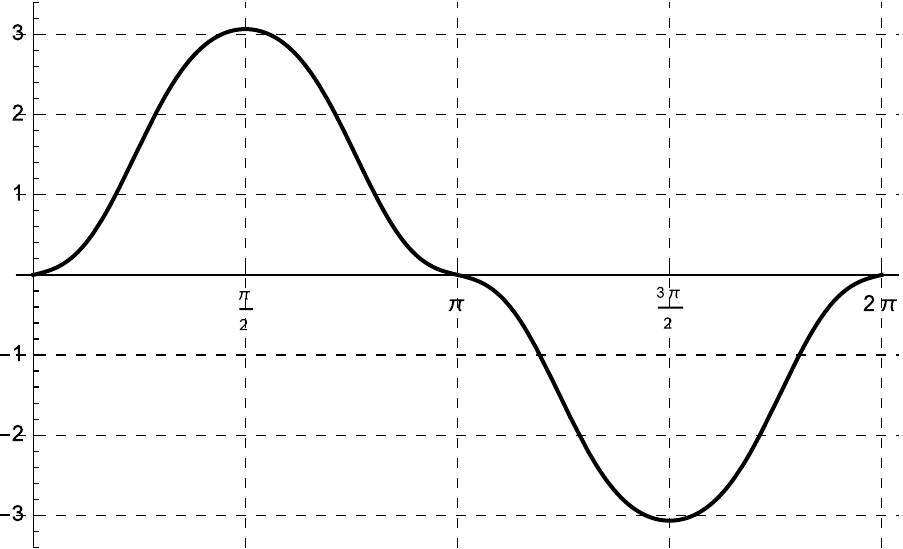}} b \end{minipage}
		
		\caption{Projection on $Ox_1$ (a) and on $Ox_3$ (b) of resultant force of light pressure upon specular-diffuse cylinder, $N_{\max}=6$. Values are divided by $P(R)$.} \label{fig:cylinder_fx_fz} \end{figure}

\begin{figure} \includegraphics[width=.5\linewidth]{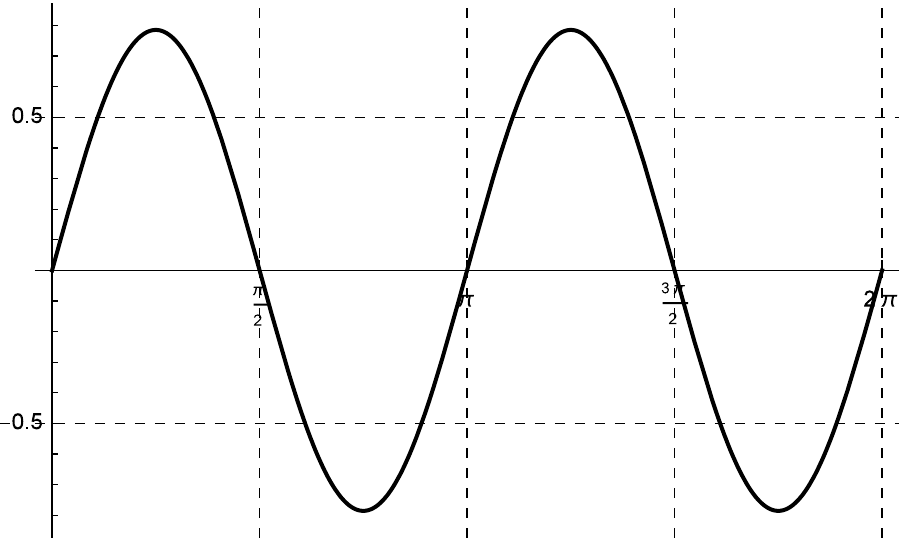} \caption{Projection on $Ox_2$ of principal moment of light pressure upon specular-diffuse cylinder, $N_{\max}=6$. Values are divided by $P(R)$.} \label{fig:cylinder_my} \end{figure}

	On the figures~\ref{fig:cylinder_fx_fz} and~\ref{fig:cylinder_my} there are graphs of components of principal force and moment of light radiation pressure upon a cylinder with the following parameters: $\rho_1=\rho_2=0$, $\rho_0=1$, $s_1=s_2=0$, $s_0=1$, the height and radius are equal to 1m, depending on angle $\alpha$ considering $\beta=0$. In these figures for the angles $\alpha=\pi/2$ and $\alpha=3\pi/2$ light strikes perpendicularly to the one of the ends of cylinder.

	\section{Numerical example of $\vec{F}$ and $\vec{M}$ approximation} \label{seq:numerical}
	
	\subsection{Model spacecraft} 
	
\begin{figure} \includegraphics[width=\linewidth]{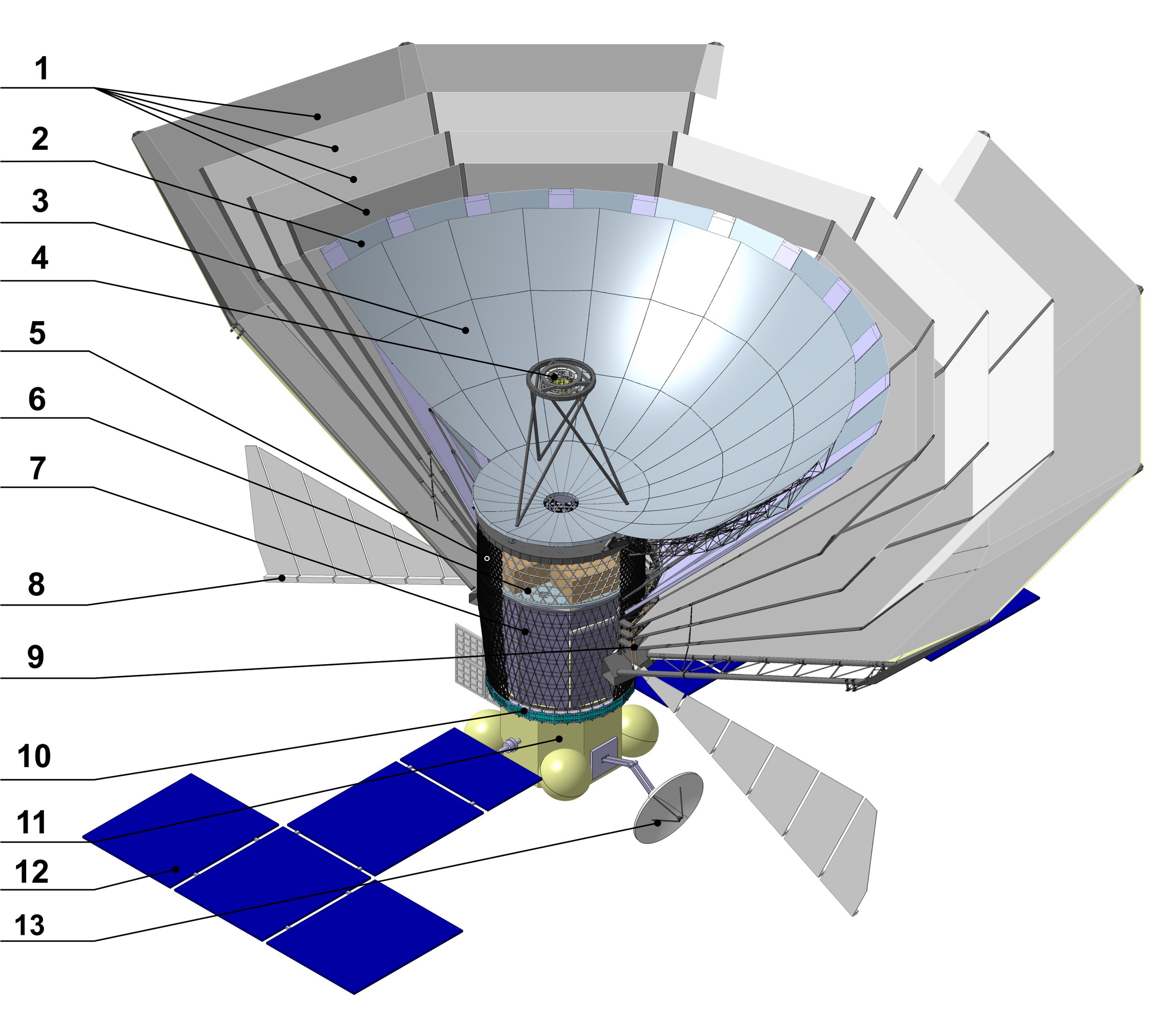} \caption{Millimentron space observatory concept. 1 -- Sun shields; 2 -- Cryo-shield; 3 -- Primary mirror's petal; 4 -- Secondary mirror; 5 -- Central part of Primary mirror; 6 -- Cryo-container; 7 -- Heat exchanger (radiator); 8 -- Warm container; 9 -- Sunshields supporting truss; 10 -- Adapter ring; 11 -- Service module; 12 -- Solar power array; 13 -- High gain antenna.} \label{fig:mm_annotated} \end{figure}

	As a numerical example we selected the problem of approximation of resultant force and moment of light radiation pressure upon a perspective space observatory ``Millimetron'' (``Spectr-M'', Fig.~\ref{fig:mm_annotated})). The purpose of the Millimetron is to study various astronomical objects in the Universe at the wavelength range 60 $\mu$m --- 20 mm with an unprecedented sensitivity in the single-dish observation mode and an extremely high spatial resolution as an element of a ground-space very long baseline interferometry (VLBI) system~\cite{kardashev_2000,wild_millimetronlarge_2008}.
	
	To achieve that, the mission will be equipped with a 10-m diameter deployable primary mirror with the high surface accuracy of the reflective surface (RMS$\leq10\mu m$), and even more large-sized sun shields for passive cooling. As well as high requirements to accuracy and stability of attitude control system are defined (1" and 0.2" respectively). Hence, а precise estimation of light pressure, as the main factor affecting the movement of the center of mass and around the center of mass, is so important.
	
	The antenna and cryogenic instruments should be cooled down to 4.5 K by a combination of passive cooling with radiation shields and active cooling with mechanical coolers. These requirements led to a bunch of engineering challenges, connecting with the development of a reliable deployment system, light weight antenna structures with high thermo-elastic distortion stability and high accuracy of the reflective surface.
	
	The deployment concept is based on the previously implemented structure in the Radioastron project~\cite{kardashev_radioastron_1997}. To achieve the required high surface accuracy of the primary mirror after the deployment and compensate reflecting surface distortions under the action of space environment factors, each petal is composed of a supporting spatial frame and three independent reflecting panels. The panels are mounted on the supporting frame via precise cryogenic actuators to adjust their reflecting surface. An active surface control system is used to control all reflecting panels of the mirror by using wave-front sensing. Additionally, a high modulus carbon fiber reinforced plastic (CFRP), providing a lightweight structure with high thermal stability, has been chosen for the material of the primary mirror.

	The Millimetron space mission is an approved project as part of Russian Space Program with the provisional launch date after 2025. It is being led by the Astro Space Center of the P.N. Lebedev Physical Institute of the Russian Academy of Science in cooperation with numerous Russian and international organizations. The main industrial partner is the Reshetnev Information Satellite Systems Company.

\begin{figure} \includegraphics[width=\linewidth]{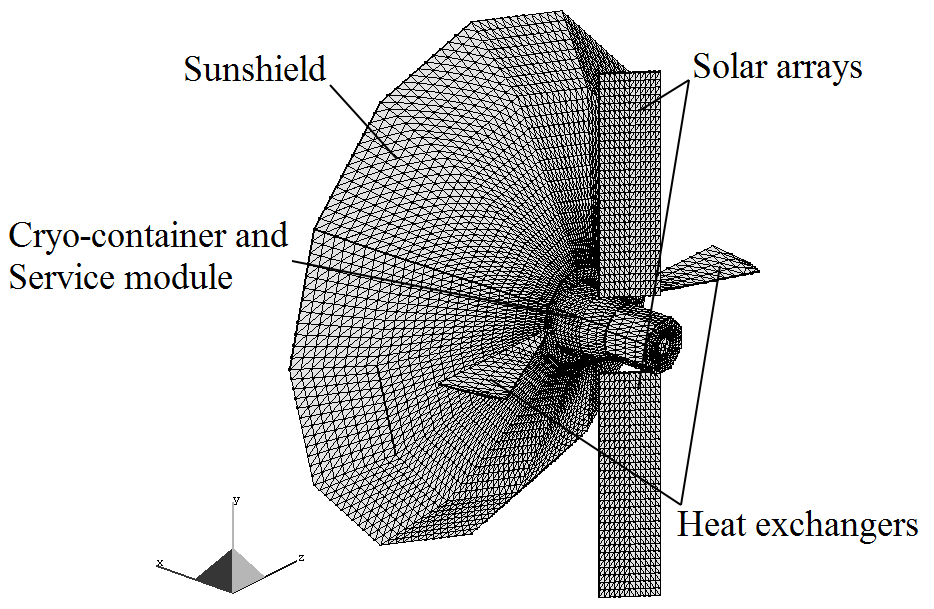} \caption{Geometrical model of Millimetron space observatory.} \label{fig:mm} \end{figure}

	The geometric model of the observatory is shown in Fig.~\ref{fig:mm}.
	It should be noted that configuration of solar panels is still undecided and thus we analyze the linear design of solar panels in the model.

	\subsection{Description of Monte--Carlo raytracing} 
	
	As a raytracing software, we used the Tracer program which was developed in Bauman   MSTU~\cite{leonov_2012_1}. 
	
	This software is used for analysis of radiation heat transfer in complex space structures. This software can use both specular, diffuse and specular--diffuse model of surface parameters.  
	
	To calculate the light pressure force we modified the modules which responsible for the processing of absorption of the light ray and its re-radiation (diffuse or specular or combined).  Geometrical model of the structure is represented as a set of triangular elements. Representation of surface by simple elements can also be useful for calculation of the matrix of inertia of structure~\cite{dobrovolskis_inertia_1996} in the case of analysis of its angular motion.
	See~\cite{simonelli_generation_1993} for the reconstruction of asteroid's shape using a mesh of simple elements.
	
	\begin{figure} \includegraphics[width=.5\linewidth]{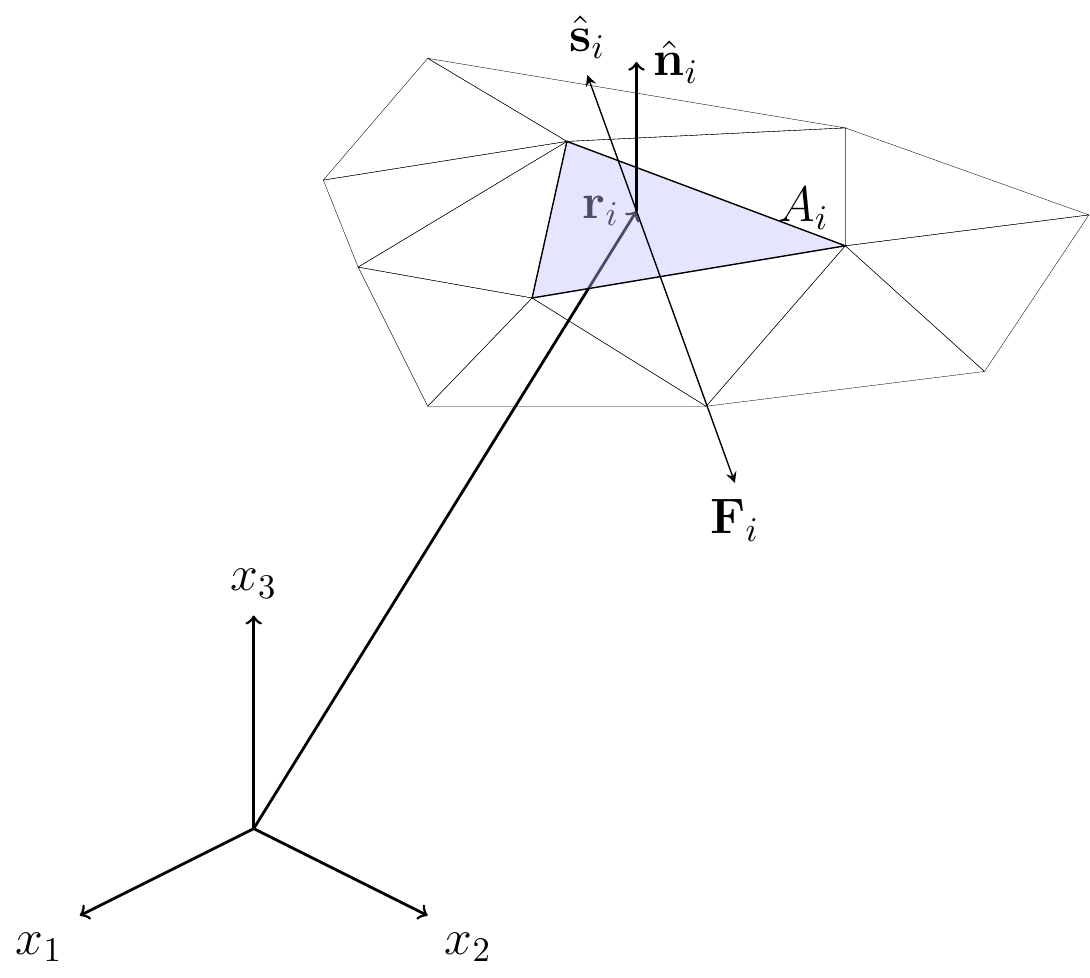} \caption{Geometrical model for determination of light radiation pressure force on the triangle element of surface (e.g. from the emitted light ray)} \label{f:calc_absorb} \end{figure} 
	
	In the case of ray absorption the program records the energy of incident beam $I_{ik}$, where $i$ and $k$ -- number of triangle element and number of incident beam for this element, respectively. Then, the program calculates the components of force and moment from the absorbed light as follows: \begin{align} &\vec{F}_{ik} = \frac{I_{ik} A_i \unit{n}_i\cdot\unit{s}_{ik}}{c}\unit{s}_{ik};\\ &\vec{M}_{ik} = \vec{r}_i\times\vec{F}_{ik}, \end{align} where $A_i$ --- area of triangle element; $\unit{n}_i$ --- normal to triangle element; $\unit{s}_{ik}$ --- direction vector for the incident beam; $\vec{r}_i$ --- position vector of the current triangle element, pointed to the center of triangle. 
	
	After processing of beam absorption event the software calculates the set of new beams which has to be emitted from the element $A_i$. The outgoing beams (fig.~\ref{f:calc_absorb}) were considered similar to the incoming beams: \begin{align} &\vec{F}_{im} = -\frac{I_{im} A_i \unit{n}_i\cdot\unit{s}_{im}}{c}\unit{s}_{im};\\ &\vec{M}_{im} = \vec{r}_i\times\vec{F}_{im}, \end{align} where $m$ --- number of outgoing beam. 
	
	The main vector and main moment of light pressure can be calculated as follows: \begin{align} &\vec{F} = \sum\limits_{i=1}^{N_i}\left( \sum\limits_{k=1}^{N_{ik}}\vec{F}_{ik} +\sum\limits_{m=1}^{N_{im}}\vec{F}_{im} \right);\\ &\vec{M} = \sum\limits_{i=1}^{N_i}\left( \sum\limits_{k=1}^{N_{ik}}\vec{M}_{ik} +\sum\limits_{m=1}^{N_{im}}\vec{M}_{im} \right), \end{align} where $N_i$ is a number of triangle elements; $N_{ik}$ --- number of beams which were absorbed by the element with number $i$; $N_{im}$ --- number of beams which were emitted by the element with number $i$. 
	
	The data on the main vector and main moment together with the original data on the direction of the external radiation vector were processed in the program implemented in GNU Octave software package \cite{gnu_octave}. The program did the numerical evaluation of expressions~\eqref{eq:approx_j} and~\eqref{eq:approx_k} and reconstructed tensors $\tensor{I}$ and $\tensor{K}$ and provided the necessary charts. 
	
	\subsection{Approximation results} 
	
	The principal force and moment of light pressure were calculated at 60 different orientations of the incident light. The resultant moment was calculated about the origin, located in the center of the outer edge of the outer heat shield. 
	
	In all cases, the estimated number of cast rays was equal to 1,000,000. It was assumed that the spacecraft radiators remain stationary when changing the orientation about the Sun, and solar panels are rotated to ensure the minimum angle of incidence of the light. All calculations neglected components of the spacecraft structure within the outer heat shield, as in these cases of orientation, when radiation from the sun falls into the bowl of the reflector, it is not compatible with the continued operation of the spacecraft as a telescope. 
	
\begin{figure} \begin{minipage}[h]{0.49\linewidth} \center{\includegraphics[width=1\linewidth]{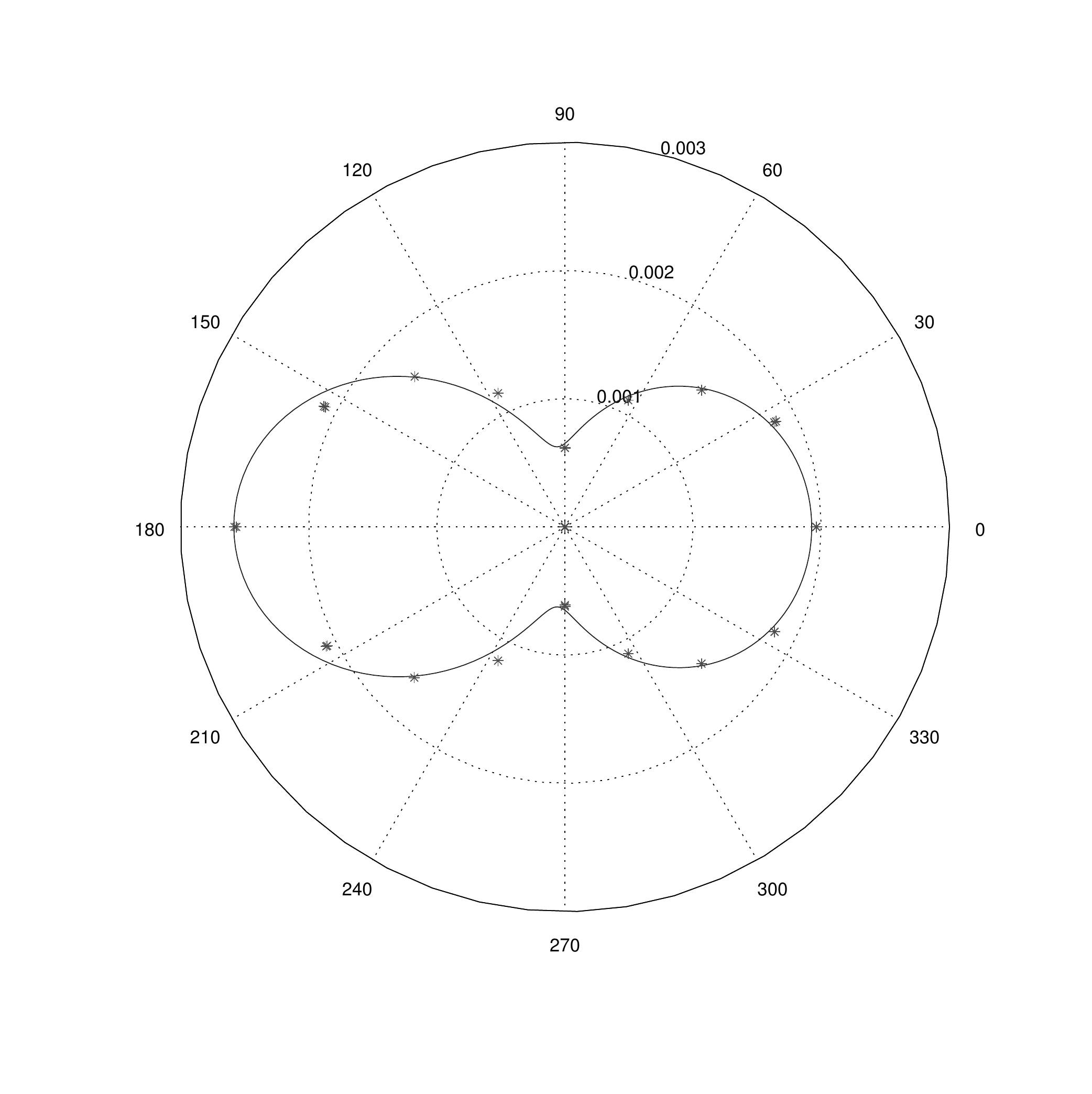}} a \end{minipage} \hfill
		\begin{minipage}[h]{0.49\linewidth} \center{\includegraphics[width=1\linewidth]{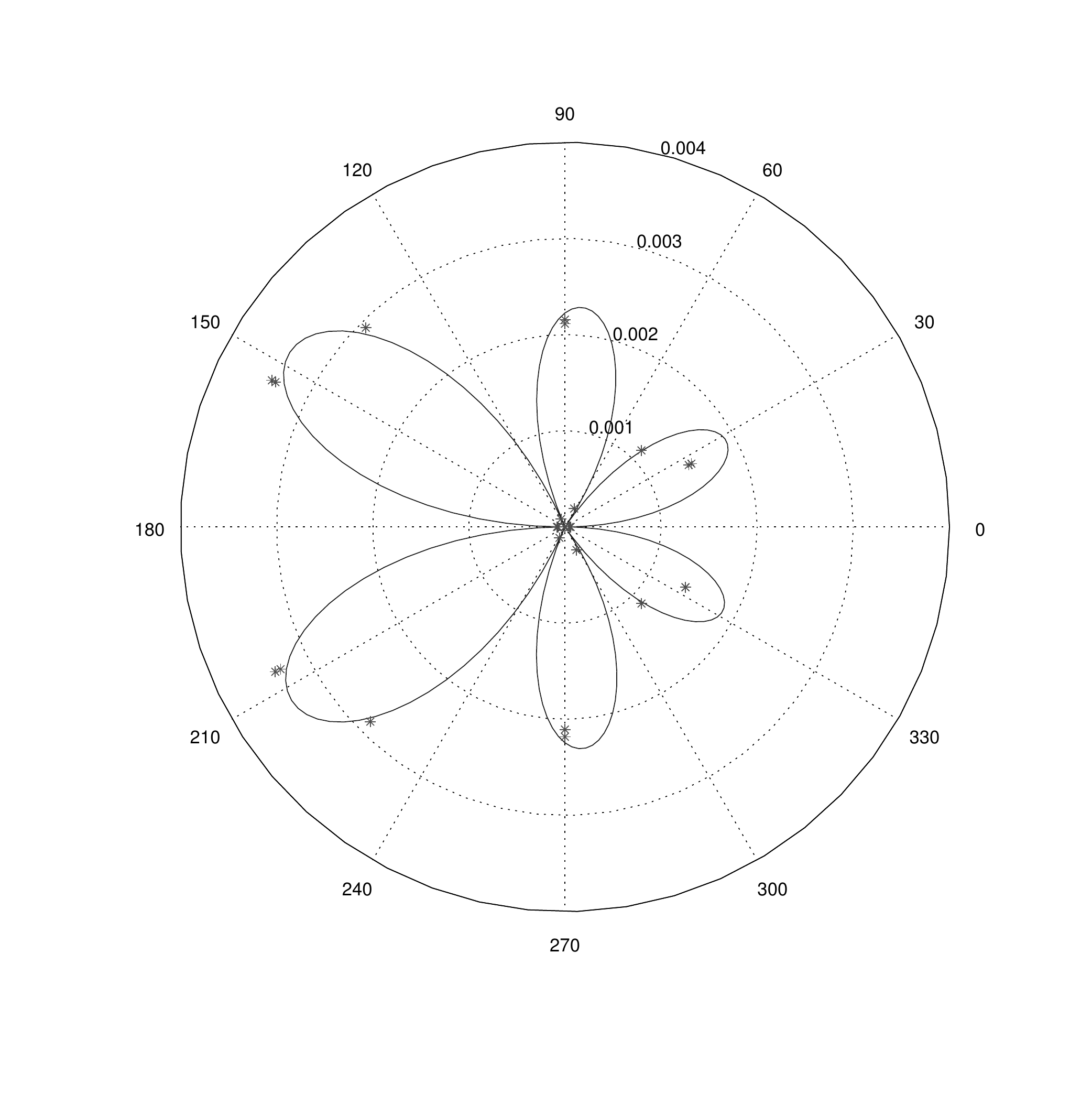}} b \end{minipage} 
		
		\caption{Dependency of the absolute value of resultant force and moment of light radiation pressure from the angle of rotation of light source in the radiators plane. Solid line -- tensor approximation ($N_{\max}=3$), dots -- Monte--Carlo simulation results which were not used for the approximation. Figure a -- absolute value of resultant force, $N$; Figure b -- absolute value of the resultant moment, $N\cdot m$. The whole surface is specular.} \label{fig:res2} \end{figure}

\begin{figure} \begin{minipage}[h]{0.49\linewidth} \center{\includegraphics[width=1\linewidth]{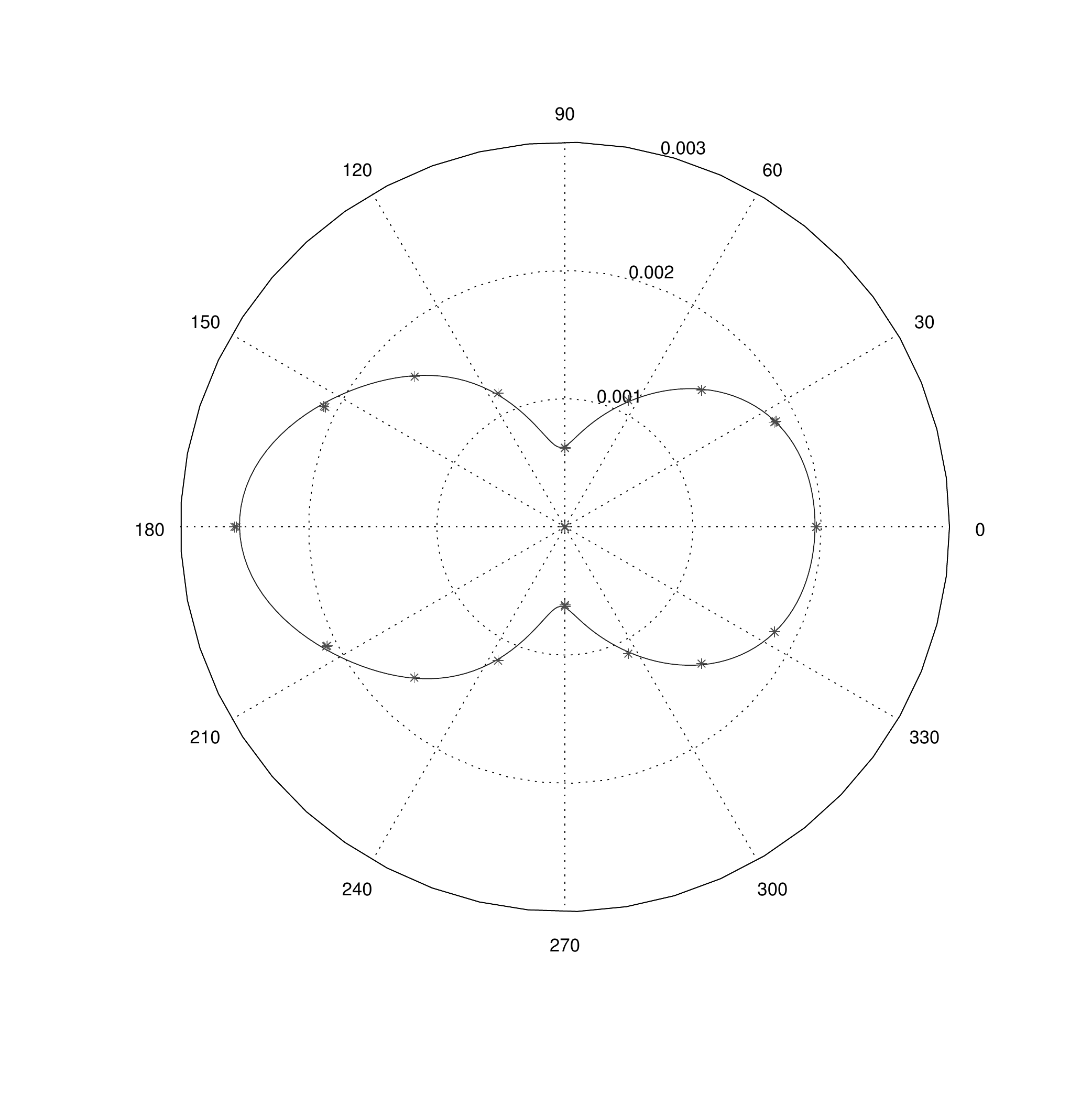}} a \end{minipage} \hfill
		\begin{minipage}[h]{0.49\linewidth} \center{\includegraphics[width=1\linewidth]{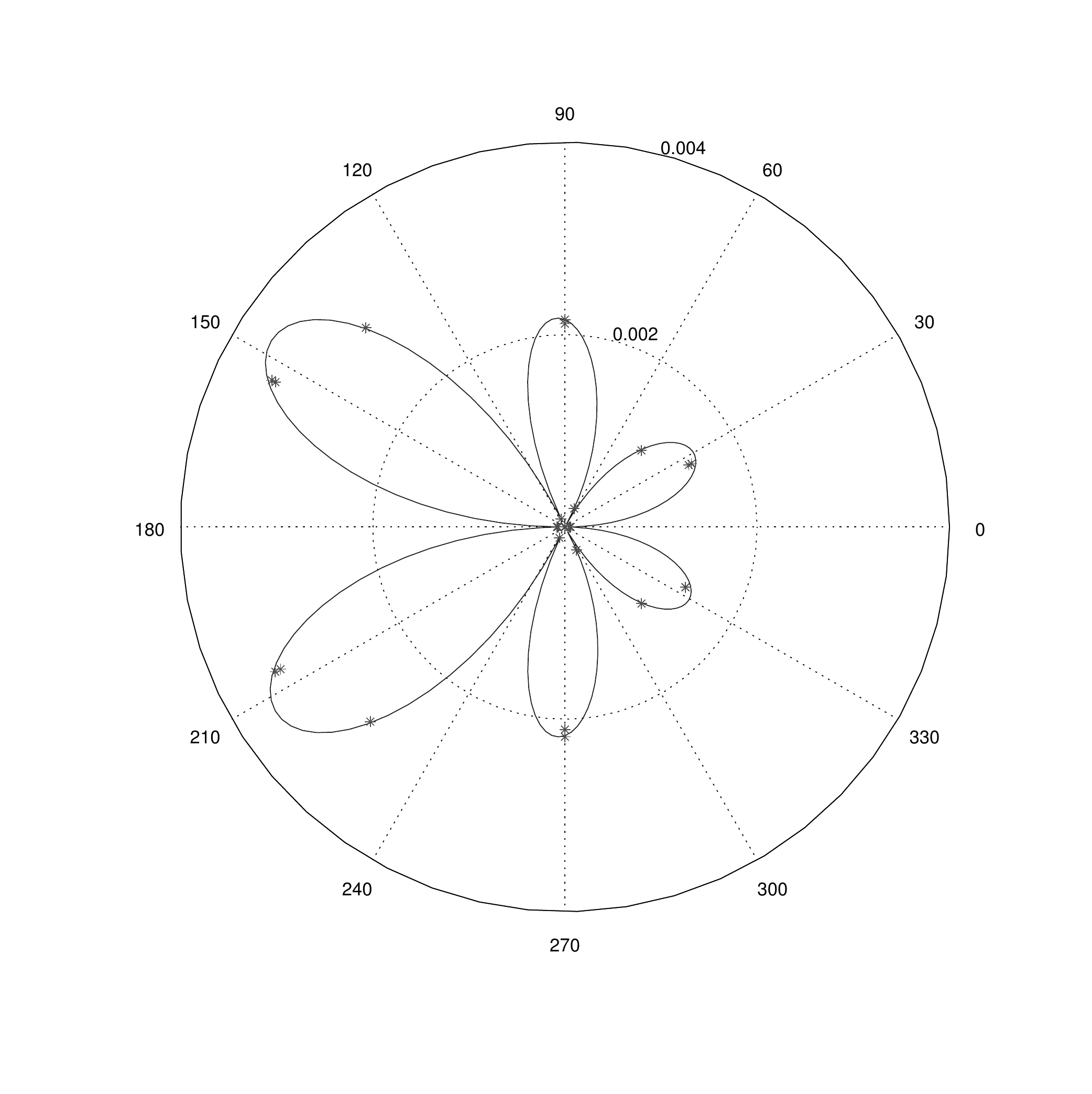}} b \end{minipage} 
		
		\caption{Dependency of the absolute value of resultant force and moment of light radiation pressure from the angle of rotation of light source in the radiators plane. Solid line -- tensor approximation ($N_{\max}=6$), dots -- Monte--Carlo simulation results which were not used for the approximation. Figure a -- absolute value of resultant force, $N$; Figure b -- absolute value of the resultant moment, $N\cdot m$. The whole surface is specular.} \label{fig:res3} \end{figure}

\begin{figure} \begin{minipage}[h]{0.49\linewidth} \center{\includegraphics[width=1\linewidth]{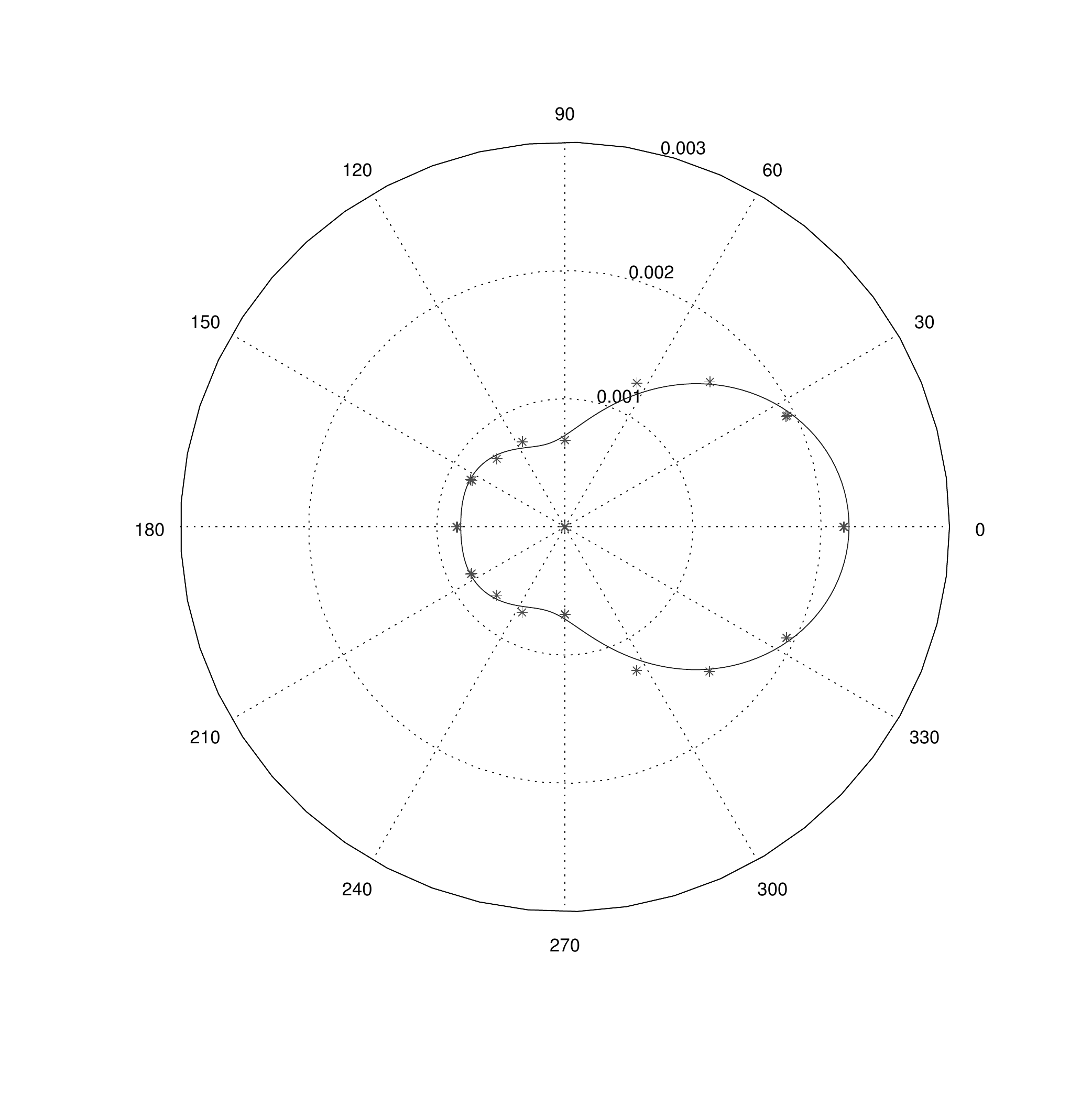}} a \end{minipage} \hfill
		\begin{minipage}[h]{0.49\linewidth} \center{\includegraphics[width=1\linewidth]{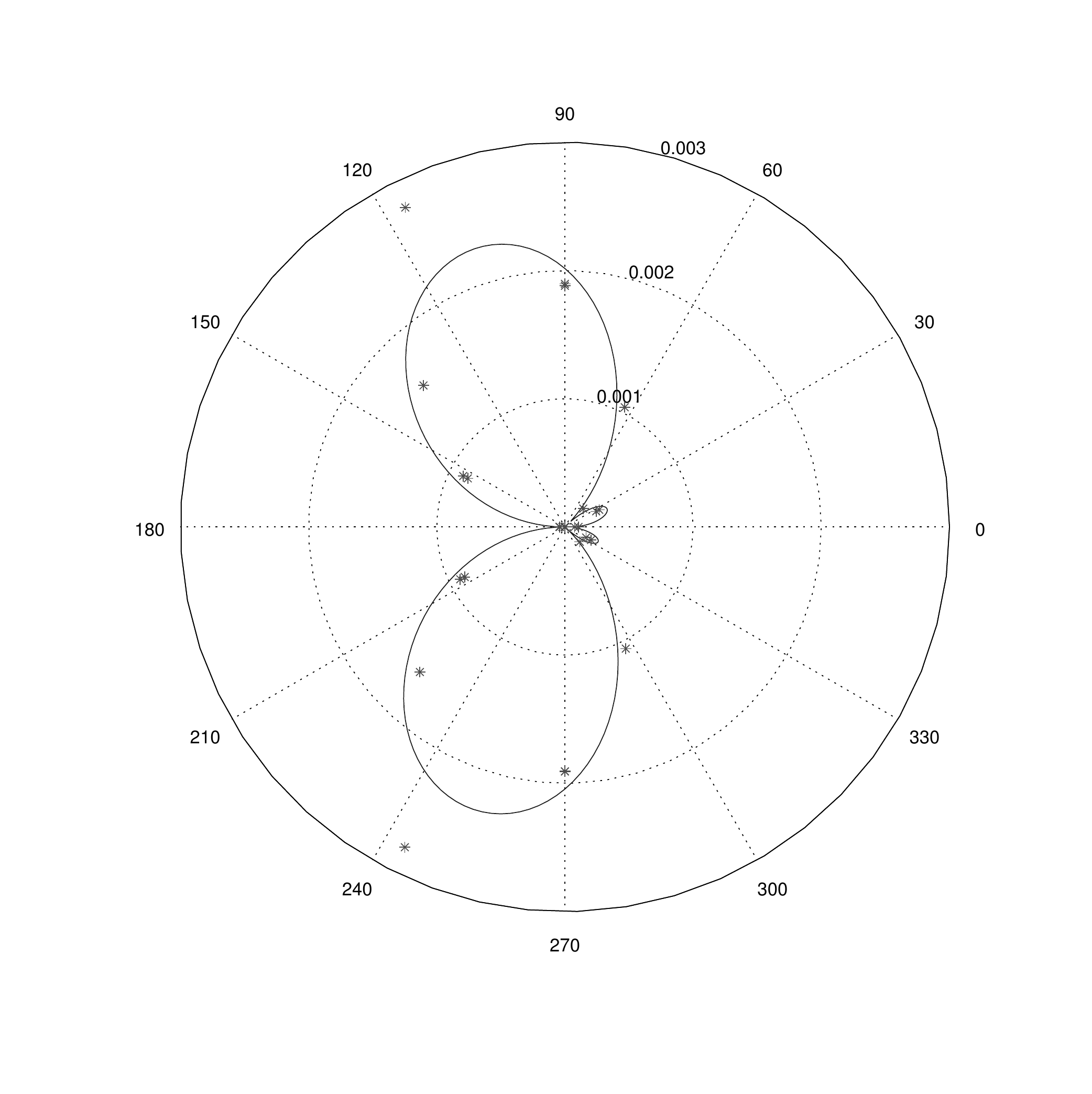}} b \end{minipage} 
		
		\caption{Dependency of the absolute value of resultant force and moment of light radiation pressure from the angle of rotation of light source in the radiators plane. Solid line -- tensor approximation ($N_{\max}=3$), dots -- Monte--Carlo simulation results which were not used for the approximation. Figure a -- absolute value of resultant force, $N$; Figure b -- absolute value of the resultant moment, $N\cdot m$. The whole surface is diffuse.} \label{fig:res5} \end{figure}

\begin{figure} \begin{minipage}[h]{0.49\linewidth} \center{\includegraphics[width=1\linewidth]{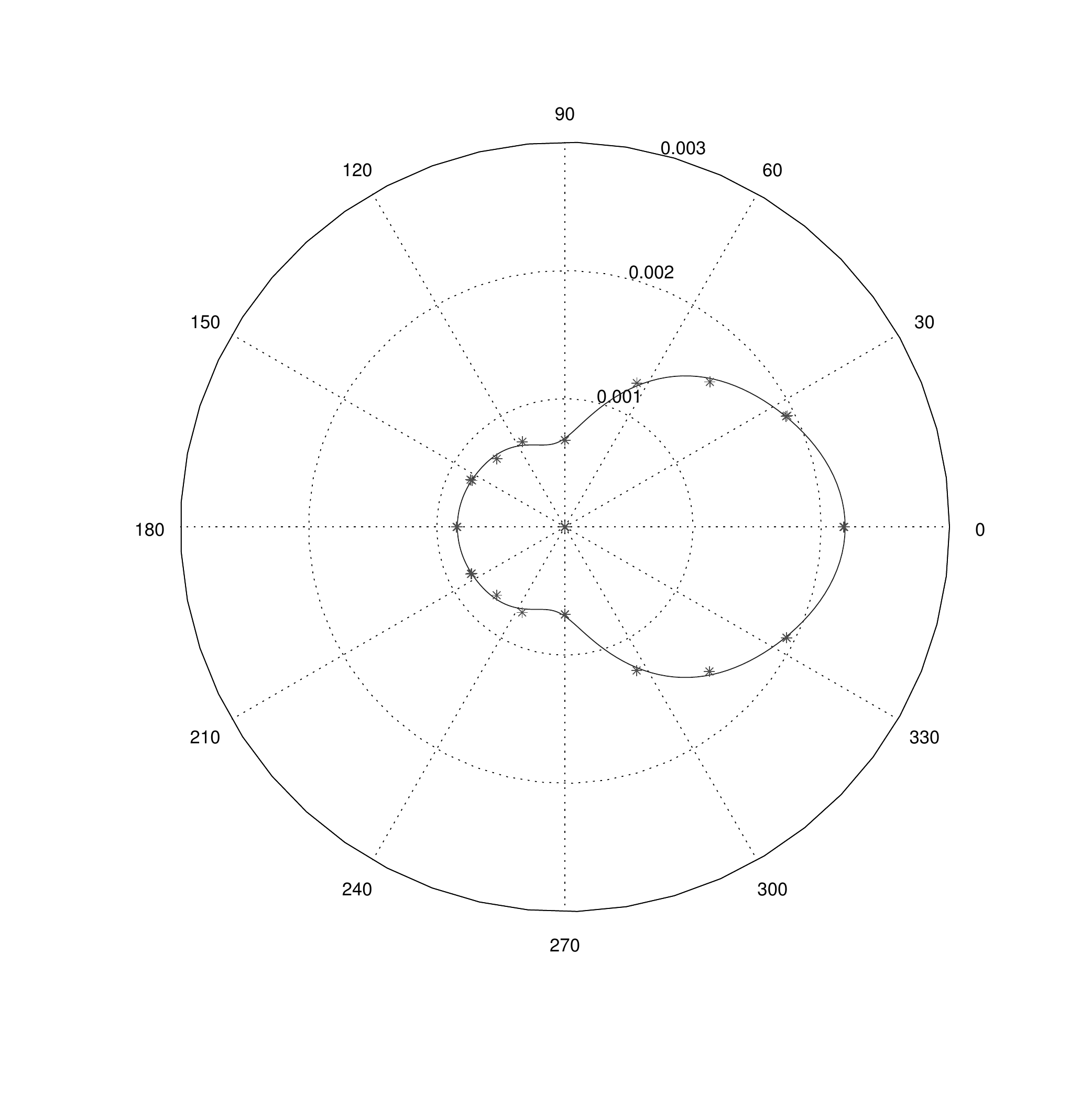}} a \end{minipage} \hfill
		\begin{minipage}[h]{0.49\linewidth} \center{\includegraphics[width=1\linewidth]{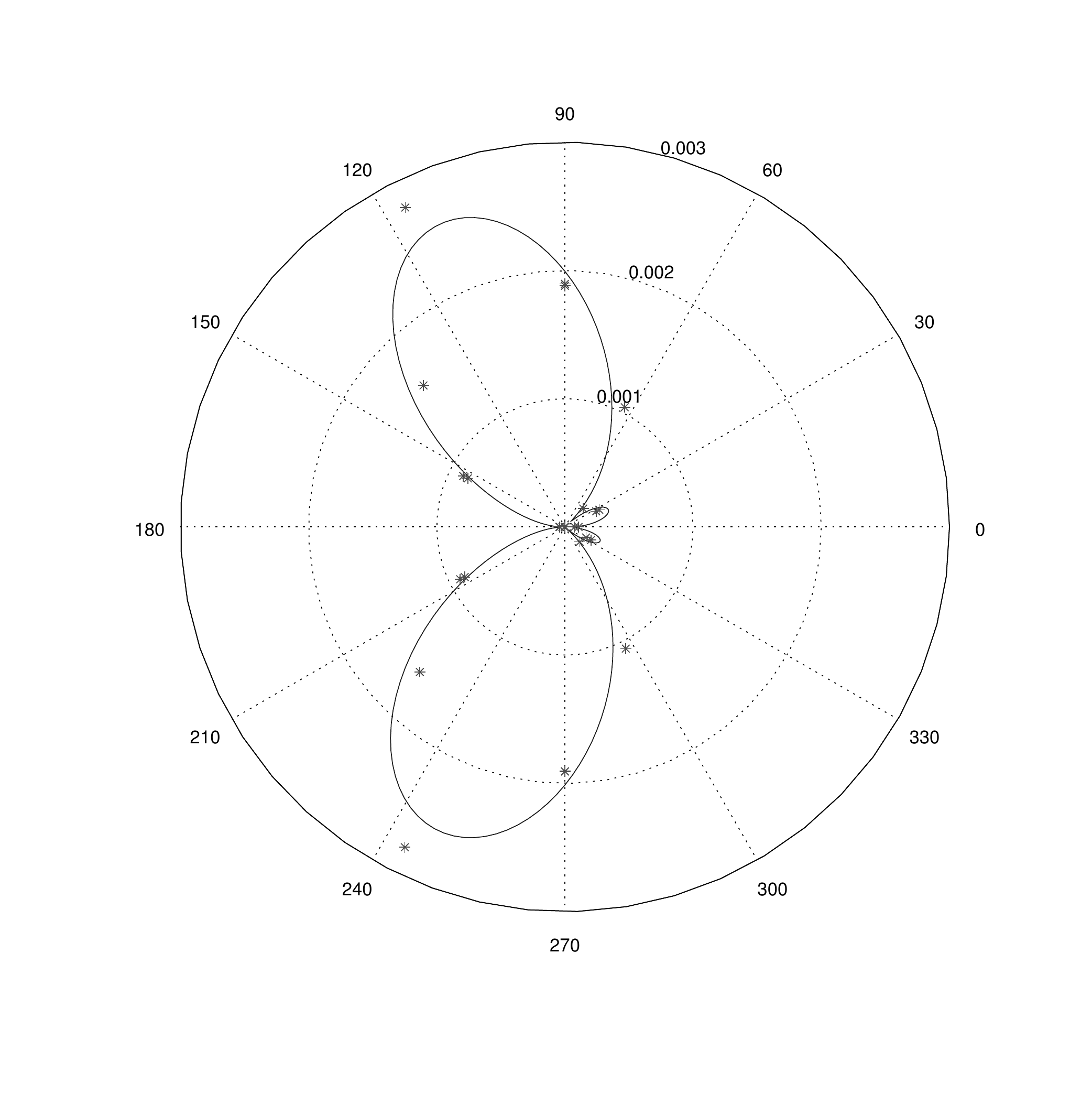}} b \end{minipage} 
		
		\caption{Dependency of the absolute value of resultant force and moment of light radiation pressure from the angle of rotation of light source in the radiators plane. Solid line -- tensor approximation ($N_{\max}=6$), dots -- Monte--Carlo simulation results which were not used for the approximation. Figure a -- absolute value of resultant force, $N$; Figure b -- absolute value of the resultant moment, $N\cdot m$. The whole surface is diffuse.} \label{fig:res6} \end{figure}

\begin{figure} \begin{minipage}[h]{0.49\linewidth} \center{\includegraphics[width=1\linewidth]{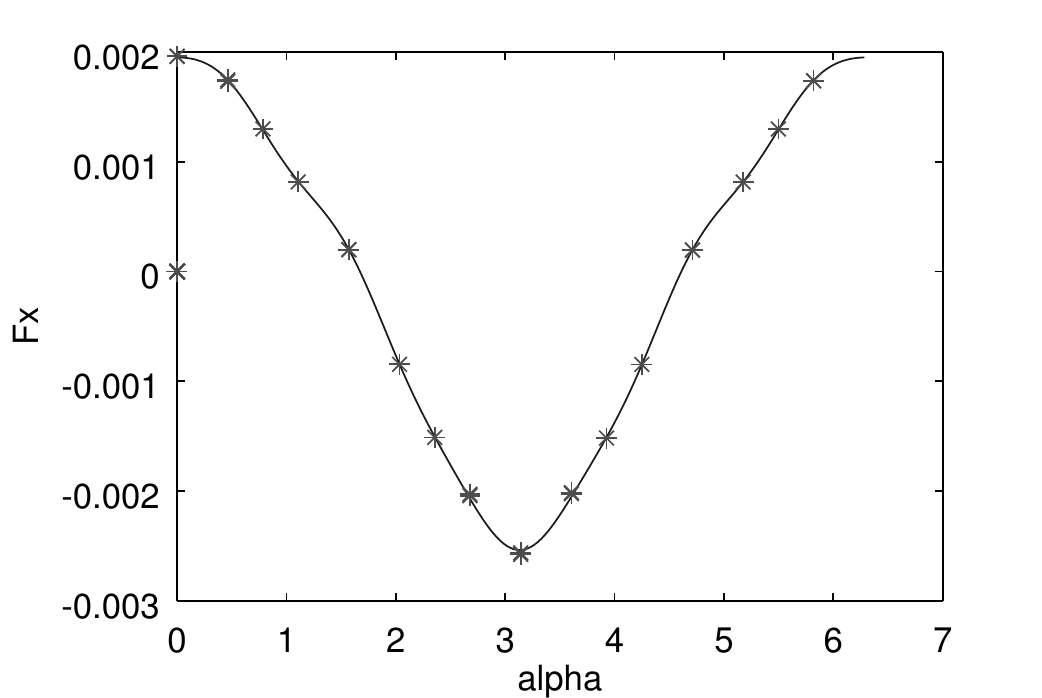}} a \end{minipage} \hfill
		\begin{minipage}[h]{0.49\linewidth} \center{\includegraphics[width=1\linewidth]{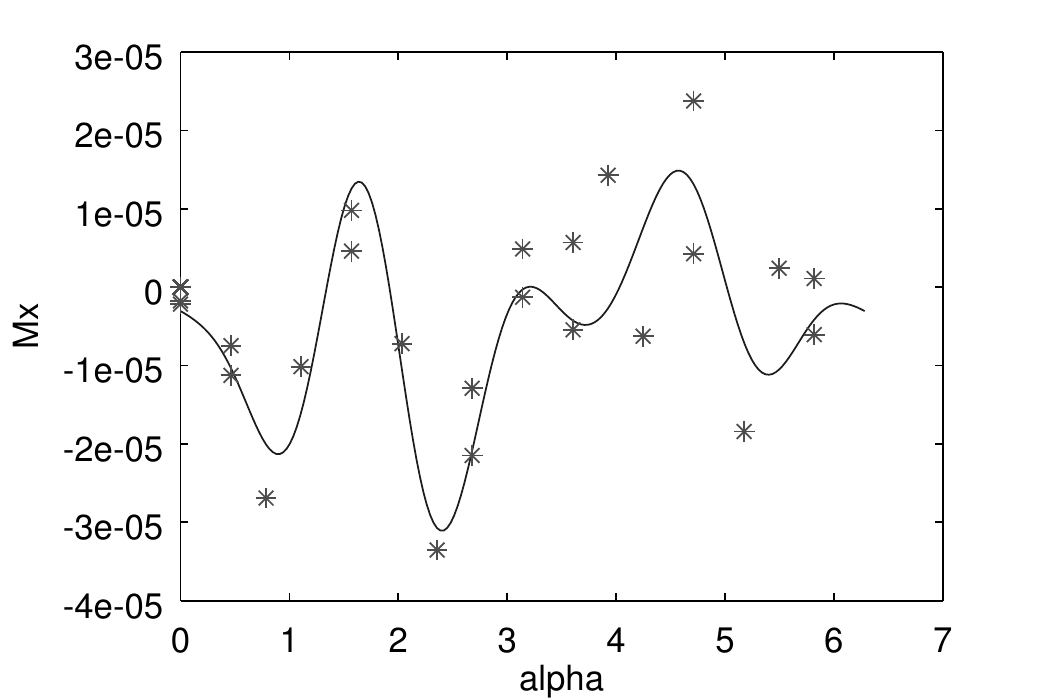}} b \end{minipage} \vfill
		\begin{minipage}[h]{0.49\linewidth} \center{\includegraphics[width=1\linewidth]{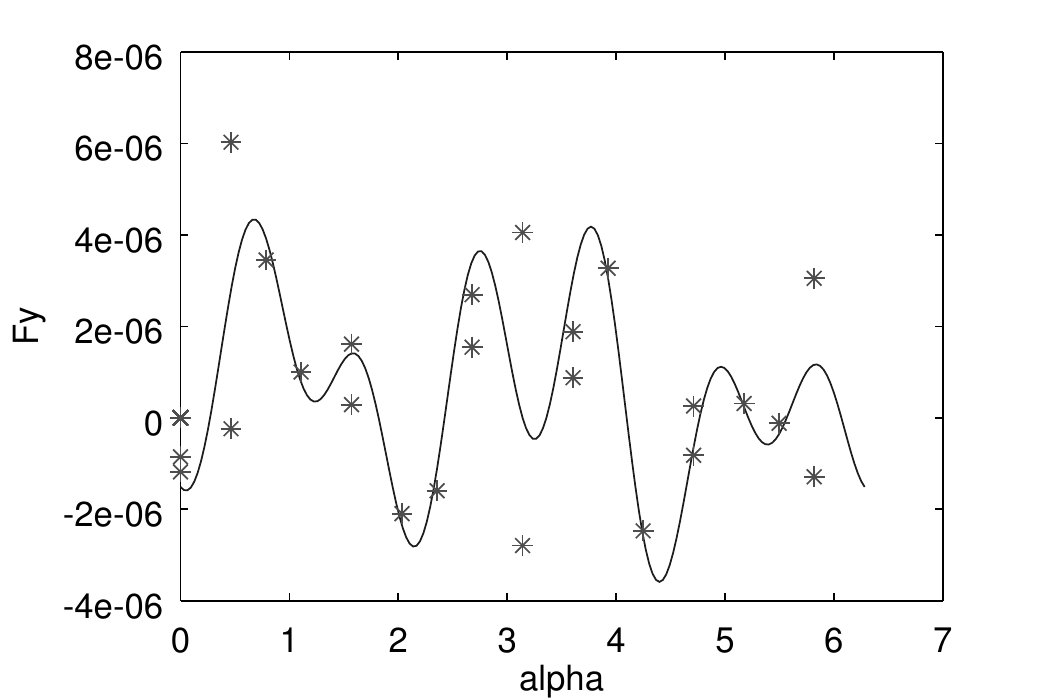}} c \end{minipage} \hfill
		\begin{minipage}[h]{0.49\linewidth} \center{\includegraphics[width=1\linewidth]{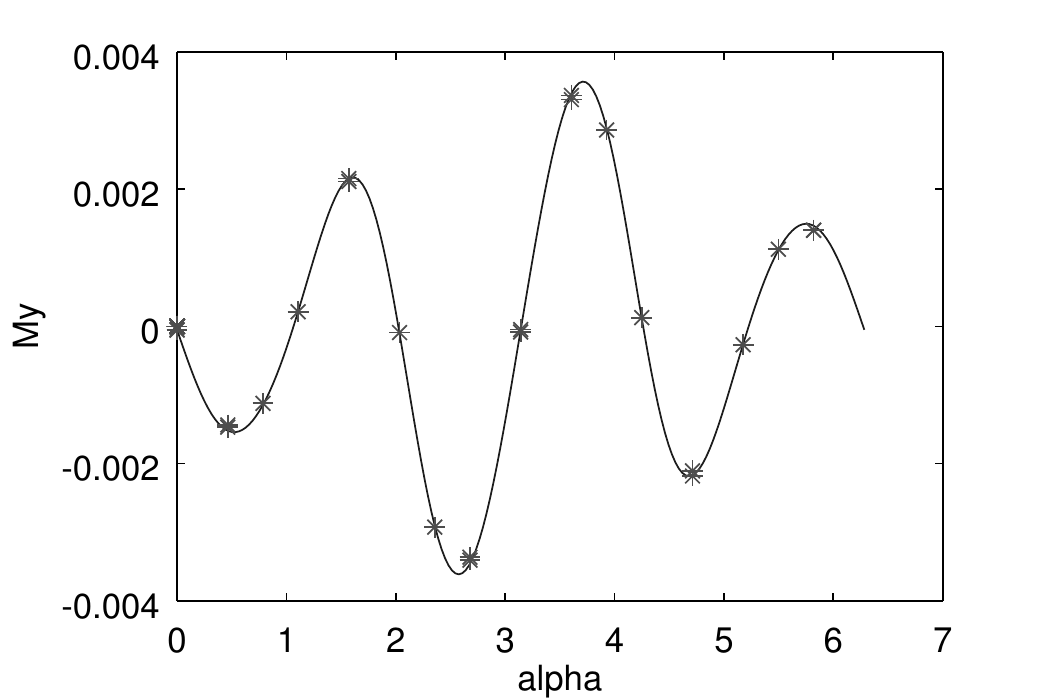}} d \end{minipage} \vfill
		\begin{minipage}[h]{0.49\linewidth} \center{\includegraphics[width=1\linewidth]{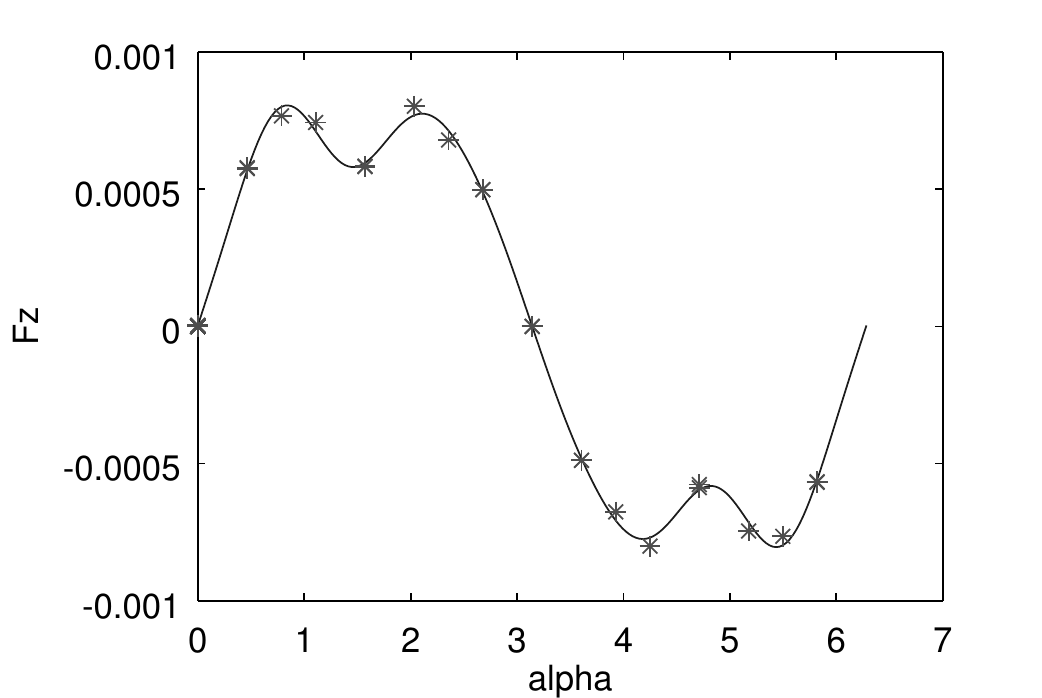}} e \end{minipage} \hfill
		\begin{minipage}[h]{0.49\linewidth} \center{\includegraphics[width=1\linewidth]{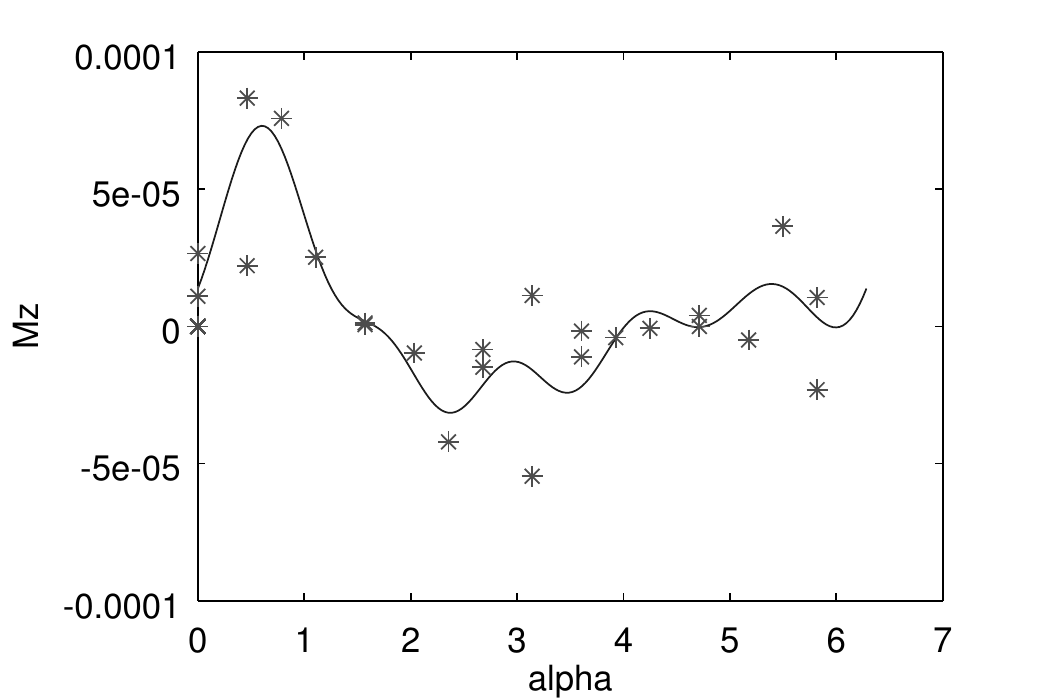}} f \end{minipage} 
		
		\caption{Approximation results ($N_{\max}=6$) for principal force (a, c, e), $N$, and for principal moment (b, c, f), $N\cdot m$, of light radiation pressure depending on angle of rotation of light source in the plane of radiators (solid line) comparing results of Monte--Carlo simulations (dots). The results in subfigures b, c, and f are non-zero because of random noise. The dotted values were not used in the construction of approximation. Specular case.} \label{fig:res_proj_1} \end{figure}

\begin{figure} \begin{minipage}[h]{0.49\linewidth} \center{\includegraphics[width=1\linewidth]{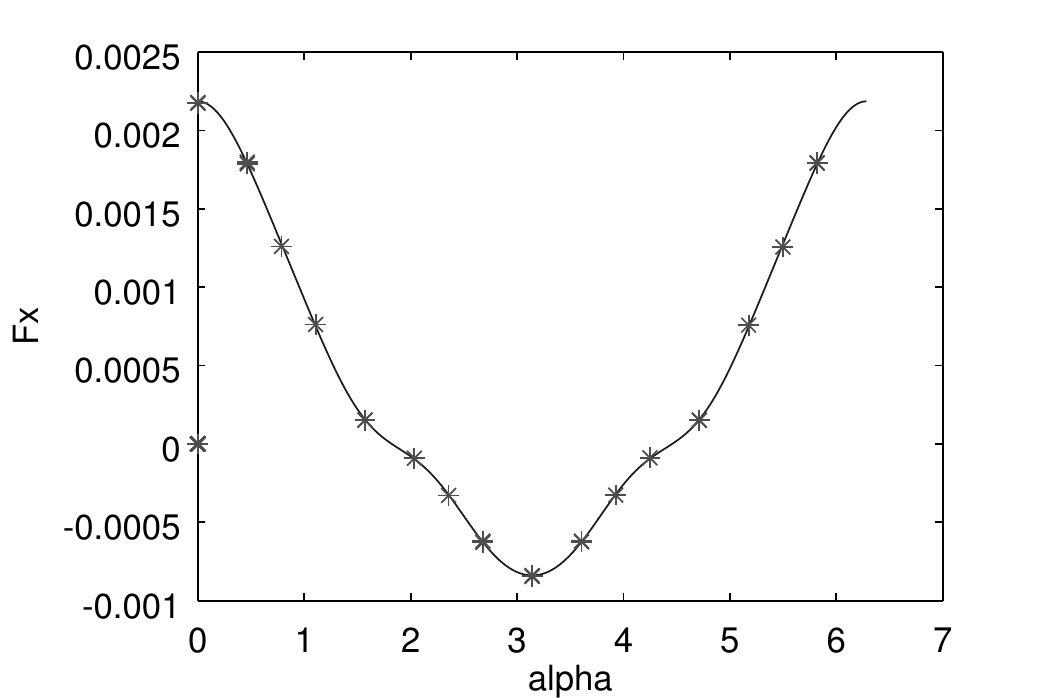}} a \end{minipage} \hfill
		\begin{minipage}[h]{0.49\linewidth} \center{\includegraphics[width=1\linewidth]{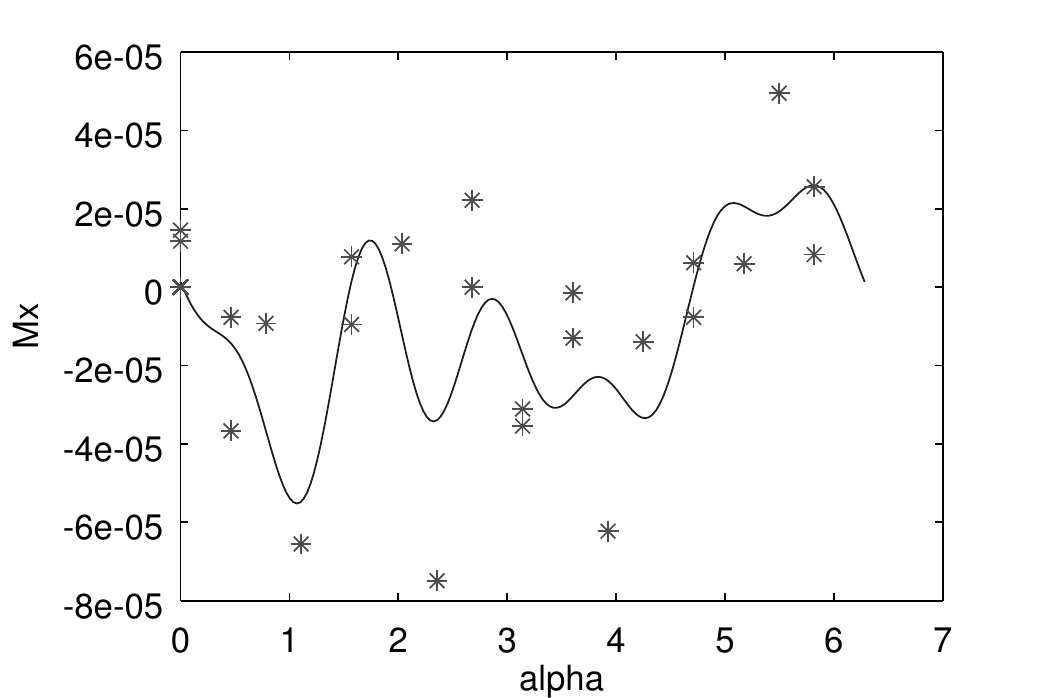}} b \end{minipage} \vfill
		\begin{minipage}[h]{0.49\linewidth} \center{\includegraphics[width=1\linewidth]{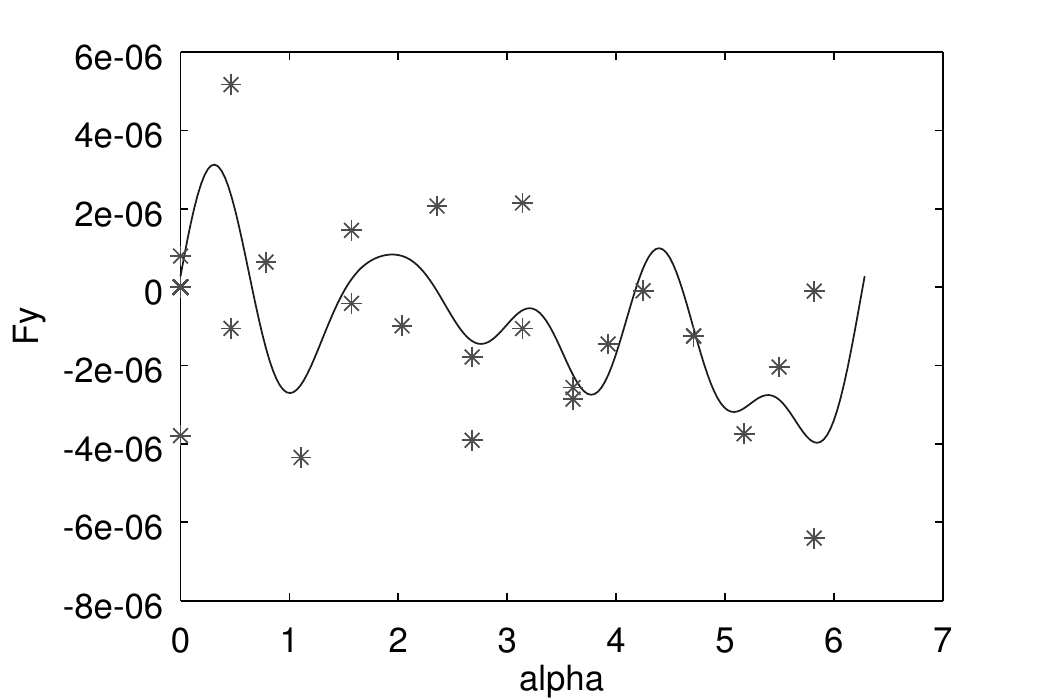}} c \end{minipage} \hfill
		\begin{minipage}[h]{0.49\linewidth} \center{\includegraphics[width=1\linewidth]{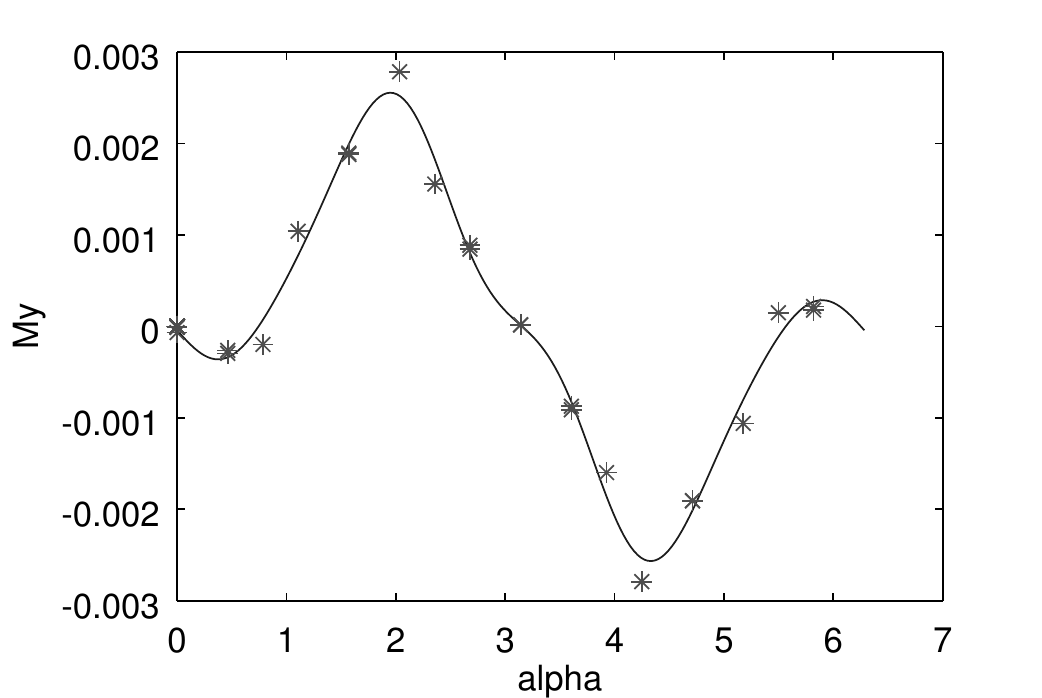}} d \end{minipage} \vfill
		\begin{minipage}[h]{0.49\linewidth} \center{\includegraphics[width=1\linewidth]{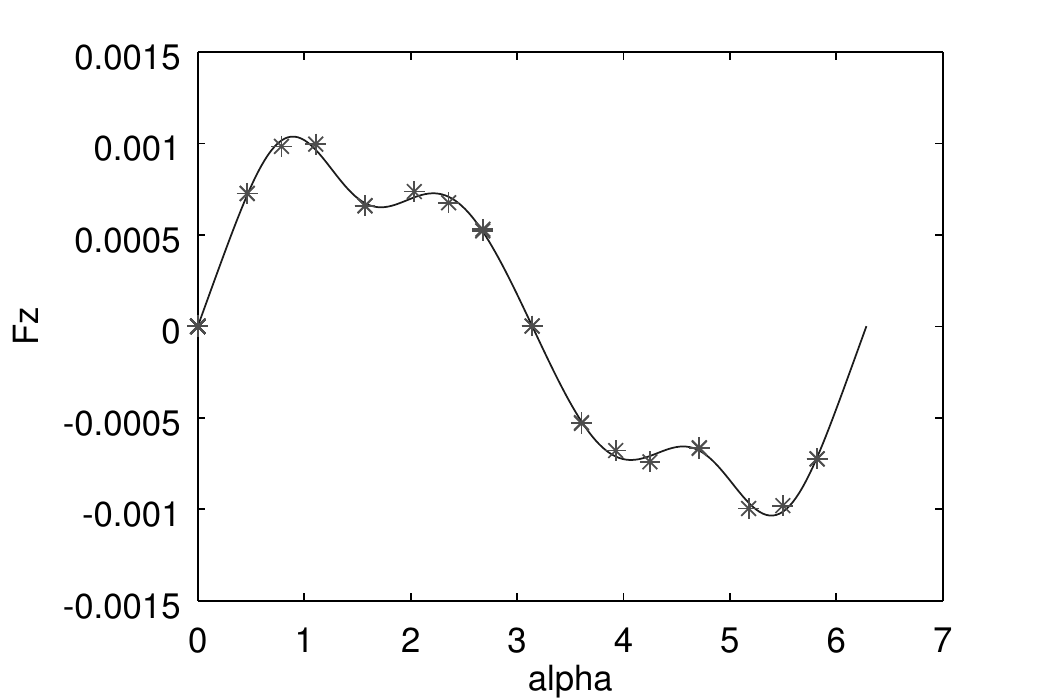}} e \end{minipage} \hfill
		\begin{minipage}[h]{0.49\linewidth} \center{\includegraphics[width=1\linewidth]{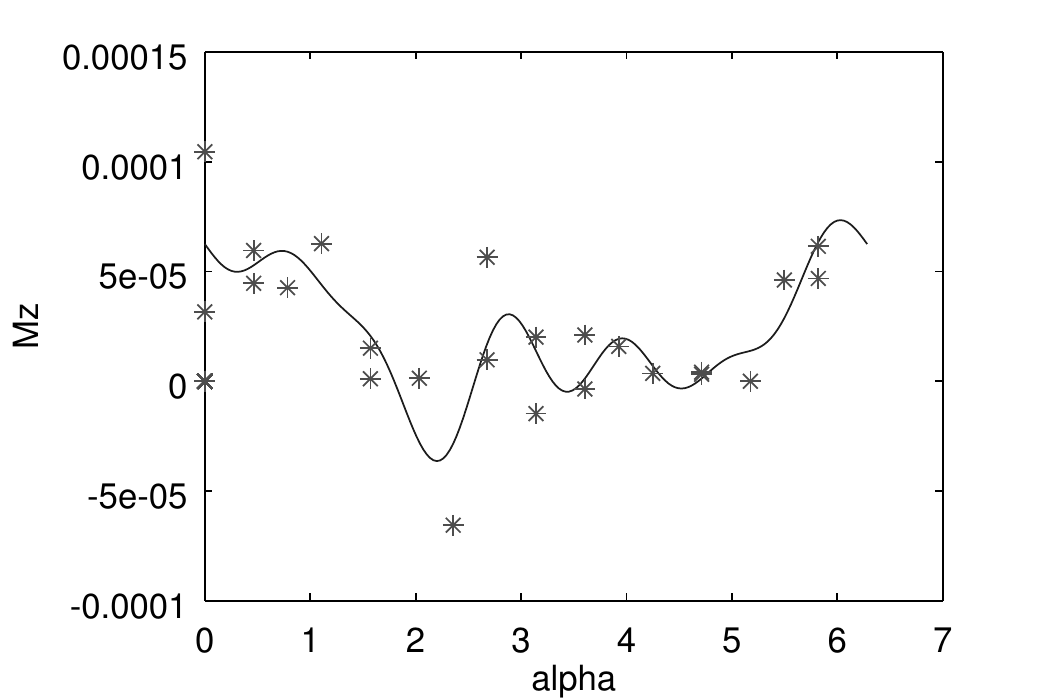}} f \end{minipage} 
		
		\caption{Approximation results ($N_{\max}=6$) for principal force (a, c, e), $N$, and for principal moment (b, c, f), $N\cdot m$, of light radiation pressure depending on angle of rotation of light source in the plane of radiators (solid line) comparing results of Monte--Carlo simulations (dots). The results in subfigures b, c, and f are non-zero because of random noise. The dotted values were not used in the construction of approximation. Diffuse case.} \label{fig:res_proj_2} \end{figure}

	Fig.~\ref{fig:res2} --- \ref{fig:res6} represent the calculation results for absolute value of resultant force and moment. Fig.~\ref{fig:res2} --- \ref{fig:res3} correspond to the case when the entire spacecraft is fully specular, and Fig.~\ref{fig:res5} --- \ref{fig:res6} --- when the entire surface of spacecraft is diffuse. The angle ranges $[-90\degree,+90\degree]$  corresponds to the ordinary orientation of space observatory. As it follows from the figures, with the increasing of numbers of terms in series~\eqref{eq:F_final} and~\eqref{eq:M_final} the force and moment approximation accuracy are also increasing. Depending on the task the number of terms in the series \eqref{eq:F_final} and~\eqref{eq:M_final} may be restricted to achieve the acceptable level of approximation accuracy. Fig.~\ref{fig:res_proj_1} and~\ref{fig:res_proj_2} represent the components of principal force and moment in the specular and diffuse cases, respectively. In this model, the approximation of the resultant force converges fast enough to the known values ​​obtained by stochastic simulation. In the case of the approximation of the moment, it may require a greater number of terms for a more precise approximation. It is worth noting that the largest error of approximation of the principal moment is observed in the areas corresponding to the abnormal orientation of the test spacecraft.

\begin{figure} \begin{minipage}[h]{0.49\linewidth} \center{\includegraphics[width=1\linewidth]{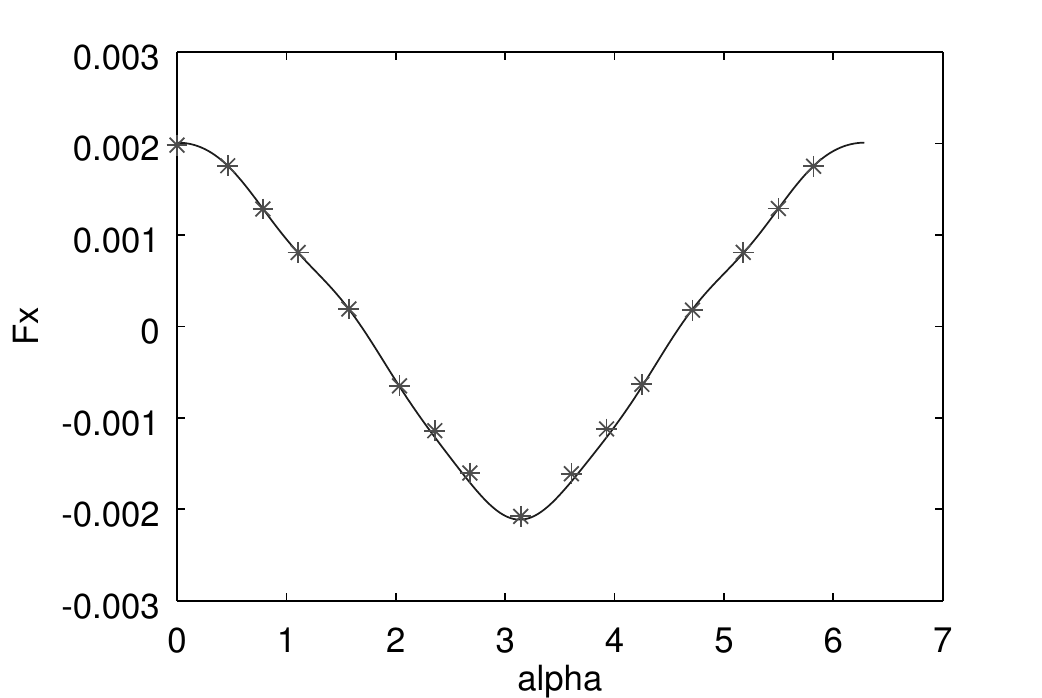}} a \end{minipage} \hfill
		\begin{minipage}[h]{0.49\linewidth} \center{\includegraphics[width=1\linewidth]{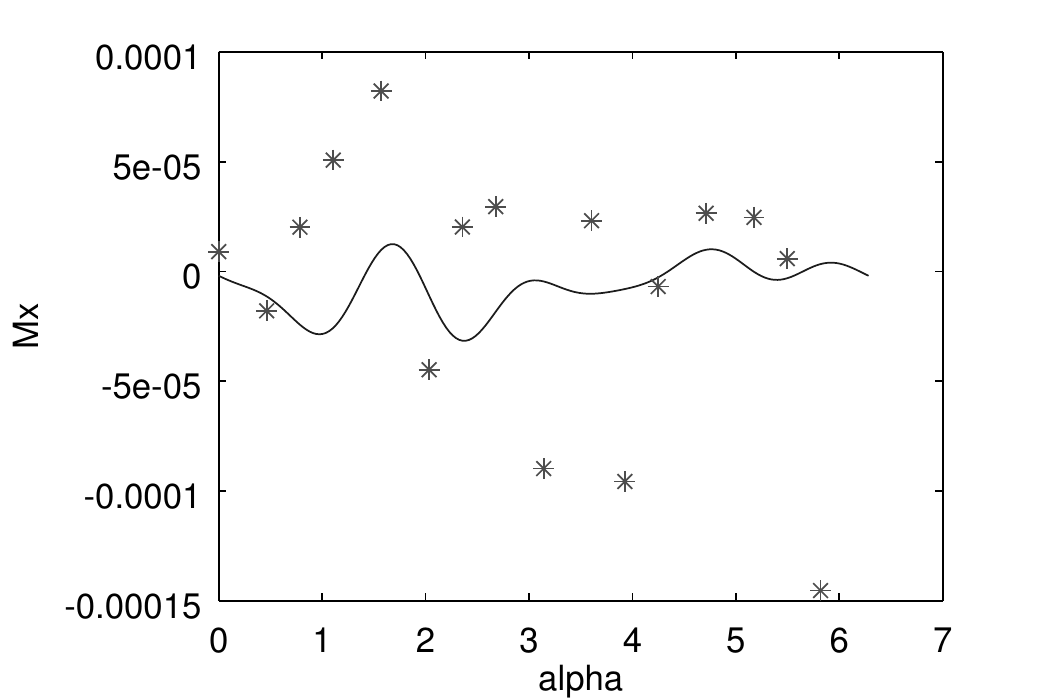}} b \end{minipage} \vfill
		\begin{minipage}[h]{0.49\linewidth} \center{\includegraphics[width=1\linewidth]{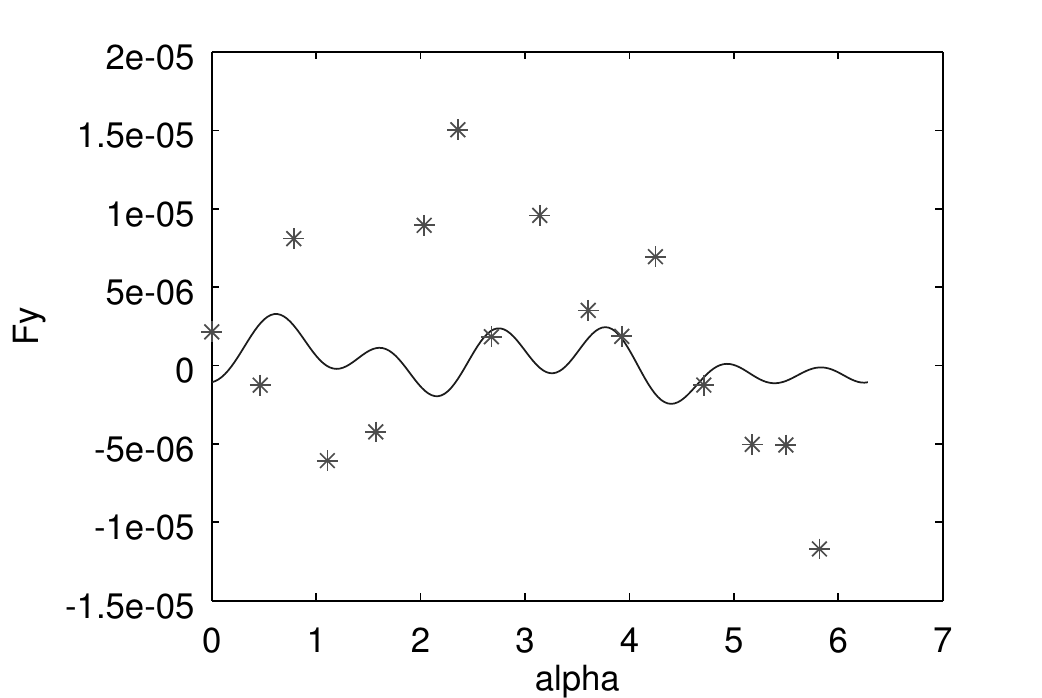}} c \end{minipage} \hfill
		\begin{minipage}[h]{0.49\linewidth} \center{\includegraphics[width=1\linewidth]{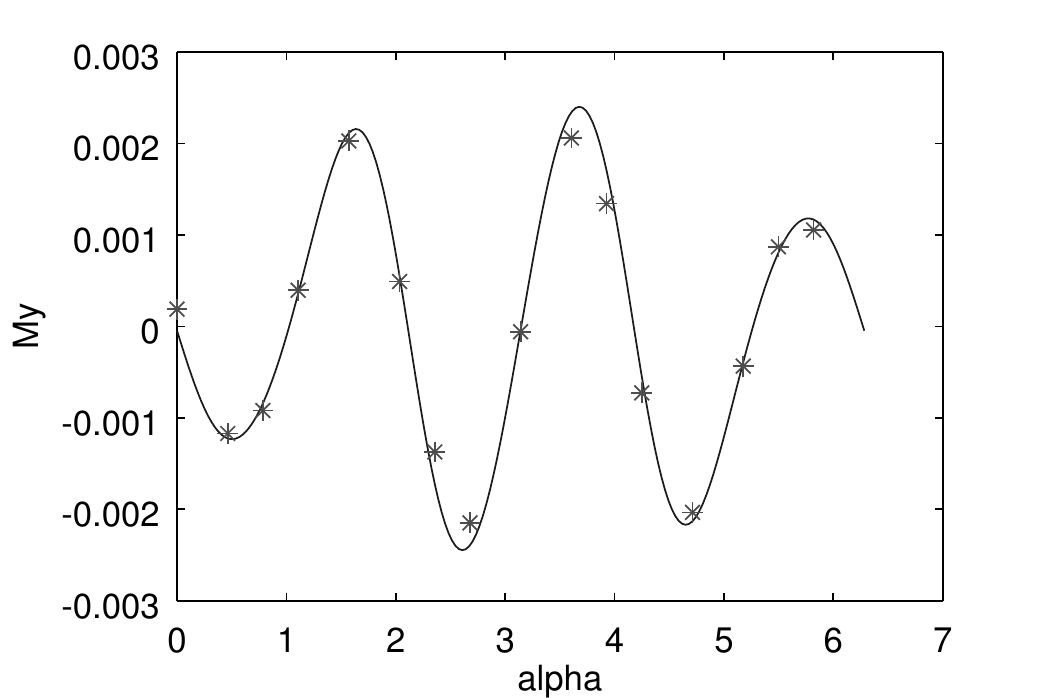}} d \end{minipage} \vfill
		\begin{minipage}[h]{0.49\linewidth} \center{\includegraphics[width=1\linewidth]{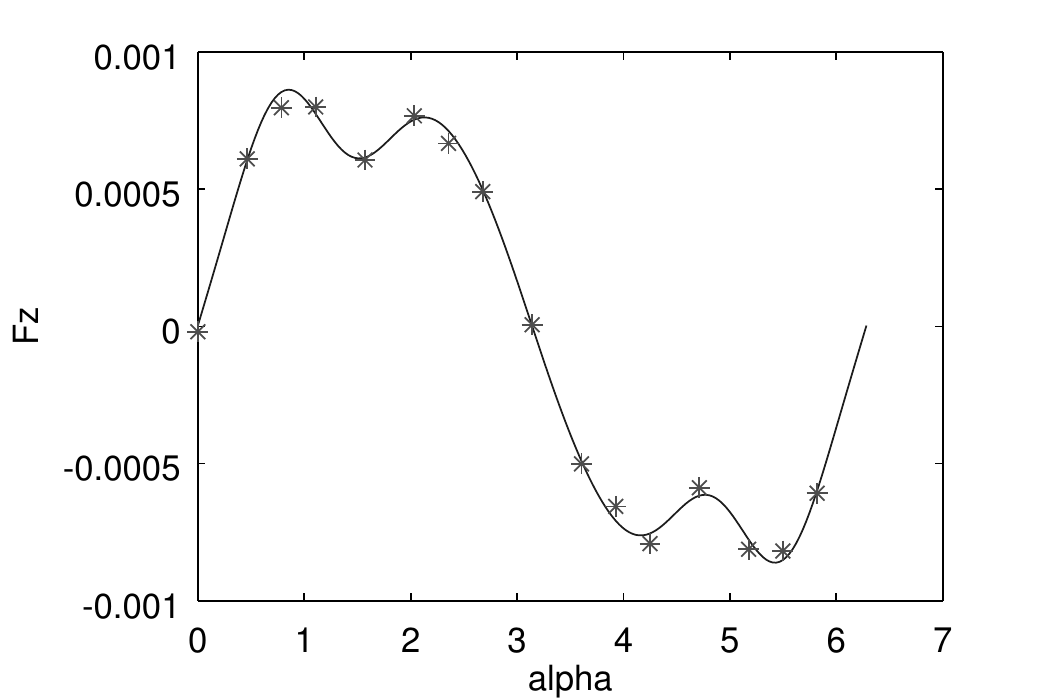}} e \end{minipage} \hfill
		\begin{minipage}[h]{0.49\linewidth} \center{\includegraphics[width=1\linewidth]{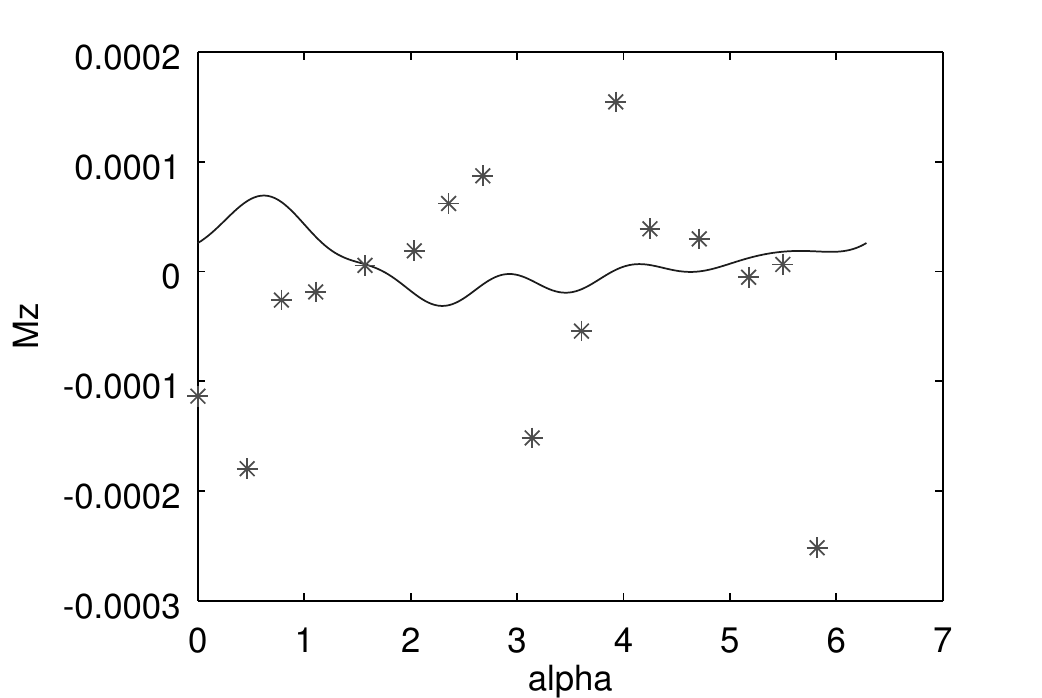}} f \end{minipage} 
		
		\caption{Approximation results ($N_{\max}=6$) for principal force (a, c, e), $N$, and for principal moment (b, c, f), $N\cdot m$, of light radiation pressure depending on angle of rotation of light source in the plane of radiators (solid line) comparing results of Monte--Carlo simulations (dots). The results in subfigures b, c and f are non-zero because of random noise. The approximation was constructed as a linear combination of specular and diffuse cases. Specular--diffuse case, $s=0{.}75$.} \label{fig:res_calc_075} \end{figure}

\begin{figure} \begin{minipage}[h]{0.49\linewidth} \center{\includegraphics[width=1\linewidth]{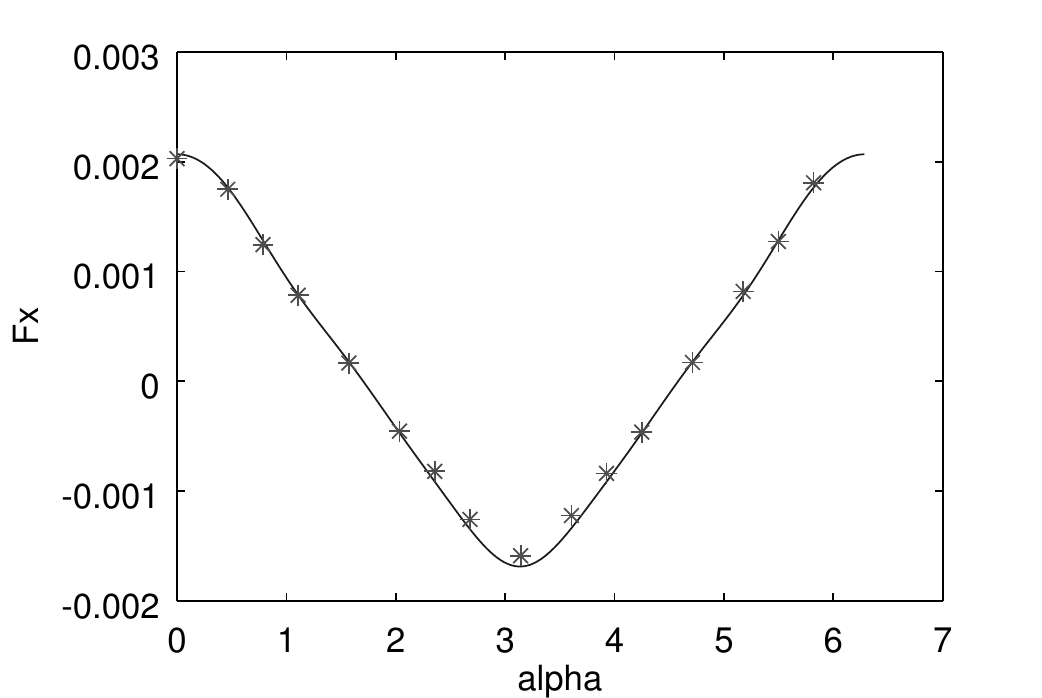}} a \end{minipage} \hfill
		\begin{minipage}[h]{0.49\linewidth} \center{\includegraphics[width=1\linewidth]{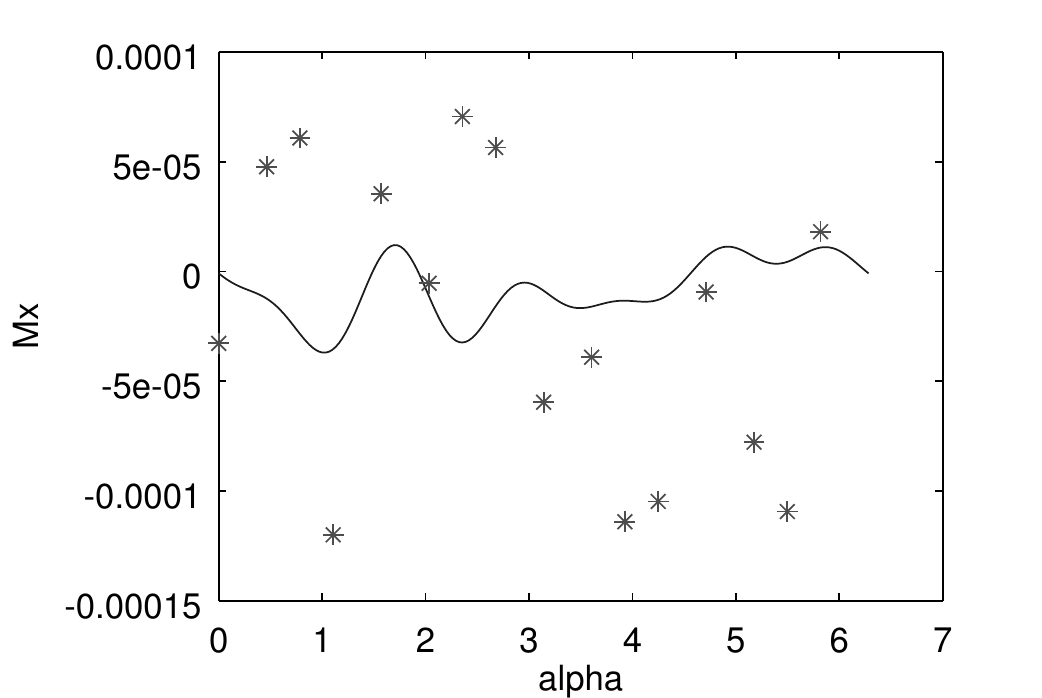}} b \end{minipage} \vfill
		\begin{minipage}[h]{0.49\linewidth} \center{\includegraphics[width=1\linewidth]{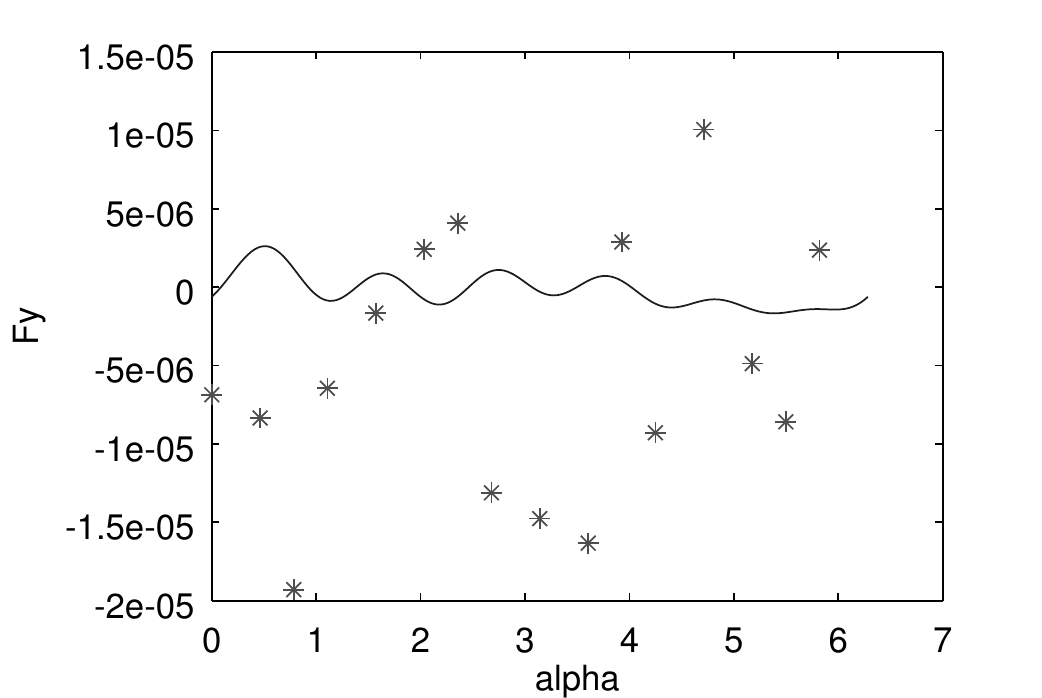}} c \end{minipage} \hfill
		\begin{minipage}[h]{0.49\linewidth} \center{\includegraphics[width=1\linewidth]{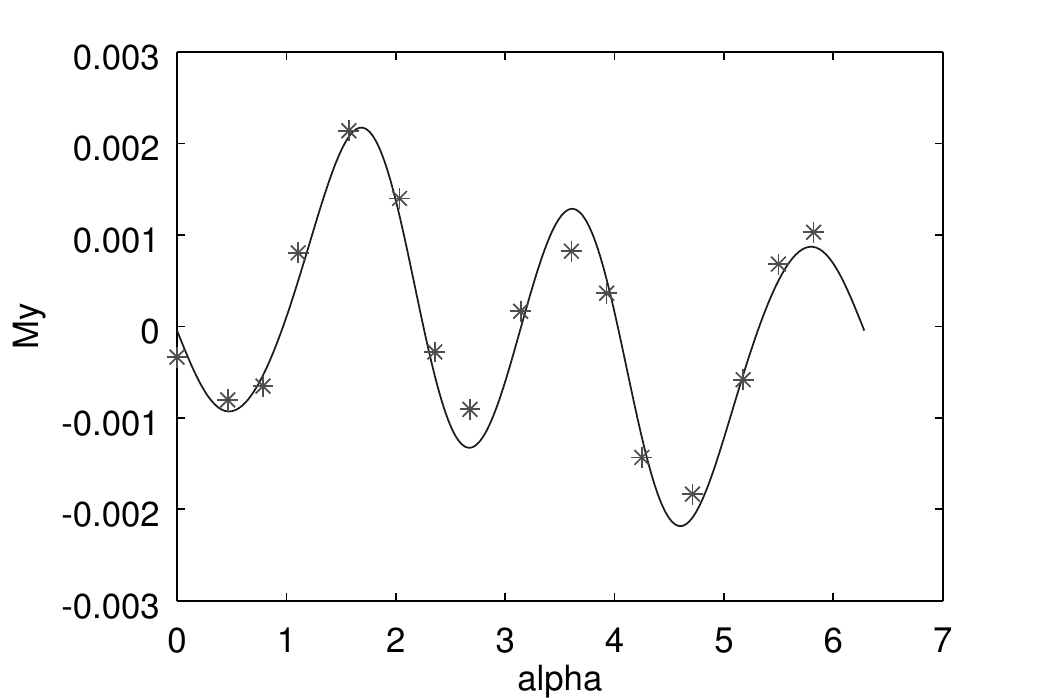}} d \end{minipage} \vfill
		\begin{minipage}[h]{0.49\linewidth} \center{\includegraphics[width=1\linewidth]{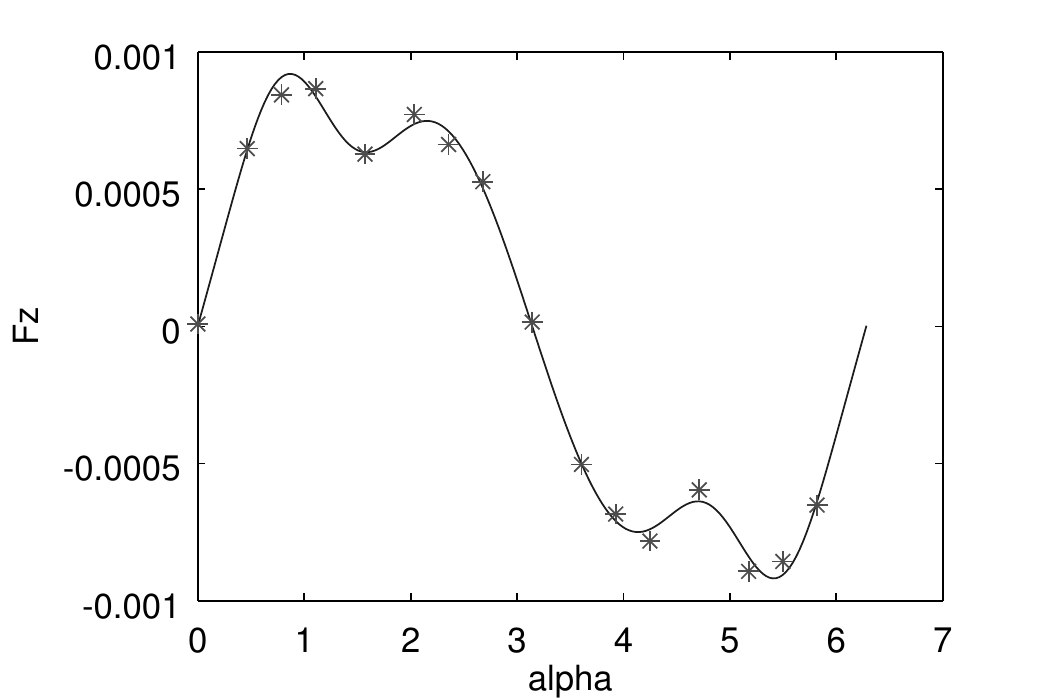}} e \end{minipage} \hfill
		\begin{minipage}[h]{0.49\linewidth} \center{\includegraphics[width=1\linewidth]{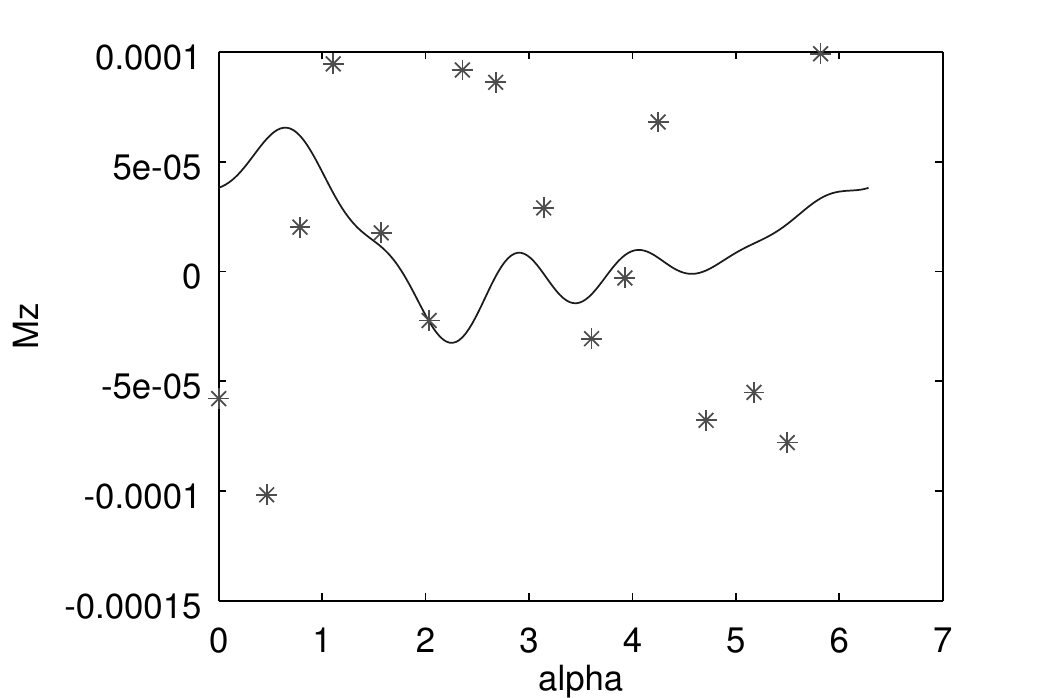}} f \end{minipage} 
		
		\caption{Approximation results ($N_{\max}=6$) for principal force (a, c, e), $N$, and for principal moment (b, c, f), $N\cdot m$, of light radiation pressure depending from angle of rotation of light source in the plane of radiators (solid line) comparing results of Monte--Carlo simulations (dots). The results in subfigures b, c and f are non-zero because of random noise. The approximation was constructed as a linear combination of specular and diffuse cases. Specular--diffuse case, $s=0{.}5$.} \label{fig:res_calc_050} \end{figure}

\begin{figure} \begin{minipage}[h]{0.49\linewidth} \center{\includegraphics[width=1\linewidth]{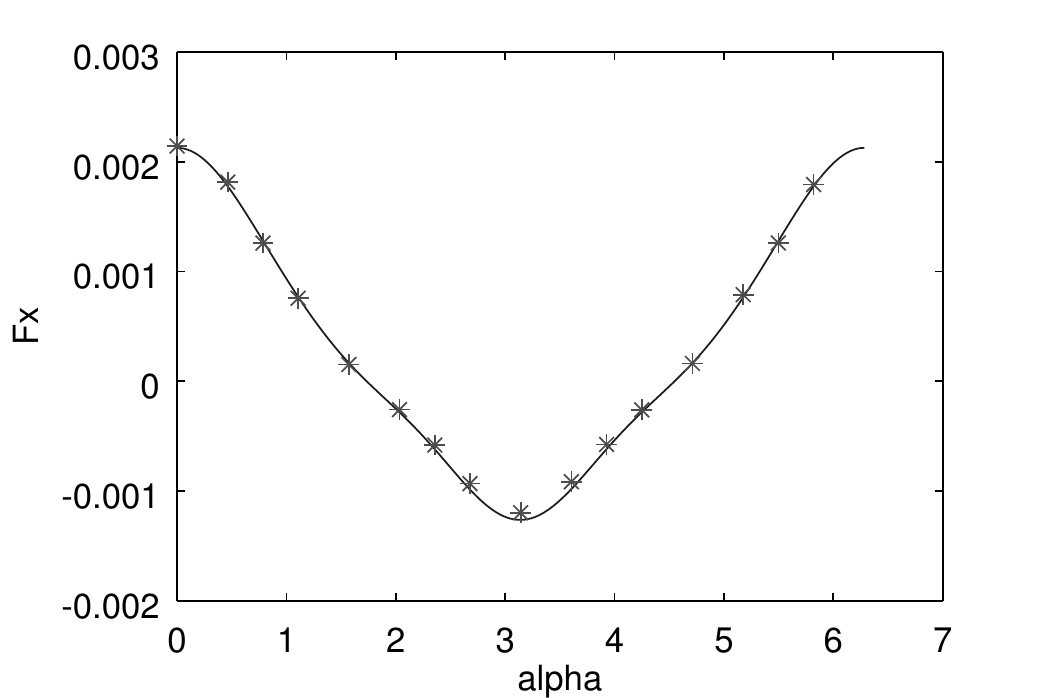}} a \end{minipage} \hfill
		\begin{minipage}[h]{0.49\linewidth} \center{\includegraphics[width=1\linewidth]{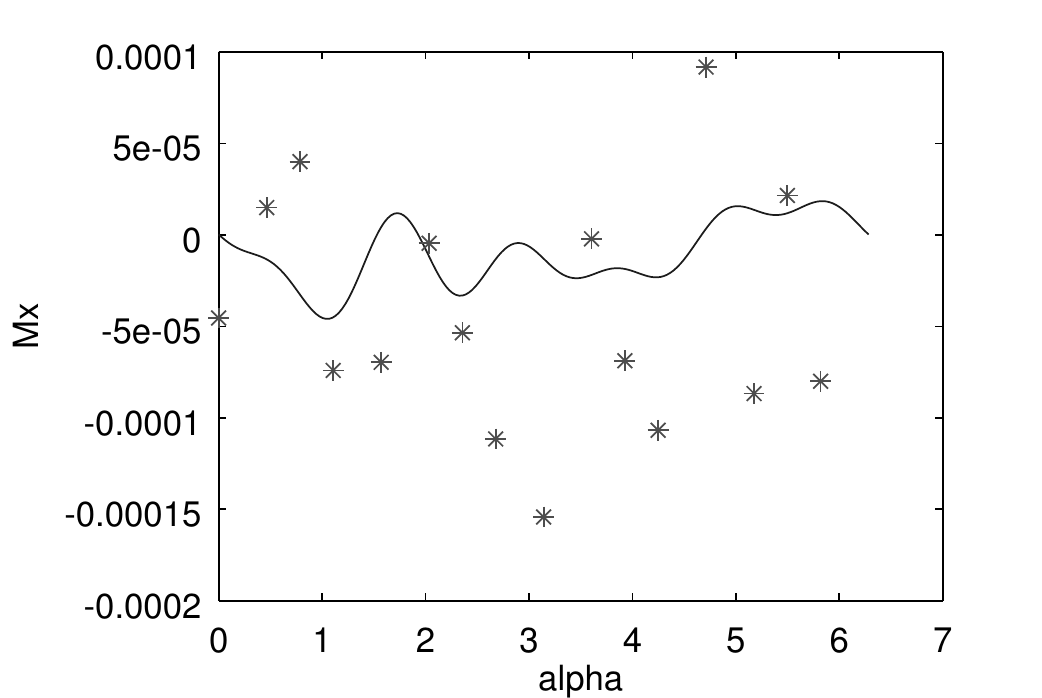}} b \end{minipage} \vfill
		\begin{minipage}[h]{0.49\linewidth} \center{\includegraphics[width=1\linewidth]{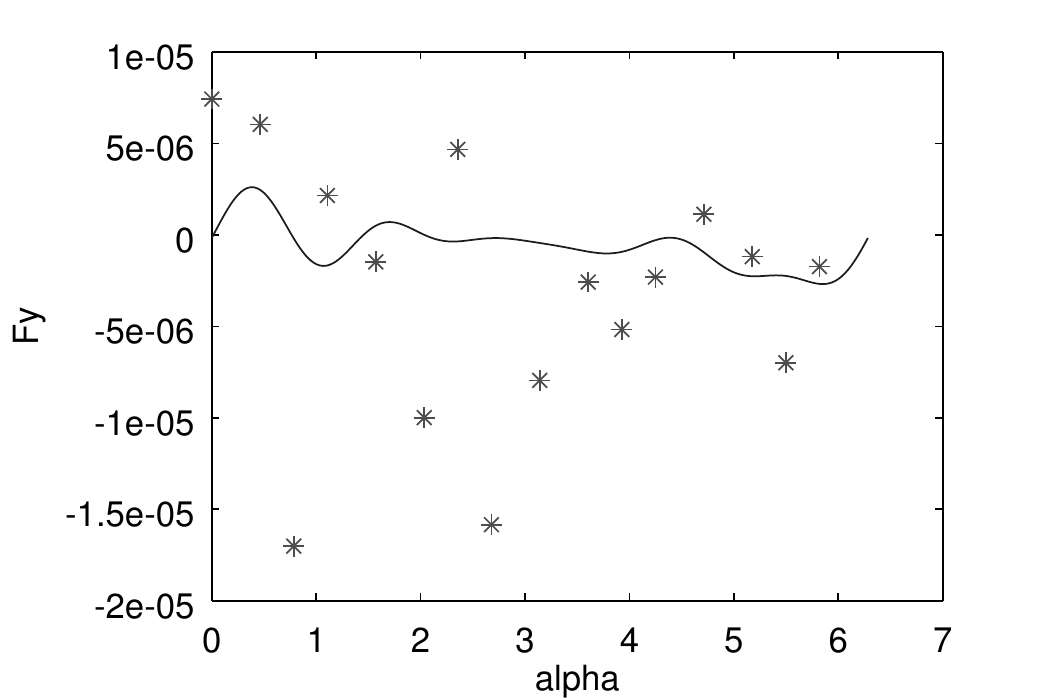}} c \end{minipage} \hfill
		\begin{minipage}[h]{0.49\linewidth} \center{\includegraphics[width=1\linewidth]{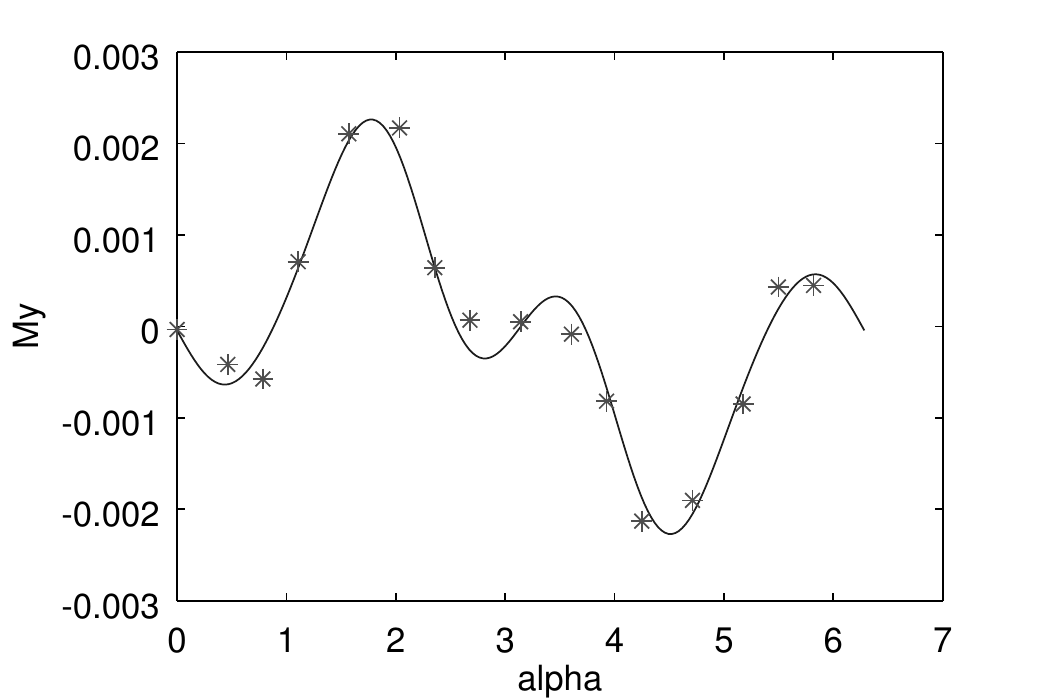}} d \end{minipage} \vfill
		\begin{minipage}[h]{0.49\linewidth} \center{\includegraphics[width=1\linewidth]{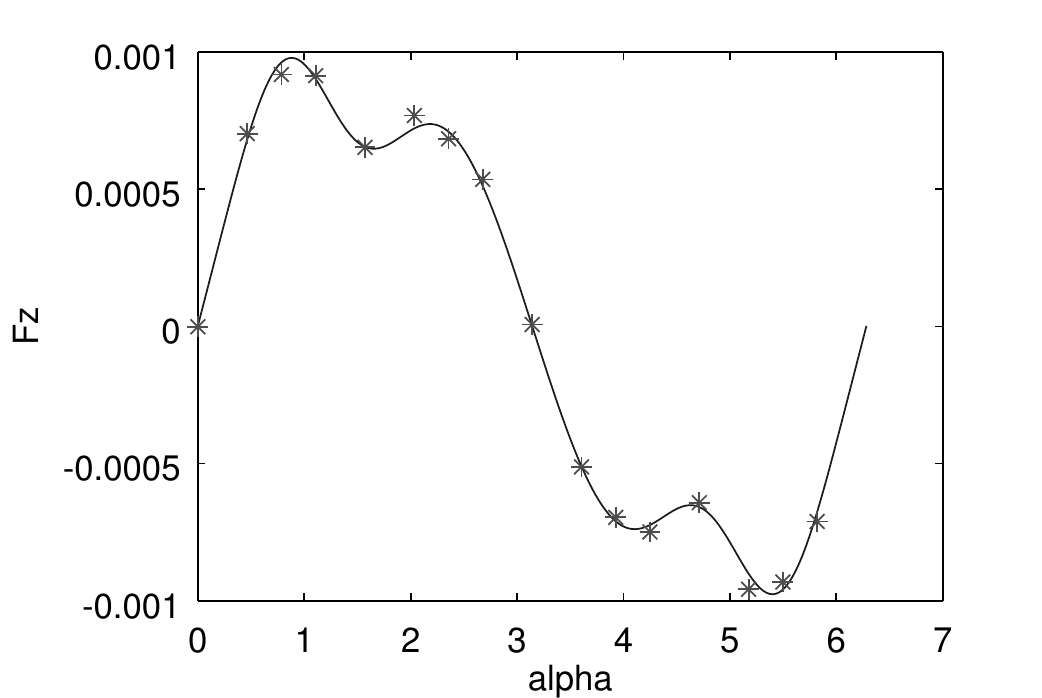}} e \end{minipage} \hfill
		\begin{minipage}[h]{0.49\linewidth} \center{\includegraphics[width=1\linewidth]{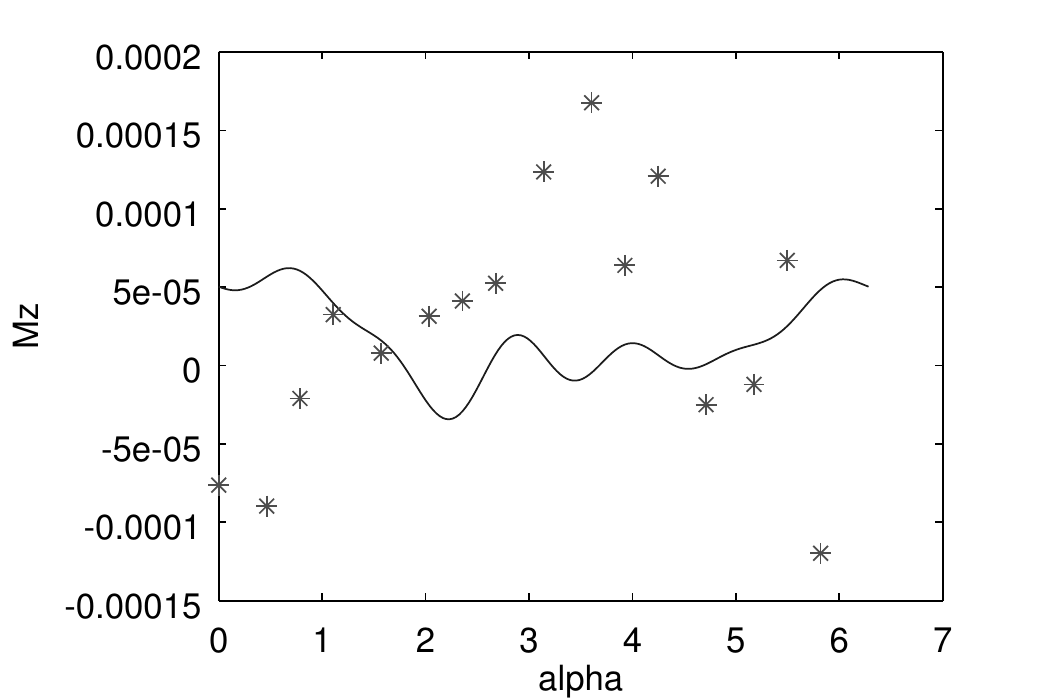}} f \end{minipage} 
		
		\caption{Approximation results ($N_{\max}=6$) for principal force (a, c, e), $N$, and for principal moment (b, c, f), $N\cdot m$, of light radiation pressure depending from angle of rotation of light source in the plane of radiators (solid line) comparing results of Monte--Carlo simulations (dots). The results in subfigures b, c and f are non-zero because of random noise. The approximation was constructed as a linear combination of specular and diffuse cases. Specular--diffuse case, $s=0{.}25$.} \label{fig:res_calc_025} \end{figure}

	Using the results of the calculation for specular and diffuse cases, we can obtain approximations for the following three specular--diffuse cases, namely $s=0{.}75$ (Fig.~\ref{fig:res_calc_075}), $s=0{.}5$ (Fig.~\ref{fig:res_calc_050}) moreover, $s=0{.}25$ (Fig.~\ref{fig:res_calc_025}), and compare them with the results from the raytracing The calculation results are presented in the figures below. As can be seen from Fig. \ref{fig:res_calc_075} --- \ref{fig:res_calc_025}, this method provides a good approximation comparing to the Monte - Carlo results, and it may be used in the calculation of the principal force and moment of the light pressure on the spacecraft with arbitrary shape, whose surface has specular--diffusive properties.

	\section{Conclusion} \label{sec:6} 
	
	We got the analytical expression for the radiation pressure force on a body with the arbitrary geometrical shape. The proposed method can be used to analyze the dynamics of the center of mass movement in and around the center of mass of space objects, which are significantly affected by the radiation pressure. The proposed method allows a parameterization of the properties of the body, at least by the degree of specularity. 
	
	\section{Acknowledgements} 
		
		Authors would like to thank assistant professor Marchevsky I.K. from department ``Applied Mathematics'' of BMSTU, assistant Kotsur O.S. from department ``Aerospace Systems'' of BMSTU and assistant Goncharov D.A. from department ``Theoretical Mechanics'' of BMSTU for their valuable advice and discussions which lead to the creation of explained method.
		
	
	%
	%
	\bibliographystyle{spbasic} 
	
	\bibliography{nerovny} 

\begin{thebibliography}{64}
\providecommand{\natexlab}[1]{#1}
\providecommand{\url}[1]{{#1}}
\providecommand{\urlprefix}{URL }
\expandafter\ifx\csname urlstyle\endcsname\relax
  \providecommand{\doi}[1]{DOI~\discretionary{}{}{}#1}\else
  \providecommand{\doi}{DOI~\discretionary{}{}{}\begingroup
  \urlstyle{rm}\Url}\fi
\providecommand{\eprint}[2][]{\url{#2}}

\bibitem[{bms(2014)}]{bmstu_knts}
 (2014) Experiment parus-mgtu (in russian).
  \urlprefix\url{http://knts.tsniimash.ru/ru/site/Experiment_q.aspx?idE=255}

\bibitem[{Alhorn et~al(2011)Alhorn, Casas, Agasid, Adams, Laue, Kitts, and
  O’Brien}]{alhorn_nanosail-d_2011}
Alhorn D, Casas J, Agasid E, Adams C, Laue G, Kitts C, O’Brien S (2011)
  {{NanoSail-D}}: {{The Small Satellite That Could}}! AIAA/USU Conference on
  Small Satellites
  \urlprefix\url{http://digitalcommons.usu.edu/smallsat/2011/all2011/37}

\bibitem[{Bar-Sever and Kuang(2004)}]{bar-sever_new_2004}
Bar-Sever Y, Kuang D (2004) New {{Empirically Derived Solar Radiation Pressure
  Model}} for {{Global Positioning}}. In: System {{Satellites}}”, {{IPN
  Progress Report}}, pp 42--159

\bibitem[{Bar-Sever and Russ(1997)}]{bar-sever_new_1997}
Bar-Sever YE, Russ KM (1997) New and {{Improved Solar Radiation Models}} for
  {{GPS Satellites Based}} on {{Flight Data}}. Tech. rep.

\bibitem[{Beekman(2006)}]{beekman_i.o._2006}
Beekman G (2006) I.{{O}}. {{Yarkovsky}} and the {{Discovery}} of'his'
  {{Effect}}. Journal for the History of Astronomy 37:71--86,
  \urlprefix\url{http://adsabs.harvard.edu/full/2006JHA....37...71B}

\bibitem[{Bottke et~al(2006)Bottke, Vokrouhlick{\'y}, Rubincam, and
  Nesvorn{\'y}}]{william_f._bottke_yarkovsky_2006}
Bottke WFJ, Vokrouhlick{\'y} D, Rubincam DP, Nesvorn{\'y} D (2006) The
  {{Yarkovsky}} and {{YORP}} effects: {{Implications}} for {{Asteroid
  Dynamics}}. Annual Review of Earth and Planetary Sciences 34(1):157--191,
  \doi{10.1146/annurev.earth.34.031405.125154},
  \urlprefix\url{http://dx.doi.org/10.1146/annurev.earth.34.031405.125154}

\bibitem[{Burns et~al(1979)Burns, Lamy, and Soter}]{burns_radiation_1979}
Burns JA, Lamy PL, Soter S (1979) Radiation forces on small particles in the
  solar system. Icarus 40(1):1--48, \doi{10.1016/0019-1035(79)90050-2},
  \urlprefix\url{http://linkinghub.elsevier.com/retrieve/pii/0019103579900502}

\bibitem[{Burns et~al(2014)Burns, Lamy, and Soter}]{burns_radiation_2014}
Burns JA, Lamy PL, Soter S (2014) Radiation forces on small particles in the
  {{Solar System}}: {{A}} re-consideration. Icarus 232:263--265,
  \urlprefix\url{http://www.sciencedirect.com/science/article/pii/S0019103514000104}

\bibitem[{Dobrovolskis(1996)}]{dobrovolskis_inertia_1996}
Dobrovolskis AR (1996) Inertia of {{Any Polyhedron}}. Icarus 124(2):698--704,
  \doi{10.1006/icar.1996.0243},
  \urlprefix\url{http://www.sciencedirect.com/science/article/pii/S0019103596902432}

\bibitem[{Eaton et~al(2015)Eaton, Bateman, Hauberg, and Wehbring}]{gnu_octave}
Eaton JW, Bateman D, Hauberg S, Wehbring R (2015) {{GNU Octave}} version 4.0.0
  manual: a high-level interactive language for numerical computations.
  \urlprefix\url{http://www.gnu.org/software/octave/doc/interpreter}

\bibitem[{Fliegel and Gallini(1996)}]{fliegel_solar_1996}
Fliegel HF, Gallini TE (1996) Solar force modeling of block {{IIR Global
  Positioning System}} satellites. Journal of Spacecraft and Rockets
  33(6):863--866, \doi{10.2514/3.26851},
  \urlprefix\url{http://dx.doi.org/10.2514/3.26851}

\bibitem[{Fliegel et~al(1992)Fliegel, Gallini, and Swift}]{fliegel_global_1992}
Fliegel HF, Gallini TE, Swift ER (1992) Global {{Positioning System Radiation
  Force Model}} for geodetic applications. Journal of Geophysical Research
  97(B1):559, \doi{10.1029/91JB02564},
  \urlprefix\url{http://doi.wiley.com/10.1029/91JB02564}

\bibitem[{Forward(1989)}]{forward_1989}
Forward R (1989) Grey solar sails. {American Institute of Aeronautics and
  Astronautics}, pp 1--12, \doi{10.2514/6.1989-2343},
  \urlprefix\url{http://arc.aiaa.org/doi/abs/10.2514/6.1989-2343}

\bibitem[{Hartmann et~al(1999)Hartmann, Farinella, Vokrouhlick{\'y},
  Weidenschilling, Morbidelli, Marzari, Davis, and
  Ryan}]{hartmann_reviewing_1999}
Hartmann WK, Farinella P, Vokrouhlick{\'y} D, Weidenschilling SJ, Morbidelli A,
  Marzari F, Davis DR, Ryan E (1999) Reviewing the {{Yarkovsky}} effect:
  {{New}} light on the delivery of stone and iron meteorites from the asteroid
  belt. Meteoritics \& Planetary Science 34(S4):A161--A167,
  \doi{10.1111/j.1945-5100.1999.tb01761.x},
  \urlprefix\url{http://doi.wiley.com/10.1111/j.1945-5100.1999.tb01761.x}

\bibitem[{Howell et~al(2015)Howell, Menguc, and Siegel}]{howell_thermal_2015}
Howell JR, Menguc MP, Siegel R (2015) Thermal {Radiation} {Heat} {Transfer},
  6th {Edition}, 6th edn. CRC Press

\bibitem[{Jing et~al(2012)Jing, ShengPing, and JunFeng}]{jing_curved_2012}
Jing H, ShengPing G, JunFeng L (2012) A curved surface solar radiation pressure
  force model for solar sail deformation. Science China Physics, Mechanics and
  Astronomy 55(1):141--155, \doi{10.1007/s11433-011-4575-7},
  \urlprefix\url{http://link.springer.com/10.1007/s11433-011-4575-7}

\bibitem[{Jing et~al(2014)Jing, Shengping, Junfeng, and
  Yufei}]{jing_solar_2014}
Jing H, Shengping G, Junfeng L, Yufei L (2014) The {Solar} {Radiation}
  {Pressure} {Force} {Models} for a {General} {Sail} {Surface} {Shape}. In:
  Macdonald M (ed) Advances in Solar Sailing, Springer {Praxis} {Books},
  Springer Berlin Heidelberg, pp 469--488,
  \urlprefix\url{http://link.springer.com/chapter/10.1007/978-3-642-34907-2_30}

\bibitem[{Johnson et~al(2011)Johnson, Whorton, Heaton, Pinson, Laue, and
  Adams}]{johnson_nanosail-d:_2011}
Johnson L, Whorton M, Heaton A, Pinson R, Laue G, Adams C (2011)
  {{NanoSail-D}}: {{A}} solar sail demonstration mission. Acta Astronautica
  68(5–6):571--575, \doi{10.1016/j.actaastro.2010.02.008},
  \urlprefix\url{http://www.sciencedirect.com/science/article/pii/S0094576510000597}

\bibitem[{Kardashev(1997)}]{kardashev_radioastron_1997}
Kardashev NS (1997) Radioastron - a {{Radio Telescope Much Greater}} than the
  {{Earth}}. Experimental Astronomy 7(4):329--343,
  \doi{10.1023/A:1007937203880},
  \urlprefix\url{http://link.springer.com/article/10.1023/A:1007937203880}

\bibitem[{Kardashev et~al(2000)Kardashev, Andreyanov, and
  Buyakas}]{kardashev_2000}
Kardashev NS, Andreyanov VV, Buyakas VI (2000) Project “{{Millimetron}}”
  (in russian). Proceedings of PN Lebedev Physical Institute 228:112--128

\bibitem[{Katasev and Kulikova(1981)}]{katasev_physical_1981}
Katasev LA, Kulikova NV (1981) Physical and mathematical modeling of the
  formation and evolution of meteor streams. {{II}}. Astronomicheskii Vestnik
  14:179--183,
  \urlprefix\url{http://adsabs.harvard.edu/abs/1981AVest..14..225K}

\bibitem[{Kawaguchi(2014)}]{ikaros1}
Kawaguchi J (2014) An {{Overview}} of {{Solar Sail Related Activities}} at
  {{JAXA}}. In: Macdonald M (ed) Advances in {{Solar Sailing}}, Springer Praxis
  Books, {Springer Berlin Heidelberg}, pp 3--14,
  \urlprefix\url{http://link.springer.com/chapter/10.1007/978-3-642-34907-2_1}

\bibitem[{Kinzel(2005)}]{kinzel_managing_2005}
Kinzel WM (2005) Managing {{Angular Momentum Accumulation}} by {{Visit
  Sequencing}} and {{Visit Date}}: {{Roll Selection}}. Technical Report
  JWST-STScI-000713, SM-12, JWST Science and Operations Center Configuration
  Management Office

\bibitem[{Kristensen et~al(2005)Kristensen, Akenine-M{\"o}ller, and
  Jensen}]{kristensen_precomputed_2005}
Kristensen AW, Akenine-M{\"o}ller T, Jensen HW (2005) Precomputed local
  radiance transfer for real-time lighting design. In: ACM Transactions on
  Graphics (TOG), ACM, vol~24, pp 1208--1215,
  \urlprefix\url{http://dl.acm.org/citation.cfm?id=1073334}

\bibitem[{Lebedew(1901)}]{lebedev_1901}
Lebedew P (1901) Untersuchungen {\"u}ber die druckkr{\"a}fte des lichtes.
  Annalen der Physik 311(11):433--458, \doi{10.1002/andp.19013111102}

\bibitem[{Leonov(2012)}]{leonov_2012_1}
Leonov VV (2012) Radiation heat transfer in mirror concentrator systems (in
  Russian). {LAP LAMBERT Academic Publishing GmbH \& Co. KG}

\bibitem[{Liu et~al(2004)Liu, Sloan, Shum, and Snyder}]{liu_all-frequency_2004}
Liu X, Sloan PPJ, Shum HY, Snyder J (2004) All-{Frequency} {Precomputed}
  {Radiance} {Transfer} for {Glossy} {Objects}. Rendering Techniques
  \urlprefix\url{http://research.microsoft.com/en-us/um/people/johnsny/papers/allfreq.pdf}

\bibitem[{M{\"a}ki-Patola(2003)}]{maki-patola_precomputed_2003}
M{\"a}ki-Patola T (2003) Precomputed radiance transfer. Tik-111500 Seminar on
  computer graphics \urlprefix\url{http://www.mcpatola.com/prtf_fixed.pdf}

\bibitem[{Maxwell(1873)}]{maxwell_1873}
Maxwell JC (1873) A treatise on electricity and magnetism, vol~2. Clarendon
  Press, Oxford

\bibitem[{McInnes(2004)}]{mcinnes_solar_2004}
McInnes CR (2004) Solar Sailing: Technology, Dynamics and Mission Applications.
  Springer Science \& Business Media

\bibitem[{McMahon and Scheeres(2014)}]{mcmahon_general_2014}
McMahon J, Scheeres DJ (2014) General {Solar} {Radiation} {Pressure} {Model}
  for {Global} {Positioning} {System} {Orbit} {Determination}. Journal of
  Guidance, Control, and Dynamics 37(1):325--330, \doi{10.2514/1.61113},
  \urlprefix\url{http://arc.aiaa.org/doi/abs/10.2514/1.61113}

\bibitem[{McMahon and Scheeres(2010)}]{mcmahon_new_2010}
McMahon JW, Scheeres DJ (2010) New {Solar} {Radiation} {Pressure} {Force}
  {Model} for {Navigation}. Journal of Guidance, Control, and Dynamics
  33(5):1418--1428, \doi{10.2514/1.48434},
  \urlprefix\url{http://arc.aiaa.org/doi/abs/10.2514/1.48434}

\bibitem[{McMahon and Scheeres(2015)}]{mcmahon_improving_2015}
McMahon JW, Scheeres DJ (2015) Improving {Space} {Object} {Catalog}
  {Maintenance} {Through} {Advances} in {Solar} {Radiation} {Pressure}
  {Modeling}. Journal of Guidance, Control, and Dynamics pp 1--16,
  \doi{10.2514/1.G000666},
  \urlprefix\url{http://arc.aiaa.org/doi/10.2514/1.G000666}

\bibitem[{Neiman et~al(1965)Neiman, Romanov, and Chernov}]{neiman_ivan_1965}
Neiman VB, Romanov EM, Chernov VM (1965) Ivan {{Osipovich Yarkovsky}}. Earth
  Univ 4:63--64

\bibitem[{Nerovnyi and Zimin(2014)}]{Nerovny_Zimin_2014}
Nerovnyi N, Zimin V (2014) Determination of the radiation pressure force acting
  on a solar sail taking into account stress-dependent optical parameters of
  sail material (in russian). Herald of the Bauman Moscow State Technical
  University Series Mechanical Engineering 96(3):61--78,
  \urlprefix\url{http://vestnikmach.ru/eng/catalog/simul/hidden/486.html}

\bibitem[{{\"O}pik(1951)}]{opik_collision_1951}
{\"O}pik EJ (1951) Collision probabilities with the planets and the
  distribution of interplanetary matter. In: Proceedings of the {{Royal Irish
  Academy}}. {{Section A}}: {{Mathematical}} and {{Physical Sciences}},
  {JSTOR}, vol~54, pp 165--199,
  \urlprefix\url{http://www.jstor.org/stable/20488532}

\bibitem[{Paddack(1969)}]{paddack_rotational_1969}
Paddack SJ (1969) Rotational bursting of small celestial bodies: {{Effects}} of
  radiation pressure. Journal of Geophysical Research 74(17):4379--4381,
  \doi{10.1029/JB074i017p04379},
  \urlprefix\url{http://onlinelibrary.wiley.com/doi/10.1029/JB074i017p04379/abstract}

\bibitem[{Polyakhova(2011)}]{polyakhova_2011}
Polyakhova EN (2011) Space flight with solar sail (in Russian), 2nd edn.
  {URSS}, Moscow

\bibitem[{Rachkin et~al(2011)Rachkin, Tenenbaum, Dmitriev, Nerovnyy, Kotsur,
  and Vorobyov}]{bmstu_iac_2011}
Rachkin D, Tenenbaum S, Dmitriev A, Nerovnyy N, Kotsur O, Vorobyov A (2011)
  2-blades deploying by centrifugal force solar sail experiment
  ({{IAC}}-11,{{E2}},3,8,x9437). In: Proceedings of 62nd {{International
  Astronautical Congress}}, Cape Town, SA, pp 9128--9142

\bibitem[{Radzievskii(1952)}]{radzievskii_about_1952}
Radzievskii VV (1952) About the influence of the anisotropically reemited solar
  radiation on the orbits of asteroids and meteoroids. Astron Zh 29:162--170

\bibitem[{Radzievskii(1954)}]{radzievskii_mechanism_1954}
Radzievskii VV (1954) A mechanism for the disintegration of asteroids and
  meteorites. Dokl Akad Nauk SSSR 97:49--52

\bibitem[{Raykunov et~al(2009)Raykunov, Komkov, Melnikov, and
  Kharlov}]{raikunov_2009}
Raykunov GG, Komkov VA, Melnikov VM, Kharlov BN (2009) Tsentrobezhnye
  beskarkasnye krupnogabaritnye kosmicheskie konstruktsii (in Russian,
  Centrifugal frameless large space structures). {ANO Fizmatlit}, Moscow

\bibitem[{Ridenoure et~al(2015)Ridenoure, Munakata, Diaz, Wong, Plante,
  Stetson, Spencer, and Foley}]{ridenoure2015lightsail}
Ridenoure R, Munakata R, Diaz A, Wong S, Plante B, Stetson D, Spencer D, Foley
  J (2015) {{LightSail Program Status}}: {{One Down}}, {{One}} to {{Go}}.
  AIAA/USU Conference on Small Satellites
  \urlprefix\url{http://digitalcommons.usu.edu/smallsat/2015/all2015/32}

\bibitem[{Rios-Reyes(2006)}]{rios-reyes_2006}
Rios-Reyes L (2006) Solar sails: Modeling, estimation, and trajectory control.
  PhD thesis, University of Michigan

\bibitem[{Rios-Reyes and Scheeres(2004)}]{rios-reyes_applications_2004}
Rios-Reyes L, Scheeres DJ (2004) Applications of the generalized model for
  solar sails. In: {AIAA} {Guidance}, {Navigation}, and {Control} {Conference}
  and {Exhibit},
  \urlprefix\url{http://arc.aiaa.org/doi/pdf/10.2514/6.2004-5434}

\bibitem[{Rios-Reyes and Scheeres(2005)}]{rios_reyes_generalized_2005}
Rios-Reyes L, Scheeres DJ (2005) Generalized {Model} for {Solar} {Sails}.
  Journal of Spacecraft and Rockets 42(1):182--185, \doi{10.2514/1.9054},
  \urlprefix\url{http://arc.aiaa.org/doi/abs/10.2514/1.9054}

\bibitem[{Rios-Reyes and Scheeres(2007)}]{rios-reyes_solar-sail_2007}
Rios-Reyes L, Scheeres DJ (2007) Solar-{Sail} {Navigation}: {Estimation} of
  {Force}, {Moments}, and {Optical} {Parameters}. Journal of Guidance, Control,
  and Dynamics 30(3):660--668, \doi{10.2514/1.24340},
  \urlprefix\url{http://arc.aiaa.org/doi/abs/10.2514/1.24340}

\bibitem[{Rodriguez-Solano et~al(2012)Rodriguez-Solano, Hugentobler, and
  Steigenberger}]{rodriguez-solano_adjustable_2012}
Rodriguez-Solano CJ, Hugentobler U, Steigenberger P (2012) Adjustable box-wing
  model for solar radiation pressure impacting {{GPS}} satellites. Advances in
  Space Research 49(7):1113--1128, \doi{10.1016/j.asr.2012.01.016},
  \urlprefix\url{http://www.sciencedirect.com/science/article/pii/S0273117712000579}

\bibitem[{Rubincam(2000)}]{rubincam_radiative_2000}
Rubincam D (2000) Radiative {{Spin}}-up and {{Spin}}-down of {{Small
  Asteroids}}. Icarus 148(1):2--11, \doi{10.1006/icar.2000.6485},
  \urlprefix\url{http://linkinghub.elsevier.com/retrieve/doi/10.1006/icar.2000.6485}

\bibitem[{Rubincam(1995)}]{rubincam_asteroid_1995}
Rubincam DP (1995) Asteroid orbit evolution due to thermal drag. Journal of
  Geophysical Research: Planets 100(E1):1585--1594,
  \urlprefix\url{http://onlinelibrary.wiley.com/doi/10.1029/94JE02411/full}

\bibitem[{Scheeres(2007)}]{scheeres_dynamical_2007}
Scheeres DJ (2007) The dynamical evolution of uniformly rotating asteroids
  subject to {YORP}. Icarus 188(2):430--450,
  \doi{10.1016/j.icarus.2006.12.015},
  \urlprefix\url{http://www.sciencedirect.com/science/article/pii/S0019103506004441}

\bibitem[{Simonelli et~al(1993)Simonelli, Thomas, Carcich, and
  Veverka}]{simonelli_generation_1993}
Simonelli DP, Thomas PC, Carcich BT, Veverka J (1993) The {{Generation}} and
  {{Use}} of {{Numerical Shape Models}} for {{Irregular Solar System Objects}}.
  Icarus 103(1):49--61, \doi{10.1006/icar.1993.1057},
  \urlprefix\url{http://www.sciencedirect.com/science/article/pii/S0019103583710572}

\bibitem[{Sloan et~al(2002)Sloan, Kautz, and Snyder}]{sloan_precomputed_2002}
Sloan PP, Kautz J, Snyder J (2002) Precomputed radiance transfer for real-time
  rendering in dynamic, low-frequency lighting environments. In: ACM
  Transactions on Graphics (TOG), ACM, vol~21, pp 527--536,
  \urlprefix\url{http://dl.acm.org/citation.cfm?id=566612}

\bibitem[{Springer et~al(1999)Springer, Beutler, and
  Rothacher}]{springer_new_1999}
Springer TA, Beutler G, Rothacher M (1999) A {{New Solar Radiation Pressure
  Model}} for {{GPS Satellites}}. GPS Solutions 2(3):50--62,
  \doi{10.1007/PL00012757},
  \urlprefix\url{http://link.springer.com/article/10.1007/PL00012757}

\bibitem[{Tsai and Shih(2006)}]{tsai_all-frequency_2006}
Tsai YT, Shih ZC (2006) All-frequency precomputed radiance transfer using
  spherical radial basis functions and clustered tensor approximation. In: ACM
  Transactions on Graphics (TOG), ACM, vol~25, pp 967--976,
  \urlprefix\url{http://dl.acm.org/citation.cfm?id=1141981}

\bibitem[{Tsander(1969)}]{tsander}
Tsander FA (1969) From a scientific heritage. Technical translation by NASA,
  Washington D.C.: NASA.

\bibitem[{Tsuda et~al(2011)Tsuda, Mori, Funase, Sawada, Yamamoto, Saiki, Endo,
  and Kawaguchi}]{tsuda_flight_2011}
Tsuda Y, Mori O, Funase R, Sawada H, Yamamoto T, Saiki T, Endo T, Kawaguchi J
  (2011) Flight status of {{IKAROS}} deep space solar sail demonstrator. Acta
  Astronautica 69(9–10):833--840, \doi{10.1016/j.actaastro.2011.06.005},
  \urlprefix\url{http://www.sciencedirect.com/science/article/pii/S0094576511001822}

\bibitem[{Vokrouhlick{\'y}(1999)}]{vokrouhlicky_complete_1999}
Vokrouhlick{\'y} D (1999) A complete linear model for the {{Yarkovsky}} thermal
  force on spherical asteroid fragments. Astronomy and Astrophysics
  344:362--366,
  \urlprefix\url{http://adsabs.harvard.edu/full/1999A\%26A...344..362V}

\bibitem[{Vokrouhlick{\'y} and {\v
  C}apek(2002)}]{vokrouhlicky_yorp-induced_2002}
Vokrouhlick{\'y} D, {\v C}apek D (2002) {{YORP}}-{{Induced Long}}-{{Term
  Evolution}} of the {{Spin State}} of {{Small Asteroids}} and {{Meteoroids}}:
  {{Rubincam}}'s {{Approximation}}. Icarus 159(2):449--467,
  \doi{10.1006/icar.2002.6918},
  \urlprefix\url{http://www.sciencedirect.com/science/article/pii/S0019103502969186}

\bibitem[{Vokrouhlick{\'y} and Farinella(1998)}]{vokrouhlicky_yarkovsky_1998}
Vokrouhlick{\'y} D, Farinella P (1998) The {{Yarkovsky}} seasonal effect on
  asteroidal fragments: {{A}} nonlinearized theory for the plane-parallel case.
  The Astronomical Journal 116(4):2032,
  \urlprefix\url{http://iopscience.iop.org/article/10.1086/300565/meta}

\bibitem[{Wild et~al(2008)Wild, Kardashev, Likhachev, Babakin, Arkhipov,
  Vinogradov, Andreyanov, Fedorchuk, Myshonkova, Alexsandrov, Novokov,
  Goltsman, Cherepaschuk, Shustov, Vystavkin, Koshelets, Vdovin, Graauw,
  Helmich, Tak, Shipman, Baryshev, Gao, Khosropanah, Roelfsema, Barthel,
  Spaans, Mendez, Klapwijk, Israel, Hogerheijde, Werf, Cernicharo,
  Martin-Pintado, Planesas, Gallego, Beaudin, Krieg, Gerin, Pagani, Saraceno,
  Giorgio, Cerulli, Orfei, Spinoglio, Piazzo, Liseau, Belitsky, Cherednichenko,
  Poglitsch, Raab, Guesten, Klein, Stutzki, Honingh, Benz, Murphy, Trappe, and
  Räisänen}]{wild_millimetronlarge_2008}
Wild W, Kardashev NS, Likhachev SF, Babakin NG, Arkhipov VY, Vinogradov IS,
  Andreyanov VV, Fedorchuk SD, Myshonkova NV, Alexsandrov YA, Novokov ID,
  Goltsman GN, Cherepaschuk AM, Shustov BM, Vystavkin AN, Koshelets VP, Vdovin
  VF, Graauw Td, Helmich F, Tak Fv, Shipman R, Baryshev A, Gao JR, Khosropanah
  P, Roelfsema P, Barthel P, Spaans M, Mendez M, Klapwijk T, Israel F,
  Hogerheijde M, Werf Pv, Cernicharo J, Martin-Pintado J, Planesas P, Gallego
  JD, Beaudin G, Krieg JM, Gerin M, Pagani L, Saraceno P, Giorgio AMD, Cerulli
  R, Orfei R, Spinoglio L, Piazzo L, Liseau R, Belitsky V, Cherednichenko S,
  Poglitsch A, Raab W, Guesten R, Klein B, Stutzki J, Honingh N, Benz A, Murphy
  A, Trappe N, Räisänen A (2008) Millimetron -- a large {{Russian-European}}
  submillimeter space observatory. Experimental Astronomy 23(1):221--244,
  \doi{10.1007/s10686-008-9097-6},
  \urlprefix\url{http://link.springer.com/article/10.1007/s10686-008-9097-6}

\bibitem[{Ziebart(2004)}]{ziebart_generalized_2004}
Ziebart M (2004) Generalized {{Analytical Solar Radiation Pressure Modeling
  Algorithm}} for {{Spacecraft}} of {{Complex Shape}}. Journal of Spacecraft
  and Rockets 41(5):840--848, \doi{10.2514/1.13097},
  \urlprefix\url{http://arc.aiaa.org/doi/abs/10.2514/1.13097}

\bibitem[{Zimin and Nerovnyi(2016)}]{Zimin_Nerovny_2016_bmstu}
Zimin VN, Nerovnyi NA (2016) To the calculation of the main vector and the main
  momentum of light pressure force on a solar sail (in russian). Herald of the
  Bauman Moscow State Technical University Series Mechanical Engineering
  106(1):17--28, \doi{10.18698/0236-3941-2016-1-17-28},
  \urlprefix\url{http://vestnikmach.ru/eng/catalog/avroc/airdyn/1055.html}

\bibitem[{Zimin and Nerovnyy(2015)}]{Zimin_Nerovny_2015_izvuz}
Zimin VN, Nerovnyy NA (2015) Analysis of the deformed shape of a heliogyro
  solar sail blade taking into account stress-dependent reflectivity of the
  material (in russian). Proceedings of Higher Educational Institutions
  Маchine Building 658(1):18--23,
  \urlprefix\url{http://izvuzmash.ru/eng/catalog/calcmach/hidden/1125.html}

\end{thebibliography}
	
	\newpage
	
\end{document}